\documentclass[12pt]{article}
\usepackage{amsfonts}
\usepackage[dvips]{graphicx}
\usepackage{latexsym,amsmath,amssymb,graphics,stmaryrd}
\usepackage{cite}
\usepackage{array}
\usepackage{subfigure}
\usepackage{multirow}
\usepackage{epsfig}
\usepackage[dvips]{graphicx}
\usepackage[dvips]{color}
\usepackage{pstricks}
\usepackage{pst-node}
\usepackage{pst-plot}
\usepackage{dsfont}
\usepackage{amsthm}
\usepackage{graphics}
\usepackage{fancyhdr}
\usepackage[dvips]{color}

\newcommand{\col}{~,}
\newcommand{\pnt}{~.}
\newcommand{\AdS}{\text{AdS}}

\newcommand{\twob}{{\text{II}\,\text{B}}}
\newcommand{\DBI}{\mathrm{DBI}}
\newcommand{\CS}{\mathrm{CS}}

\newcommand{\preparderiv}[1]{\frac{\partial}{\partial #1}}

\newcommand{\de}{\operatorname{d}\!}
\newcommand{\e}{\operatorname{e}}
\newlength{\neglength}
\newlength{\diameter}
\newcommand{\olett}[1]{
  \ensuremath{
  \text{$\settowidth{\diameter}{$\bigcirc$}%
  \bigcirc%
  \hspace{-0.5\diameter}%
  \makebox[0pt][c]{$\scriptstyle{#1}$}%
  \hspace{0.5\diameter}$}}}    
\DeclareMathOperator{\tr}{tr}

\DeclareMathOperator{\re}{Re}
\DeclareMathOperator{\im}{Im}
\DeclareMathOperator{\vol}{vol}
\numberwithin{equation}{section}
\newlength{\unit}
\psset{xunit=\unit,yunit=\unit,runit=\unit}
\newlength{\linew}
\psset{linewidth=\linew}
\begin{document}
\begin{titlepage}
\begin{flushright}
IFUM 891-FT\\
\end{flushright}
%\mbox{ }  \hfill hep-th/yymmnnn
\vspace{5ex}
\Large
\begin {center}     
{\bf Holographic flavour in the $\mathcal{N}=1$ Polchinski-Strassler 
background
}
\end {center}
\large
\vspace{1ex}
\begin{center}
Christoph Sieg\footnote{\label{mi} csieg@mi.infn.it} 
\end{center}
\begin{center}
%\ref{mi}
Universit\`a degli Studi di Milano \\
Via Celoria 16, 
20133 Milano, 
Italy \\[2mm]
\end{center}
\vspace{4ex}
\rm
\abstract
%\begin{center}
%{\bf Abstract}
To endow the $\mathcal{N}=1^\star$ SYM theory with quarks, we embed 
$\text{D}7$-brane probes into its gravity dual, known as the 
Polchinski-Strassler background. The non-vanishing $3$-form 
flux $G_3$ in the background 
is dual to mass terms for the three 
adjoint chiral superfields, deforming the 
$\mathcal{N}=4$ SYM theory to the $\mathcal{N}=1^\star$ SYM theory. 
%We construct the gravity dual of the $\mathcal{N}=1^\star$ SYM
%theory with quarks by embedding $\text{D}7$-brane probes into the 
%Polchinski-Strassler background. 
We keep its three mass parameters independent. This generalizes our analysis 
in hep-th/0610276 for the $\mathcal{N}=2^\star$ SYM theory. 
%which is obtained
%when two masses are identical and one is vanishing.
We work at second order in the mass perturbation, i.e. 
$G_3$ and its backreaction on the
background are considered perturbatively up to this order.  
%At order $\mathcal{O}(m^2)$ 
We find analytic solutions for the embeddings
which in general depend also on angular variables. 
%In detail we treat special embeddings in $\mathcal{N}=2^\star$ SYM
%and in $\mathcal{N}=1^\star$ SYM with three equal masses.
%We generalize the analysis of $\text{D}7$-brane probe embeddings 
%into the $\mathcal{N}=2$ Polchinski-Strassler background in hep-th/0610276
%to embeddings into a whole class of backgrounds that also include 
%the $\mathcal{N}=1$ case with three arbitary mass deformation parameters. 
%We reexamine the computation of the gauge field on 
%the probe brane, and find that in all cases it can be consistently set to zero
%in our approximation. We derive analytic solutions for the generalized 
%embeddings that depend also on the angular coordinates. 
%The general analysis shows the peculiarity of the Polchinski-Strassler 
%case in which the radial embedding coordinate remains a function of 
%the worldvolume radial coordinate only. 
We discuss the properties of the solutions and give error estimates on our 
approximation. By applying the method of holographic renormalization,
we show that in all cases the embeddings are at least 
consistent with supersymmetry.
%\end{center} 
\normalsize 

\vfill
\end{titlepage} 
\tableofcontents

\pagebreak

\section{Introduction}

The $\AdS/\text{CFT}$ correspondence \cite{Maldacena:1997re}
allows us to study $\mathcal{N}=4$ supersymmetric Yang-Mills (SYM) theory 
with gauge group $SU(N)$ in the large $N$ limit and at 
large 't Hooft coupling constant $\lambda=g_\text{YM}^2N$ by 
analyzing its conjectured gravity dual, given by 
type $\twob$ supergravity in 
$\AdS_5\times\text{S}^5$ with $N$ units of Ramond-Ramond $5$-form flux. 

The correspondence has been extended to cases in which the dual gauge theory
is not maximally supersymmetric and conformal, but preserves less
supersymmetries and is confining.
Several backgrounds 
for the dual gravity description of such gauge theories have 
been proposed e.g.\ in  
\cite{Witten:1998zw,Klebanov:2000hb,Constable:1999ch,Maldacena:2000yy}. 

A particularly interesting example for a supergravity 
background was discussed by
Polchinski and Strassler \cite{Polchinski:2000uf}.
It is based on the observation that from the $\mathcal{N}=4$ SYM 
theory one can obtain a confining gauge theory with 
less supersymmetry by adding mass terms for the three adjoint
chiral $\mathcal{N}=1$ supermultiplets.
In the dual gravity description this mass perturbation corresponds 
to certain non-vanishing $3$-form flux components.
In the underlying brane picture this flux polarizes the background
generating $\text{D}3$-branes due to the Myers effect \cite{Myers:1999ps} into
their transverse directions in which they extend to rotational ellipsoids 
\cite{Apreda:2006bu}. The full effective description of the brane 
configuration is not known. 
The Polchinski-Strassler background which should be obtained as its
near horizon limit therefore is not given in a closed form. 
At sufficiently large distance from the extended $\text{D}3$-brane sources 
the near horizon limit of the configuration is given as perturbative 
expansion around $\AdS_5\times\text{S}^5$. 
The $3$-form flux is considered as a perturbation. Its backreaction 
on the geometry corrects the background order by order 
in the mass parameters 
\cite{Polchinski:2000uf,Freedman:2000xb,LopesCardoso:2004ni}. 

Generically the dual gauge theory is the so called $\mathcal{N}=1^*$ theory.
Its special case when all three masses are identical has been discussed 
by Polchinski and Strassler \cite{Polchinski:2000uf}. 
%The resulting dual gauge theory depends on the concrete choice of the 
%three mass parameters. In the generic case one obtains the so called 
%$\mathcal{N}=1^*$ theory for which Polchinski and Strassler 
%discussed the special case when all three masses are identical.
If instead two masses are identical and non-vanishing, while the third one 
is zero, the theory is the $\mathcal{N}=2^*$ theory.
By introducing a 
fourth mass for the gravitino into the $\mathcal{N}=1^*$ theory,
supersymmetry can be completely broken.
This case has been addressed in \cite{Taylor-Robinson:2001pp}.  

All the fields in the above mentioned gauge theories transform in the 
adjoint representation of the gauge group. To approach a dual gravity
description of QCD we should extend the field content by adding fields
that transform in the fundamental representation (henceforth denoted as 
quarks).
It was proposed by Karch and Katz \cite{Karch:2002sh}
that $\mathcal{N}=4$ SYM can be endowed with $N_\text{f}$ quark flavours 
by embedding $N_\text{f}$ spacetime-filling $\text{D}7$-branes into 
$\AdS_5\times\text{S}^5$. 
In the brane picture, the $N_\text{f}$ quark 
flavours correspond to open strings that connect the 
stack of $N$ $\text{D}3$-branes with the $N_\text{f}$ 
$\text{D}7$-branes. Taking the near horizon limit to
obtain the gravity background for the correspondence, the gauge 
symmetry on the $\text{D}7$-branes becomes the global flavour symmetry.
The choice $N_\text{f}\ll N$ thereby allows one to 
neglect the backreaction of the $\text{D}7$-branes on the background, 
considering them as brane probes.
Each $\text{D}7$-brane probe spans an $\AdS_5\times\text{S}^3$ 
inside $\AdS_5\times\text{S}^5$.
It fills all of $\AdS_5$ down to a minimal value $r=\hat u$ of the radial
coordinate $r$, at which it terminates. 
Since $r$ has the interpretation of an energy scale with small 
and large $r$ corresponding to the IR and UV regimes in the dual gauge 
theory, the value $r=\hat u$ is related to the quark mass $m_\text{q}$ 
via $m_\text{q}=\frac{1}{2\pi\alpha'}\hat u$. The 
termination of the $\text{D}7$-brane at $r<u$ means that at energies 
$E<m_\text{q}$ the corresponding quark degree of freedom freezes out.
Furthermore, the fluctuations of the $\text{D}7$-brane embedding 
coordinates around the found solution 
determine the meson spectrum in the dual gauge theory
\cite{Kruczenski:2003be}.
 
In the context of the $\AdS/\text{CFT}$ correspondence the 
embeddings of $\text{D}p$-brane probes into various supergravity
backgrounds have been studied extensively in the literature 
\cite{Kruczenski:2003uq,Sakai:2004cn,Casero:2005se,Sakai:2003wu,
Kuperstein:2004hy,Ouyang:2003df,Evans:2005ti,Arean:2007nh,Arean:2004mm,
Nunez:2003cf,Babington:2003vm,Bertolini:2001qa,Wang:2003yc}.
Analyses beyond the probe approximation have also been performed 
\cite{Casero:2006pt,Casero:2005se,Burrington:2004id,Kirsch:2005uy,
Paredes:2006wb,Benini:2006hh,Cherkis:2002ir}.
Embeddings of $\text{D}p$-branes in backgrounds with flux have been 
treated in \cite{Bandos:2006wb,Martucci:2006ij,Apreda:2006bu}.

To add flavour to the mass perturbed $\mathcal{N}=1^\star$ and 
$\mathcal{N}=2^\star$ theories, we study in this paper the embedding of 
$\text{D}7$-brane probes into the order $\mathcal{O}(m^2)$
Polchinski-Strassler background with generic mass parameters. 
This generalizes our analysis in \cite{Apreda:2006bu} in which
we restricted ourselves to two specific embeddings in the 
$\mathcal{N}=2$ background. In the generic case the presence of anti-selfdual 
source terms in the equation of motion for $F$ disposes us to revisit the 
solution and also the treatment of the gauge field in \cite{Apreda:2006bu}. 
%The equations of 
%motion for $F$ and for the embedding coordinates $y^m$ can be 
%decoupled by an expansion of $y^m$ into an unperturbed constant 
%part $\hat y^m$ and a non-constant order $\mathcal{O}(m^2)$ correction 
%$\tilde y^m$.
%We then find that $F=0$ is a solution for all embeddings to this order. 
%Regarding the gauge field as an independent field, its solution has
%to be inserted into the order $\mathcal{O}(m^2)$ equations of motion for the 
%embedding coordinates $y^m$. The order $\mathcal{O}(m^2)$ result for $F$ hence
%suffices. This corrects the treatment of the gauge field in 
%\cite{Apreda:2006bu}, where we did not regard $F$ as independent field. 
%We inserted its solution into the action. This required that a solution for 
%$F$ was found without making use of the expansion of the embedding.  
Before studying the embedding coordinates themselves, we
introduce additional parameters into the underlying action
which allow us to reproduce the results of \cite{Apreda:2006bu} even after 
modifying the treatment of $F$. Furthermore, we can directly see how 
the individual contributions to the 
action influence the behaviour of the embeddings.

The paper is organized as follows.
In section \ref{sec:PSbackground} we review in brief the Polchinski-Strassler 
background with arbitrary mass parameters up to order $\mathcal{O}(m^2)$. 
In section \ref{sec:action} we present the expanded generalized form of
the $\text{D}7$-brane action on which the whole analysis is based.
In section \ref{sec:gaugefield} we revisit the equation of motion for the 
the $\text{D}7$-brane worldvolume gauge field and its solution. 
%This becomes necessary, since in the generic
%mass case self-dual source terms for the gauge field can be present that 
%are absent for the peculiar embeddings in \cite{Apreda:2006bu}.
In section \ref{sec:embedding} we evaluate the action for the expanded 
embeddings and discuss the resulting equations of motion and their 
general regular solutions. We also revisit the case of \cite{Apreda:2006bu}
with our new treatment of the gauge field and analyze additional embeddings
in the $\mathcal{N}=2$ and in the $\mathcal{N}=1$ background with three equal 
masses in more detail. Moreover, we give error estimates for 
the analytic solutions. 
In section \ref{sec:holoren} we apply the method of holographic renormalization
\cite{Skenderis:2002wp,deHaro:2000xn,Karch:2005ms}
to the on-shell action of arbitrary $\text{D}7$-brane embeddings in the 
generalized background and show that in all cases the subtracted action 
can be made vanish by adding appropriate finite counterterms. 
Various detailed computations that include the generalization of the
background to arbitrary mass parameters, the derivation of the 
explicit form of the equations of motion and of the action as well as the
derivation of their solutions can be found in a series of appendices.

\section{Polchinski-Strassler background to order $\mathcal{O}(m^2)$}
\label{sec:PSbackground}
\renewcommand\thefootnote{\arabic{footnote}}
\setlength{\skip\footins}{8mm}
In the following we work in the regime in which the Polchinski-Strassler 
background \cite{Polchinski:2000uf} can be described as a perturbative 
expansion around $\AdS_5\times\text{S}^5$.
The corrections are determined by the  backreaction of the 
$3$-form flux on the geometry. 
In the Einstein frame of \cite{Polchinski:2000uf,Apreda:2006bu}
the unperturbed metric of $\AdS_5\times\text{S}^5$ reads 
\begin{equation}
\begin{aligned}\label{AdSSmetric}
\de s^2&= Z^{-\frac{1}{2}}\eta_{\mu\nu}\de x^\mu\de x^\nu
+ Z^{\frac{1}{2}}\delta_{ij}\de y^i\de y^j\col\\
 Z(r)&=\frac{R^4}{r^4}\col\qquad 
r^2=y^i y^i\col\qquad
R^4= 4 \pi g_\text{s}N\alpha'^2\col
\end{aligned}
\end{equation}
where $\mu,\nu = 0,1,2,3$ and $i,j=4, \dots, 9$. The radius $R$ 
depends on the string coupling constant 
$g_\text{s}$ and on the number $N$ of $\text{D}3$-branes which in the near 
horizon limit generate the $\AdS_5\times\text{S}^5$ background.
The unperturbed background also contains the complex combination 
of the axion and dilaton and the $4$-form potential which are defined as
\begin{equation}\label{tauC4def}
\begin{aligned}
\hat\tau&=\hat C_0+i\e^{-\hat\phi}=\text{const.}\col\qquad
%F_{0123i}&=\partial_i\hat C_{0123}\col\qquad
\hat C_{0123}=\e^{-\hat\phi} Z^{-1}\col
\end{aligned}
\end{equation}
where in the following a `hat' 
always denotes an unperturbed quantity. The unperturbed dilaton is
related to the string coupling constant as $\e^{\hat\phi}=g_\text{s}$.

Polchinski and Strassler \cite{Polchinski:2000uf} have considered a 
perturbation in the form of a non-vanishing $3$-form flux $G_3$ given 
by \begin{equation}\label{G3expl}
G_3=\tilde F_3-\hat\tau H_3
=\e^{-\hat\phi}\frac{\zeta}{3}\de(ZS_2)\col
\end{equation}
where $\tilde F_3$ and $H_3$ are the $3$-form field strengths, which are 
respectively obtained from the potentials
\begin{equation}\label{tildeC2B2inS2}
\tilde C_2=C_2-\hat C_0B=\e^{-\hat\phi}\frac{\zeta}{3}Z\re S_2\col\qquad
B=-\frac{\zeta}{3}Z\im S_2\pnt
\end{equation}
The constant $\zeta$ assumes the value $\zeta=-3\sqrt{2}$ 
in a proper normalization scheme \cite{Polchinski:2000uf}.
The $2$-form $S_2$ has the component expression
\begin{equation}\label{S2def}
S_2=\frac{1}{2}T_{ijk}y^i\de y^j\wedge\de y^k\col
\end{equation}
where the $3$-tensor $T_3$ is imaginary anti-selfdual (IASD), i.e.\ it 
fulfills
\begin{equation}\label{T3iasdrel}
(\star_6+i)T_3=0\pnt
\end{equation}
Thereby $\star_6$ is the Hodge star operator in flat space with components 
in the directions $y^i$ of \eqref{AdSSmetric}.

To present the explicit form of $T_3$ it is
advantageous to work in a basis of three complex coordinates $z^p$ 
and their complex conjugates $\bar z^p$ for the transverse directions $y^i$. 
It is defined as 
\begin{equation}\label{cplxbasis}
z^p=\frac{1}{\sqrt{2}}(y^{p+3}+iy^{p+6})\col\qquad p=1,2,3 \pnt
\end{equation}
The components of the tensor $T_3$ then read
\begin{equation}\label{Tcplx}
T_{pqr}= T_{\bar{p}\bar{q}\bar{r}}=T_{\bar{p}qr} =0 \col \qquad
T_{p\bar{q}\bar{r}}= \epsilon_{pqr} m_p \col
\end{equation}
where in the dual gauge theory the three parameters $m_p$  
are the masses of the three adjoint chiral  $\mathcal{N}=1$ 
multiplets $\Phi_p$ of $\mathcal{N}=4$ SYM. To be more 
precise, the $G_3$ perturbation \eqref{G3expl} is dual to 
a deformation of $\mathcal{N}=4$ SYM by a mass-term superpotential
\begin{equation} \label{DeltaW}
\Delta W= \frac{1}{g^2_{\text{YM}}} (m_1 \tr \Phi_1^2+m_2 \tr
\Phi_2^2+m_3 \tr \Phi_3^2)
\col
\end{equation}
where $g_\text{YM}^2=4\pi g_\text{s}$.
For generic masses the theory is $\mathcal{N}=1$ supersymmetric, while
for one mass vanishing and the other two being equal 
it preserves $\mathcal{N}=2$ supersymmetries.

The mass perturbation \eqref{G3expl} backreacts on the geometry. 
Up to linear order in the masses only $6$-form potentials 
\cite{Polchinski:2000uf} are induced.
At quadratic order in the masses, the metric and $4$-form potential
\cite{Freedman:2000xb} as well as the complex axion-dilaton
\cite{Polchinski:2000uf,Freedman:2000xb,LopesCardoso:2004ni}
acquire corrections.\footnote{See \cite{Apreda:2006bu} for remarks about some 
typos in the original papers.} 
Furthermore, a non-vanishing $8$-form RR potential 
$C_8$ is induced \cite{Apreda:2006bu}.
The linear combination \eqref{G3expl} still contains only the constant 
unperturbed $\hat\tau$, which changes at order $\mathcal{O}(m^3)$, 
at which a component $G_{(3,0)}$ is generated that is dual to a 
non-vanishing gaugino condensate \cite{LopesCardoso:2004ni}.

The deformations at quadratic order for the metric, $C_4$ and $\tau$ 
have been computed in \cite{Freedman:2000xb} with an appropriate gauge choice.
At this order, the deformed metric reads
\begin{equation}\label{metriccorr}
\de s^2=(Z^{-\frac{1}{2}}+h_0)\eta_{\mu\nu}\de x^\mu\de x^\nu
+\Big[(5 Z^{\frac{1}{2}}+p)I_{ij}
+(Z^{\frac{1}{2}}+q)\frac{y^iy^j}{r^2}
+w W_{ij}\Big]\de y^i\de y^j\col
\end{equation}
where the tensors $I_{ij}$ and $W_{ij}$ are given by
\begin{equation}\label{IWdef}
\begin{aligned}
I_{ij}&=\frac{1}{5}\Big(\delta_{ij}-\frac{y^iy^j}{r^2}\Big)\col\qquad
W_{ij}=\frac{1}{|T_3|^2}\re(T_{ipk}\bar T_{jpl})\frac{y^ky^l}{r^2}-I_{ij}
\col\qquad
|T_3|^2=\frac{1}{3!}T_{ijk}\bar T_{ijk}\pnt
\end{aligned}
\end{equation}
It is important to remark that our definition of $|T_3|^2$ deviates from 
the one in \cite{Freedman:2000xb} by an extra factor $\frac{1}{3!}$, such that
we have the relation
\begin{equation}\label{T3abssquare}
|T_3|^2=m_1^2+m_2^2+m_3^2=M^2
\pnt
\end{equation}  
The functions $h$, $w$, $p$, $q$ are given by \cite{Freedman:2000xb}
%\footnote{ 
%We note two misprints in 
%\cite{Freedman:2000xb}: Their eq.\ (125) to determine $w$ is ill 
%written, though the final result matches; moreover their eq.\ (126) has an 
%extra factor of 4 which contradicts their explicit results in
%eqs. (71) and (143). With the latter two equations we 
%coincide.}
\begin{equation}\label{wpqh0}
\begin{aligned}
w&%=-\e^{2\hat\phi}\frac{\alpha^2M^2}{18R^6}Z
=-\frac{\zeta^2M^2R^2}{18}Z\col\qquad
p&%=-\e^{2\hat\phi}\frac{\alpha^2M^2}{48R^6}Z
=-\frac{\zeta^2M^2R^2}{48}Z\col\qquad
q&%=\e^{2\hat\phi}\frac{\alpha^2M^2}{1296R^6}Z
=\frac{\zeta^2M^2R^2}{1296}Z\col\qquad
h_0&%=\e^{2\hat\phi}\frac{7\alpha^2M^2}{1296R^6}
=\frac{7\zeta^2M^2R^2}{1296}\col
\end{aligned}
\end{equation}
and they satisfy
\begin{equation}\label{h0pqrel}
4h_0Z=q-p\pnt
\end{equation}
The correction to the dilaton $\tilde\phi=\varphi Y_+$ is given as a product 
of a purely radial dependent part $\varphi$ and an $SO(6)$ spherical 
harmonic $Y_+$, which
explicitly read%\footnote{\cite{Freedman:2000xb} has the correct factor. 
%\cite{LopesCardoso:2004ni} finds 18 times $\varphi(r)$ instead.}
\begin{equation}\label{tildephi}
\begin{aligned}
\varphi
&%=\e^{2\hat\phi}\frac{\alpha^2M^2}{108R^6}Z^{\frac{1}{2}}
=\frac{\zeta^2M^2R^2}{108}Z^{\frac{1}{2}}\col\qquad%\\
Y_+=\frac{3}{M^2r^2}\big(m_2m_3(y_4^2-y_7^2)
+m_1m_3(y_5^2-y_8^2)+m_1m_2(y_6^2-y_9^2)\big)\pnt
\end{aligned}
\end{equation}
The (backreacted) forms to order $\mathcal{O}(m^2)$ relevant here are given by 
the $2$-form potentials \eqref{tildeC2B2inS2} and by
\begin{equation}
\begin{aligned}\label{bgforms}
C_4
&=\e^{-\hat\phi}\Big(Z^{-1}+\frac{\zeta^2M^2R^2}{3^42^3}Z^{-\frac{1}{2}}\Big)
\de\vol(\mathds{R}^{1,3})+\frac{1}{2}B\wedge C_2\col\\
C_6&=\frac{2}{3}B\wedge\hat C_4\col\\
C_8&=-\frac{1}{6}\big(
\e^{2\hat\phi}\tilde C_2\wedge\tilde C_2+B\wedge B\big)\wedge\hat C_4 \col
\end{aligned} 
\end{equation}
where again $\hat C_4$ denotes the unperturbed $4$-form potential in
\eqref{tauC4def} which is the first term in the expression for $C_4$ above. 
It turns out that only $\hat C_4$ of $C_4$ 
is relevant for a $\text{D}7$-brane embedding up to 
order $\mathcal{O}(m^2)$. 

%%%%%%%%%%%%%%%%%%%%%%%%%%%%%%%

\section{The action}
\label{sec:action}

The action for a $\text{D}7$-brane is given by the
sum of the Dirac-Born-Infeld (DBI) and Chern-Simons (CS) action, i.e.\
\begin{align} \label{action}
S&=S_\DBI+S_\CS\col\\
%S_\mathrm{DBI}&=-T_7\e^{-2\hat\phi}
% \int\de^8\xi\e^{\phi}
%\sqrt{\big|\det P\big[g-\e^{-\frac{\phi-\hat\phi}{2}}B\big]
%+2\pi\alpha'\e^{-\frac{\phi-\hat\phi}{2}}F\big|}\col\\
S_\DBI&=-\frac{T_7}{\e^{2\hat\phi}}
 \int\de^8\xi\e^{\phi}
\sqrt{\big|\det\big(P[g]
+2\pi\alpha'\e^{-\frac{\phi-\hat\phi}{2}}\mathcal{F}\big)\big|}
\col\label{DBIaction} \\
S_\CS & = -
  \mu_7 \int \sum\limits_{r=1}^{4}P[C_{2r}]\wedge
\e^{2\pi\alpha'\mathcal{F}}\label{CSaction} \col
\end{align}
where $T_7=\mu_7$ and the expressions are given in the Einstein frame which is
related to the string frame by using only the non-constant part 
$\tilde\phi=\phi-\hat\phi$ of the dilaton $\phi$.
The field strength $\mathcal{F}$ is 
a linear combination of the field strength $F=\de A$ of the 
worldvolume gauge potential $A$ and the pullback of $B$ as 
\begin{equation}\label{mathcalFdef}
2\pi\alpha'\mathcal{F}=2\pi\alpha'F-P[B]\pnt
\end{equation}

In \cite{Apreda:2006bu} we have introduced the minus sign in 
\eqref{CSaction} for physical reasons. We also assumed there that it should 
be the right choice to preserve some supersymmetries of the background. 
Here our modified treatment of the gauge field in general 
alters the embeddings. However, based on the unaffected $y^4$ embeddings, we 
can still favour this sign choice. At some points we nevertheless also 
discuss the effects of the alternative choice. 
For a final decision, a check of the $\kappa$-symmetry on the 
worldvolume of the $\text{D}7$-brane is required, which we leave as 
an open problem. 

\setlength{\extrarowheight}{+1pt}
\begin{table}[t]
\begin{center}
%\begin{tabular}{|c|c|c|c|c|c|c|c|c|c|c|}
%\hline
%% & \multicolumn{4}{c|}{$x^\mu$} & \multicolumn{6}{c|}{$y^i$} \\
%& $x^0$ & $x^1$ & $x^2$ & $x^3$ 
%& $y^4$ & $y^5$ & $y^6$ & $y^7$ & $y^8$ & $y^9$ \\
%\hline
%$\text{D}3$ & $+$ & $+$ & $+$ & $+$ & $-$ & $-$ & $-$ & $-$ & $-$ & $-$ \\
%$\text{D}7$ & $+$ & $+$ & $+$ & $+$ & $-$ & $+$ & $+$ & $-$ & $+$ & $+$ \\
%\hline
%\end{tabular}
\begin{tabular}{cc|c|c|}
%\hline
%& & \multicolumn{2}{c|}{$y^i$, $i=4,5,6,7,8,9$} \\
\cline{3-4}
& & \multicolumn{2}{c|}{$y^i$, $i=4,5,6,7,8,9$} \\ 
\cline{2-4}
& \multicolumn{1}{|c|}{$x^\mu$, $\mu=0,1,2,3$} & $y^a$, $a=5,6,8,9$ & $y^m$, $m=4,7$ \\
\hline
\multicolumn{1}{|c|}{$\text{D}3$} & $-$ & $\cdot$ &  $\cdot$ \\
\hline
\multicolumn{1}{|c|}{$\text{D}7$} & $-$ & $-$ & $\cdot$ \\
\hline
\end{tabular}
\caption{Orientation of the $\text{D}7$-brane probe w.r.t.\ the background 
generating stack of $\text{D}3$-branes in absence of the mass perturbation.
The $\text{D}3$-branes are then localized (denoted by `$\cdot$') in the 
six transverse directions, while the $\text{D}7$-branes fill four of these 
directions (denoted by `$-$'). }
\label{tab:D3D7orient}
\end{center}
\end{table}
In the coordinate system used in \eqref{AdSSmetric}
the background generating stack of  $\text{D}3$-branes and the 
$\text{D}7$-brane probe are oriented as shown in figure \ref{tab:D3D7orient}.
For embedding coordinates that do not depend on the four worldvolume 
directions $x^\mu$, the pullbacks are non-trivial only in the additional 
four directions labeled by $y^a$. In static gauge, 
the pullback of a generic $2$-tensor $E_{ij}$ on these directions reads
\begin{equation}\label{PofEstatic}
\begin{aligned}
P[E]_{ab}&=E_{ab}+\partial_aX^mE_{mb}+\partial_bX^nE_{an}
+\partial_aX^m\partial_bX^nE_{mn}\pnt
\end{aligned}
\end{equation}

Expanding the complete $\text{D}7$-brane action
to quadratic order in the mass perturbation around the
unperturbed background \eqref{AdSSmetric} and \eqref{tauC4def}, we find
\begin{equation}\label{Sexpand}
\begin{aligned}
S
&=-\frac{T_7}{\e^{\hat\phi}}\int\de^8\xi
\sqrt{\det P[\delta]}\Big[1+\tilde\phi
+\frac{1}{2}Z^{\frac{1}{2}}\tilde g_{\mu\mu}
+\frac{1}{2}Z^{-\frac{1}{2}}P[\delta]^{ab}P[\tilde g]_{ab}\\
&\phantom{{}={}-\frac{T_7}{\e^{\hat\phi}}\int\de^8\xi
\sqrt{\det P[\delta]}\Big[}
+\frac{1}{2}Z^{-1}\big((\alpha-\beta\star_4)P[B]\cdot P[B]%\\
%&\phantom{{}={}-\frac{T_7}{\e^{\hat\phi}}\int\de^8\xi
%\sqrt{\det P[\delta]}\Big[+\frac{1}{2}Z^{-1}\big(}
-4\pi\alpha'(\mu-\nu\star_4)F\cdot P[B]\\
&\phantom{{}={}-\frac{T_7}{\e^{\hat\phi}}\int\de^8\xi
\sqrt{\det P[\delta]}\Big[+\frac{1}{2}Z^{-1}\big(}
+4\pi^2\alpha'^2(1+\star_4)F\cdot F%\\
%&\phantom{={}-\frac{T_7}{\e^{\hat\phi}}\int\de^8\xi
%\sqrt{\det P[\delta]}\Big[+\frac{1}{2}Z^{-1}\big(}
%-P[B]\cdot\star_4P[B]
%-\frac{4}{3}P[B]\cdot\star_4F\\
%&\phantom{{}={}-\frac{T_7}{\e^{\hat\phi}}\int\de^8\xi
%\sqrt{\det P[\delta]}\Big[+\frac{1}{2}Z^{-1}\big(}
+\tau\e^{2\hat\phi}P[\tilde C_2]\cdot\star_4
P[\tilde C_2]\big)\Big]
\col
\end{aligned}
\end{equation}
where we have introduced constants which in the case of the 
Polchinski-Strassler background take values
\begin{equation}\label{const1}
\alpha=1\col\qquad
\beta=\frac{2}{3}\col\qquad
\mu=1\col\qquad
\nu=-\frac{1}{3}\col\qquad
\tau=-\frac{1}{3}
\col
\end{equation}
where $\tau$ must not be confused with the complex axion-dilation defined 
in \eqref{tauC4def}.
Furthermore, throughout the paper with a `tilde' we denote the order 
$\mathcal{O}(m^2)$ 
corrections\footnote{By notational abuse, this does not apply to 
$\tilde C_2$ and its field strengths $\tilde F_3$.}  
to the unperturbed quantities which carry a `hat'.
We should stress that here the four-dimensional inner product $\cdot$ 
as well as the Hodge star $\star_4$  in \eqref{Sexpand} 
are understood to be computed 
with the pullback of the Kronecker delta denoted by $P[\delta]_{ab}$. 
For two generic $2$-forms
$\omega_2$ and $\omega'_2$ they are defined as
\begin{equation}\label{ipdef}
\omega_2\cdot\omega'_2
=\frac{1}{2}P[\delta]^{a_1b_1}P[\delta]^{a_2b_2}\omega_{a_1a_2}\omega'_{b_1b_2}
\col\qquad
\star_4\omega_{a_1a_2}=\frac{1}{2}\sqrt{\det P[\delta]}
\epsilon_{a_1a_2}^{\hphantom{a_1a_2}b_1b_2}\omega_{b_1b_2}
\col
\end{equation}
where $\epsilon_{5689}=1$, and indices are raised with the inverse of 
$P[\delta]_{ab}$ denoted by $P[\delta]^{ab}$.

We have introduced the constants
$\alpha$, $\beta$, $\mu$, $\nu$, $\tau$ in \eqref{Sexpand}
for two reasons. First of all, we want to keep the option 
to alter the corresponding values. This turns out to be necessary after the 
original treatment of the gauge field \cite{Apreda:2006bu} has been 
modified as described in section \ref{sec:gaugefield}.
Secondly, keeping these constants makes it 
easy to identify how the individual parts in the action 
contribute to the equations of motion and thus how they take influence on the 
embeddings.

With the above values of the parameters it is obvious that the action
\eqref{Sexpand} does not longer depend on the gauge invariant combination
\eqref{mathcalFdef} of the gauge field strength $F$ and of the pullback of 
$B$. The additional dependence on $B$ arises because the explicit expressions 
for $C_6$ and $C_8$ in \eqref{bgforms} explicitly contain $B$. 
They are obtained if we
make use of \eqref{tildeC2B2inS2} which relates
the $2$-form potentials $\tilde C_2$ and $B$ to $S_2$ an hence fixes 
their gauge freedom. This allows e.g.\ 
$P[B]$ to appear explicitly outside the combination $\mathcal{F}$ defined in
\eqref{mathcalFdef}. It also implies that the equations of motion for $F$ 
do not contain $F$ and $P[B]$ only as the combination $\mathcal{F}$.

\section{The gauge field equation revisited}
\label{sec:gaugefield}

The equations of motion for the $\text{D}7$-brane embedding coordinates 
depend on $F$. To determine the embeddings, we therefore have to discuss 
also the equation of motion for $F$ and its solution. 
As we have already shown in \cite{Apreda:2006bu}, it in general contains 
source terms which come from the terms linear in $F$ 
in the action \eqref{Sexpand}. 
In this section we will extend the discussion from the 
$\mathcal{N}=2$ case to the generic $\mathcal{N}=1$ case. With an  
expansion of the embedding coordinates we will show that up to order 
$\mathcal{O}(m^2)$ no source terms for $F$ are present. 
For the $\mathcal{O}(m^2)$ result to suffice, we will 
have to modify our previous understanding \cite{Apreda:2006bu} of the role of 
$F$. 

A variation of the action \eqref{Sexpand} w.r.t.\ the gauge potential $A$ 
gives the equation of motion
\begin{equation}\label{Feom}
\begin{aligned}
\de\big(\hat C_4\wedge(2\pi\alpha'(\star_4+1)F
-(\mu\star_4-\nu)P[B])\big)=0
\col
\end{aligned}
\end{equation}
where we have transformed inner products multiplied by the volume element
into wedge products by using the Hodge star $\star_4$ and also the explicit 
unperturbed metric \eqref{AdSSmetric}.
Integrating the above expression and inserting the explicit expression for 
$\hat C_4$ in \eqref{tauC4def}, we find
\begin{equation}\label{Feomdimred}
\begin{aligned}
Z^{-1}(2\pi\alpha'(\star_4+1)F-(\mu\star_4-\nu)P[B])
=\de P[\omega_1]
\pnt
\end{aligned}
\end{equation}
We have introduced $\omega_1$ to take into consideration the freedom in
integrating the exterior derivative. 
By acting with the linear combination $1\pm\star_4$,
the above equation is separated into two equations 
according to
\begin{equation}\label{Feomdecomp}
\begin{aligned}
4\pi\alpha'Z^{-1}F_+-(\mu-\nu)Z^{-1}P[B]_+=\de P[\omega_1]_+
\col\qquad
(\mu+\nu)Z^{-1}P[B]_-=\de P[\omega_1]_-
\pnt
\end{aligned}
\end{equation}
We have thereby used that the decomposition of a $2$-form $\omega_2$ 
in its selfdual and anti-selfdual components $\omega_{2+}$ and respectively 
$\omega_{2-}$ is given by
\begin{equation}
\omega_2=\omega_{2+}+\omega_{2-}\col\qquad
\omega_{2+}=\frac{1}{2}(1+\star_4)\omega_2
%=\omega_{(1,1)}^\text{P}
\col\qquad
\omega_{2-}=\frac{1}{2}(1-\star_4)\omega_2
%=\omega_{(2,0)}+\omega_{(0,2)}
%-\frac{i}{2}\omega_{c\bar c}K
\pnt
\end{equation}
While the first equation in \eqref{Feomdecomp} contains $F_+$, 
the second one does not contain 
any degrees of freedom of $F$. For $\mu+\nu\neq0$ 
this equation in general is a non-trivial constraint for the embedding
coordinates, which enter via the pullback. It means that the pullback of 
$Z^{-1}B$ has to follow as exterior derivative of a $1$-form. 
One therefore has to solve a coupled system of differential equations that
consists of the equations of motion for the embedding 
coordinates and the two equations in 
\eqref{Feomdecomp}. In the following we describe a solution which is 
based on perturbation theory. 
%We remember that the Polchinski-Strassler background itself 
%is a perturbative expansion around $AdS_5\times S^5$ with the masses $m_p$
%serving als expansion parameters. In particular, the 
%action \eqref{Sexpand} considers the corrections up to order 
%$\mathcal{O}(m^2)$.

We assume that, as the background itself, also
the $\text{D}7$-brane embedding coordinates $y^m(y^a)$
can be treated perturbatively.
The leading contributions are constants $\hat y^m$
that describe the constant embedding of $\text{D}7$-branes in 
pure $\AdS_5\times\text{S}^5$ found in \cite{Karch:2002sh}.
In the Polchinski-Strassler background the embeddings 
are corrected at higher orders by non-constant contributions 
$\tilde y^m(y^a)$, such that we write
\begin{equation}\label{yembedexpand}
y^m(y^a)=\hat y^m+\tilde y^m(y^a)
\pnt
\end{equation}
In case of the $\mathcal{N}=2$ background \cite{Apreda:2006bu}
the decomposition was used to expand the action and equations
of motion for the embedding coordinates themselves. However, unlike here, 
it was not used for determining the gauge field.  

The correction $\tilde y^m$ is of order $\mathcal{O}(m^2)$.
Since $B$ itself is of order $\mathcal{O}(m)$, the derivative terms 
in the pullbacks \eqref{PofEstatic} in static gauge are therefore 
beyond the order $\mathcal{O}(m^2)$ up to which we consider the 
background and the equations of motion for $y^m$. The same holds for 
the pullback of the unperturbed diagonal metric
\eqref{AdSSmetric}, for which the terms 
linear in the derivatives vanish exactly. Thus, the Hodge star as defined in 
\eqref{ipdef} reduces to the one in flat space. 
It is again advantageous to work in the complex basis \eqref{cplxbasis}, 
in which the $\text{D}7$-brane embeddings oriented as in 
table \ref{tab:D3D7orient} are along $z^a$, $\bar z^a$, $a=2,3$ and the 
transverse embedding coordinates 
are given by $z^m$, $\bar z^m$, $m=1$. In this basis, 
the imaginary selfdual
and anti-selfdual components of any $2$-form $\omega$ decompose as
\begin{equation}\label{orthstar4proj}
\omega_{2+}=\omega_{(1,1)}^\text{P}\col\qquad
\omega_{2-}=\omega_{(2,0)}+\omega_{(0,2)}
+\frac{1}{2}\omega_{a\bar a}\de z^b\wedge\de\bar z^b
\col
\end{equation}
where P denotes the primitive part of $\omega_2$, i.e.\ 
$\omega_{a\bar a}^\text{P}=0$, and summations over $a$ and $b$ are understood. 
The potential $B$ is primitive. We therefore assume that so are $F$ and 
$\de\omega_1$. The equations \eqref{Feomdecomp} then reduce to
\begin{equation}\label{Feomdecompexpand}
\begin{aligned}
4\pi\alpha'Z^{-1}F_{(1,1)}
-(\mu-\nu)Z^{-1}B^\parallel_{(1,1)}=\de\omega^\parallel_{(1,1)}
\col\qquad
(\mu+\nu)Z^{-1}B^\parallel_{(2,0)}=\de\omega^\parallel_{(2,0)}
\col
\end{aligned}
\end{equation}
where by $\parallel$ we denote the components of the corresponding form 
which are parallel to the directions of the $\text{D}7$-brane. 
Up to order $\mathcal{O}(m^2)$ we write
\begin{equation}
\begin{aligned}
\im S_{(1,1)}^\parallel
&=-\frac{i}{2}(T_{\bar ma\bar b}\hat{\bar z}^m-\bar T_{m\bar ab}\hat z^m)
\de z^a\wedge\de\bar z^b
\col\qquad
\im S_{(2,0)}^\parallel
=-\frac{i}{4}\bar T_{\bar mab}\hat{\bar z}^m\de z^a\wedge\de z^b
\col
\end{aligned}
\end{equation}
which according to \eqref{tildeC2B2inS2} up to a constant factor are the 
components of $Z^{-1}B^\parallel$. The two expressions follow as exterior 
holomorphic or anti-holomorphic derivatives of $1$-form potentials. With  
$\de=\partial+\bar\partial$, we hence find that the choice 
\begin{equation}
\begin{aligned}
\omega_1^\parallel
&=-i\frac{\zeta}{12}\big(
2(\mu-\nu)(T_{\bar m\bar ab}\bar z^a\hat{\bar z}^m\de z^b
-\bar T_{m\bar ab}z^a\hat z^m\de\bar z^b)\\
&\phantom{{}={}-i\frac{\zeta}{12}\big(}
-(\mu+\nu)(\bar T_{\bar mab}z^a\hat{\bar z}^m\de z^b
-T_{m\bar a\bar b}\bar z^a\hat z^m\de\bar z^b)
\big)
\end{aligned}
\end{equation} 
allows us to gauge away all source terms for $F$ such that to order 
$\mathcal{O}(m^2)$ the equations \eqref{Feomdecompexpand} are consistently 
solved if $F$ obeys
\begin{equation}
\de(Z^{-1}F_+)=0\col\qquad\de F=0\col
\end{equation}
where the second relation is the Bianchi identity. Clearly, both equations 
are compatible with a vanishing gauge field $F=0$ on the $\text{D}7$-brane,
which is what we will assume from now on.

Up to order $\mathcal{O}(m^2)$ also the solution for $F$ found in 
\cite{Apreda:2006bu} 
in the case of the $\mathcal{N}=2$ background solves the above equation.
Moreover, it contains also $\mathcal{O}(m^3)$ terms, which we had to keep 
since we did not consider $F$ as an independent field. 
This means, we have plugged the found $F$ into the action before deriving 
the equations of motion for the embedding coordinates. The   
$\mathcal{O}(m^3)$ terms of $F$ that contained derivatives of the embedding 
coordinates then contributed to the order $\mathcal{O}(m^2)$ embedding 
equations of motion. 
However, we should have better regarded $F$ as an independent field and hence 
have inserted the result for $F$ into the equations of motion for $y^m$. 
This procedure requires the result for $F$ up to order 
$\mathcal{O}(m^2)$ only. 

\section{The expanded embeddings}
\label{sec:embedding}

\subsection{Expanded action, equations of motion and solutions}

In section \ref{sec:gaugefield} we have already made use of the 
expansion of the embedding into the constant unperturbed part $\hat y^m$
and the order $\mathcal{O}(m^2)$ correction \eqref{yembedexpand}.
Inserting this decomposition into 
\eqref{Sexpand}, the pullbacks of the 
Kronecker $\delta$ simplify to the Kronecker $\delta$ on the worldvolume of 
the $\text{D}7$-brane. Since the equations of motion are found by taking 
derivatives w.r.t.\ $\tilde y^m$ and $\partial_a\tilde y^m$, one has to 
keep those terms which contribute up to order $\mathcal{O}(m^2)$ 
to the equations, even if they are of higher order in the action.
The action \eqref{Sexpand} is then expanded as 
\begin{equation}\label{Sembed}
\begin{aligned}
S&=
-\frac{T_7}{\e^{\hat\phi}}\int\de^8\xi\Big[
1+\tilde\phi+\frac{1}{2}Z^{\frac{1}{2}}\tilde g_{\mu\mu}
+\frac{1}{2}Z^{-\frac{1}{2}}\tilde g_{aa}
+\frac{1}{2}(\partial_a\tilde y^m)^2
+Z^{-\frac{1}{2}}\partial_a\tilde y^m\tilde g_{ma}\\
&\phantom{{}={}-\frac{T_7}{\e^{\hat\phi}}\int\de^8\xi\Big[}
+\frac{1}{2}Z^{-1}\big((\alpha-\beta\star_4)B\cdot B
+4(\gamma-\delta\star_4)B\cdot\partial\tilde yB\\
&\phantom{{}={}-\frac{T_7}{\e^{\hat\phi}}\int\de^8\xi\Big(
+\frac{1}{2}Z^{-1}\big(}
-4\pi\alpha'(\mu-\nu\star_4) F\cdot(B+4\partial\tilde yB)
+4\pi^2\alpha'^2(1+\star_4)F\cdot F
\\
&\phantom{{}={}-\frac{T_7}{\e^{\hat\phi}}\int\de^8\xi\Big(
+\frac{1}{2}Z^{-1}\big(}
+\tau\e^{2\hat\phi}\star_4\tilde C\cdot
(\tilde C+4\partial\tilde y\tilde C)
\big)\Big]
\col
\end{aligned}
\end{equation}
where the inner product and the Hodge star operator are computed w.r.t.\ the 
flat $4$-dimensional metric.
We have furthermore used the abbreviations
\begin{equation}
(\partial\tilde yB)_{ab}=\partial_a\tilde y^mB_{mb}\col\qquad
(\partial\tilde y\tilde C)_{ab}=\partial_a\tilde y^m\tilde C_{mb}\col\qquad
\end{equation}
and in addition we have introduced the constants
\begin{equation}\label{const2}
\gamma=1\col\qquad
\delta=\frac{2}{3}\col
\end{equation}
which in our case take the same values as respectively $\alpha$ and $\beta$ in 
\eqref{const1}. 
As already explained at the end of section \ref{sec:gaugefield}, 
inserting the non-vanishing solution for the gauge field found in 
\cite{Apreda:2006bu} directly into the action alters some terms that 
contain derivatives of the embedding coordinates.
The values of two parameters $\gamma$ and $\delta$ then become
\begin{equation}\label{gammadeltaold}
\gamma=\frac{1}{3}\col\qquad
\delta=\frac{4}{3}\col
\end{equation}
while the $F$-dependent terms then have to be removed from \eqref{Sembed}.
To describe both cases, we keep $\gamma$ and $\delta$ as independent constants.

The equations of motions for the embedding coordinates, which follow from the 
action \eqref{Sembed}, are given by
\begin{equation}\label{tildeyeom}
\begin{aligned}
{}&\partial_a\Big(
\partial_a\tilde y^m
+Z^{-\frac{1}{2}}\tilde g_{ma}
+Z^{-1}\big((\gamma-\delta\star_4)B_{ab}B_{mb}
-2\pi\alpha'(\mu-\nu\star_4) F_{ab}B_{mb}
+\tau\e^{2\hat\phi}\star_4\tilde C_{ab}\tilde C_{mb}\big)\Big)\\
&=
\preparderiv{\tilde y^m}\Big(
\tilde\phi+\frac{1}{2}Z^{\frac{1}{2}}\tilde g_{\mu\mu}
+\frac{1}{2}Z^{-\frac{1}{2}}\tilde g_{aa}%\\
%&\phantom{{}={}\preparderiv{\tilde y^m}\Big(}
+\frac{1}{2}Z^{-1}\big((\alpha-\beta\star_4)B\cdot B%\\
%&\qquad\phantom{=\preparderiv{\tilde y^m}\Big(
%+\frac{1}{2}Z^{-1}\big(}
+\tau\e^{2\hat\phi}\star_4\tilde C_2\cdot\tilde C_2\big)
\Big)\Big|_{\tilde y^m=0}\\
&\phantom{{}={}}
+2\pi\alpha'\Big(\pi\alpha'\preparderiv{\tilde y^m}Z^{-1}(1+\star_4)F\cdot F
-(\mu-\nu\star_4) F\cdot\preparderiv{\tilde y^m}(Z^{-1}B)\Big)
\Big|_{\tilde y^m=0}
\pnt
\end{aligned}
\end{equation}
The solution $F=0$ of \eqref{Feom} is from now on inserted into the above  
equations.

It is advantageous to introduce polar coordinate systems for the four 
worldvolume directions $y^a$ of the $\text{D}7$-brane and for the two 
transverse directions $y^m$.
The radial coordinate $r$ of the full six-dimensional transverse space
as defined in \eqref{AdSSmetric} splits into the radii 
$\rho$ on the $\text{D}7$-brane worldvolume and $u$ of the two transverse
embedding directions according to 
\begin{equation}\label{rhordef}
r=\sqrt{\rho^2+u^2}\col\qquad
\rho=\sqrt{y^ay^a}\col\qquad
u=\sqrt{y^my^m}
\col
\end{equation}
where summations over $a$ and $m$ are understood. In the polar coordinate 
system with angular coordinate $\psi$ the two embedding coordinates 
$y^m$ read
\begin{equation}
\begin{aligned}
y^4&=u\cos\psi
=\hat u\cos\hat\psi-\hat u\tilde\psi\sin\hat\psi+\tilde u\cos\hat\psi\col\\
y^7&=u\sin\psi
=\hat u\sin\hat\psi+\hat u\tilde\psi\cos\hat\psi+\tilde u\sin\hat\psi\pnt
\end{aligned}
\end{equation}
In the final equalities we have expanded up to linear order in the 
corrections $\tilde u$ and $\tilde\psi$ to the unperturbed radius $\hat u$
and angle $\hat\psi$ which also present the boundary values 
at $\rho\to\infty$ of the embedding functions. In the dual gauge theory
$\hat u$ determines the mass $m_\text{q}$ of the quarks 
via $m_\text{q}=\frac{1}{2\pi\alpha'}\hat u$.

In appendix \ref{app:eomex} we derive
the equations of motion \eqref{tildeyeom} in the above coordinate system. 
They assume the same form for the radial coordinate $u$ as also for 
the angle $\psi$, such that we can compactly write
\begin{equation}\label{eomimpl}
\begin{aligned}
2\partial_a\partial_{\bar a}f
&=\frac{n_f}{\hat r^4}\Big(B_f+C_f\frac{\hat u^2}{\hat r^2}
-C_f^Iy_I\frac{\rho^2}{\hat r^2}\Big)%\\
%&=\frac{n_f}{\hat r^4}\Big(B_f+C_f
%-(C_f+C_f^{++}y_{++}+C_f^{+-}y_{+-}+C_f^{-+}y_{-+}+C_f^{--}y_{--})
%\frac{\rho^2}{\hat r^2}\Big)
\col
\end{aligned}
\end{equation}
where we set either $f=u$ or $f=\psi$ and identify
the normalization factor with $n_u=\hat u$ and $n_\psi=1$, respectively. 
The r.h.s.\ depends on $\rho$ explicitly and implicitly 
via the total radius $\hat r$
which is found from \eqref{rhordef} when only the  
unperturbed parts $\hat y^m$ of the embedding coordinates are inserted.  
The dependence on the three angles in the four worldvolume directions $y^a$ 
of the $\text{D}7$-brane is encoded within 
four of the nine $l=2$ $SO(4)$ spherical harmonics $y_I$ which 
are defined in \eqref{SO4shdef}.
The constants $B_f$, $C_f$ and $C_f^I$ which depend on the masses, 
the parameters \eqref{const1} and
\eqref{const2} and the angle $\hat\psi$ 
are given in \eqref{BuCudef} for $f=u$ and in \eqref{BpsiCpsidef} for $f=\psi$.

In appendix \ref{app:anasol} we show that, fixing the boundary value 
to $\hat f$, the above differential equation
admits a unique analytic regular solution. Together with 
its derivatives it is given in \eqref{app:freg} and reads
\begin{equation}\label{freg}
\begin{aligned}
f&=\hat f-\frac{n_f}{8}\Big(
B_f\frac{2}{\rho^2}\ln\frac{\hat r^2}{\hat u^2}
+C_f\frac{1}{\hat r^2}
-C_f^I\Big(
\frac{2}{\rho^2}
\Big(1-\frac{\hat u^2}{\rho^2}\ln\frac{\hat r^2}{\hat u^2}\Big)
-\frac{1}{\hat r^2}\Big)y_I\Big)\\
\partial_\rho f
&=\frac{n_f}{4}\Big(
B_f\frac{2}{\rho}\Big(\frac{1}{\rho^2}\ln\frac{\hat r^2}{\hat u^2}
-\frac{1}{\hat r^2}\Big)
+C_f\frac{\rho}{\hat r^4}
-C_f^I\Big(
\frac{2}{\rho^3}\Big(
1-2\frac{\hat u^2}{\rho^2}\ln\frac{\hat r^2}{\hat u^2}
+\frac{\hat u^2}{\hat r^2}\Big)
-\frac{\rho}{\hat r^4}\Big)y_I\Big)\\
\partial_\rho^2f
&=\frac{n_f}{4}\Big(
B_f\frac{2}{\rho^2}\Big(-\frac{3}{\rho^2}\ln\frac{\hat r^2}{\hat u^2}
+\frac{3}{\hat r^2}+2\frac{\rho^2}{\hat r^4}\Big)
+C_f\frac{1}{\hat r^4}\Big(1-4\frac{\rho^2}{r^2}\Big)\\
&\phantom{{}={}\frac{n_f}{4}\Big(}
-C_f^I\Big(
-\frac{20}{\rho^4}\Big(
1-\frac{\hat u^2}{\rho^2}\ln\frac{\hat r^2}{\hat u^2}\Big)
+\frac{1}{\rho^2\hat r^2}\Big(10+3\frac{\rho^2}{\hat r^2}\Big)
+4\frac{\rho^2}{\hat r^6}\Big)\Big)y_I\Big)
\col
\end{aligned}
\end{equation}
where $I$ is a summation index that runs over the four combinations 
that label the spherical harmonics $y_I$.
The solution and their derivatives have asymptotic behaviours found in 
\eqref{app:fregasymp}. The results read
\begin{equation}\label{fregasymp}
\begin{aligned}
f&=
\begin{cases}
\hat f-\frac{n_f}{8\hat u^2}(2B_f+C_f)& \rho\to0 \\
\hat f-\frac{n_f}{8\rho^2}\big(
2B_f\ln\frac{\rho^2}{\hat u^2}+C_f-C_f^Iy_I\big) & 
\rho\to\infty
\end{cases}\col\\
\partial_\rho f&=
\begin{cases}
%\frac{n_f}{12\hat u^2}(3(B_f+C_f)+C_f^Iy_I)\rho 
0 & \rho\to0 \\
\frac{n_f}{4\rho^3}\big(
2B_f(\ln\frac{\rho^2}{\hat u^2}-1)+C_f-C_f^Iy_I\big) & 
\rho\to\infty
\end{cases}\col\\
\partial_\rho^2f&=
\begin{cases}
\frac{n_f}{12\hat u^2}(3(B_f+C_f)+C_f^Iy_I) & \rho\to0 \\
\frac{n_f}{4\rho^4}(
2B_f(-3\ln\frac{\rho^2}{\hat u^2}+5)-3C_f+3C_f^Iy_I) 
& \rho\to\infty
\end{cases}\col
\end{aligned}
\end{equation}
where in case of the $\rho\to\infty$ limit we also have kept the next 
subleading contributions.
Based on the above results the monotony properties of the solutions 
and also estimates of the deviation from the full numerical results 
are discussed in the following.

\subsection{Monotony properties of the solutions}

A physical embedding should lead to a monotonically increasing function
$r(\rho)$ \cite{Babington:2003vm}. 
Taking the derivative of the total radius $r$ as defined in \eqref{rhordef}
w.r.t\ $\rho$, and expanding the result up to order $\mathcal{O}(m^2)$, 
we find the condition 
\begin{equation}
\hat r\partial_\rho r=\rho-\frac{\rho\hat u}{\hat r^2}\tilde u
+\hat u\partial_\rho\tilde u
\ge0\pnt
\end{equation}
Inserting the explicit result for $u(\rho)$ taken from \eqref{freg}, 
the result reads
\begin{equation}
\begin{aligned}
\hat r\partial_\rho r
&=\rho+\frac{\hat u^2}{8}\Big(B_u\frac{2}{\rho}\Big(
\Big(\frac{1}{\hat r^2}+\frac{2}{\rho^2}\Big)\ln\frac{\hat r^2}{\hat u^2}
-\frac{2}{\hat r^2}\Big)
+3C_u\frac{\rho}{\hat r^4}\\
&\phantom{{}={}\rho+\frac{\hat u^2}{8}\Big(}
+C_u^I\Big(3\frac{\rho}{\hat r^4}
+2\frac{\hat u^2}{\rho^3}\Big(\frac{1}{\hat r^2}+\frac{4}{\rho^2}\Big)
\ln\frac{\hat r^2}{\hat u^2}
+\frac{2}{\rho\hat r^2}-\frac{8}{\rho^3}\Big)y_I
\Big)
\pnt
\end{aligned}
\end{equation}
At large $\rho$, this expression is dominated by the first term, and therefore
$r(\rho)$ is linearly increasing with $\rho$ there. 
For $\rho\ll\hat u$ the above result expands as
\begin{equation}
\begin{aligned}
\hat r\partial_\rho r
&=\Big(1+\frac{1}{24\hat u^2}(12B_u+9C_u+2C_u^Iy_I)\Big)\rho
\pnt
\end{aligned}
\end{equation}
Since the constants $B_u$, $C_u$ and $C_u^I$ are proportional to $m^2R^4$, 
the derivative $\partial_\rho r$ becomes negative only if their linear 
combination is negative and if $\hat u\lesssim mR^2$ is sufficiently small to 
compensate the leading term.
We do not investigate this further, since in the regime 
$\hat u\lesssim mR^2$ the analytic solution based on the 
expansion \eqref{yembedexpand} cannot be trusted anyway, and therefore 
the exact result is required for a precise statement on the monotony 
properties.
In any case, for $\hat u$ sufficiently large, one finds that $r(\rho)$
is monotonically increasing and hence the embedding is physical.

\begin{figure}[t]
\begin{center}
\setlength{\unit}{0.1\textwidth}
\psset{xunit=\unit,yunit=\unit,runit=\unit}
\begin{pspicture}(-0.5,-0.5)(7.75,3.75)
%\psframe(-0.5,-0.5)(7.75,3.75)
\footnotesize
\psaxes[ticksize=2pt,tickstyle=bottom,Dx=2,Dy=1]{->}(0,0)(7.25,3.25)
\rput(0,3.5){$u(\rho)$}
\rput(7.5,0){$\rho$}
\rput(0.25,3){\olett{1}}
\rput[l](0.75,2.75){$B_u<0$, $C_u<-B_u$}
\rput(0.25,1.7){\olett{2}}
\rput[l](0.75,1.8){$B_u<0$, $C_u>-B_u$}
\rput(0.25,1.3){\olett{3}}
\rput[lt](0.75,1.3){$B_u>0$, $C_u<-B_u$}
\rput(0.25,0.375){\olett{4}}
\rput[lT](0.75,0.25){$B_u>0$, $C_u>-B_u$}
\psplot{0.01}{7}{
x 1.5 div dup mul % x=(rho/u)^2
dup dup 1 add % x^2 x^2 1+x^2
ln exch 1 exch div % x^2 ln(1+x^2) 1/x^2
-1 mul 4 mul 1 div 1.5 div -0.25 mul % x^2 ln(1+x^2) -B/(4u x^2)
mul exch 1 add % -B/(4u x^2)*ln(1+x^2) 1+x^2
1 exch div % -B/(4u x^2)*ln(1+x^2) 1/(1+x^2)
-1 mul 8 mul 1 div 1.5 div -0.125 mul % -B/(4u x^2)*ln(1+x^2) -C/(8u(1+x^2))
add 1.5 add}
\psplot{0.01}{7}{
x 1.5 div dup mul % x=(rho/u)^2
dup dup 1 add % x^2 x^2 1+x^2
ln exch 1 exch div % x^2 ln(1+x^2) 1/x^2
-1 mul 4 mul 1 div 1.5 div -0.25 mul % x^2 ln(1+x^2) -B/(4u x^2)
mul exch 1 add % -B/(4u x^2)*ln(1+x^2) 1+x^2
1 exch div % -B/(4u x^2)*ln(1+x^2) 1/(1+x^2)
1 mul 8 mul 1 div 1.5 div -0.125 mul % -B/(4u x^2)*ln(1+x^2) -C/(8u(1+x^2))
add 1.5 add}
\psplot{0.01}{7}{
x 1.5 div dup mul % x=(rho/u)^2
dup dup 1 add % x^2 x^2 1+x^2
ln exch 1 exch div % x^2 ln(1+x^2) 1/x^2
1 mul 4 mul 1 div 1.5 div -0.25 mul % x^2 ln(1+x^2) -B/(4u x^2)
mul exch 1 add % -B/(4u x^2)*ln(1+x^2) 1+x^2
1 exch div % -B/(4u x^2)*ln(1+x^2) 1/(1+x^2)
-1 mul 8 mul 1 div 1.5 div -0.125 mul % -B/(4u x^2)*ln(1+x^2) -C/(8u(1+x^2))
add 1.5 add}
\psplot{0.01}{7}{
x 1.5 div dup mul % x=(rho/u)^2
dup dup 1 add % x^2 x^2 1+x^2
ln exch 1 exch div % x^2 ln(1+x^2) 1/x^2
1 mul 4 mul 1 div 1.5 div -0.25 mul % x^2 ln(1+x^2) -B/(4u x^2)
mul exch 1 add % -B/(4u x^2)*ln(1+x^2) 1+x^2
1 exch div % -B/(4u x^2)*ln(1+x^2) 1/(1+x^2)
1 mul 8 mul 1 div 1.5 div -0.125 mul % -B/(4u x^2)*ln(1+x^2) -C/(8u(1+x^2))
add 1.5 add}
\end{pspicture}
\caption{The four possible types of embeddings, presented for illustration 
with the values $\hat u=1.5$, $mR^2=1$, $|C_u|=2|B_u|=8$.}
\label{fig:embcases}
\end{center}
\end{figure}
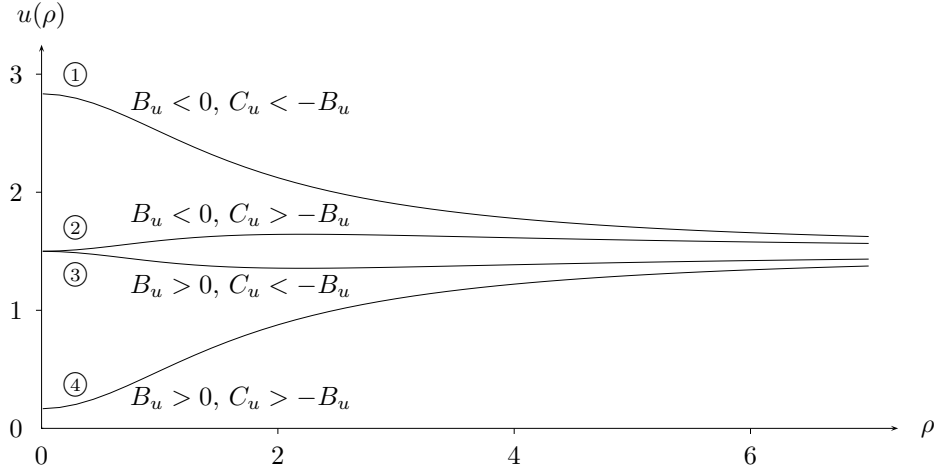
The requirement that $r(\rho)$ has to be a monotonically increasing function 
does not imply that $u(\rho)$ has to be monotonic, and in fact in
general it is not. 
As shown in figure \ref{fig:embcases}, with $C_u^I=0$ 
we find four distinct behaviours, depending on the relations
between $B_u$ and $C_u$. Two of them are monotonically increasing and 
respectively decreasing, while the other two assume an intermediate 
relative maximum or minimum. This is also visible from the asymptotic 
behaviour in \eqref{fregasymp}.
The transgression between a monotonic and non-monotonic embedding
takes place at
\begin{equation}\label{monotonycond}
C_u=-B_u
\pnt
\end{equation}

It is interesting to analyze the behaviour of $u(\rho)$ under a relative 
sign flip between the DBI and CS action \eqref{DBIaction} and 
\eqref{CSaction}. We just have to compare the relations between $B_u$ and 
$C_u$ for both choices of the relative sign. 
In the combined expanded action \eqref{Sembed} a change of the relative sign  
inverts the signs in front of all Hodge stars, i.e.\ the 
parameters $\beta$, $\delta$ and $\tau$ change their signs. 
As seen from \eqref{BuCudef}, the  
coefficient $C_u$ only depends on the combinations
$\alpha\pm\beta-(\gamma\pm\delta)$. It is therefore insensitive to such 
a sign flip as long as $\beta=\delta$.
The coefficient $B_u$ differs from $B_u^{(-)}$, which is the one found in case 
of a relative minus sign between the DBI and CS action, as
\begin{equation}\label{BuBumrel}
\begin{aligned}
B_u-B_u^{(-)}
%&=\frac{\zeta^2R^4}{36}\big(
%2(m_2+m_3)^2\tau-2(m_2-m_3)^2\beta
%+2m_1^2(\beta-\tau)
%-4m_2m_3(\beta+\tau)(1-\cos2\hat\psi)\big)\\
&=\frac{\zeta^2R^4}{18}\big(
(m_2^2+m_3^2-m_1^2)(\tau-\beta)
+2m_2m_3(\beta+\tau)\cos2\hat\psi\big)%\\
%&=\frac{\zeta^2R^4}{18}\tau\big(
%3(m_2^2+m_3^2-m_1^2)
%-2m_2m_3\cos2\hat\psi\big)\\
\col
\end{aligned}
\end{equation}
where in the Polchinski-Strassler background 
$\beta$ and $\tau$ assume the values given in \eqref{const1}. 
In particular, one has $\beta=-2\tau>0$.
%\begin{equation}
%\begin{aligned}
%B_u-B_u^{(-)}
%&=\frac{\zeta^2R^4}{18}\tau\big(
%3(m_2^2+m_3^2-m_1^2)-2m_2m_3\cos2\hat\psi\big)%\\
%\col
%\end{aligned}
%\end{equation}
If $m_2=m_3=m$ and either $m_1=0$ or $m_1=m$
we find the inequality $B_u<B_u^{(-)}$ for arbitrary values of  
the angle $\hat\psi$. Therefore, a relative minus sign between the 
DBI and CS action leads to $\text{D}7$-brane embeddings which are more 
attracted towards the center of the space. 
If $m_1=m_2=m$ and $m_3=0$, $B_u$ is 
independent of the relative sign choice, i.e.\ $B_u=B_u^{(-)}$.

In the following we keep the sign choice as in \cite{Apreda:2006bu}. 
The $y^4$ embedding in the case $m_1=0$ and $m_2=m_3=m$ then is monotonically
decreasing as a function of $\rho$. The alternative sign choice would 
alter this behaviour and lead to an intermediate minimum, as in the third  
case presented in figure \ref{fig:embcases}. 
A numerical study reveals that the allowed radial boundary values 
$\hat u$ for which the $y^4$ embeddings obey $u(\rho)>0$, differ for 
both choices of the sign. 
For the sign choice as in \cite{Apreda:2006bu} the $y^4$ embeddings
can assume all values $\hat u\ge0$, while for the alternative sign choice
$\hat u$ is restricted from below by $\hat u\ge\hat u_0>0$. 
There appears thus a gap in the allowed values for $\hat u$, separating the 
case $\hat u=0$ from the continuum $\hat u\ge\hat u_0>0$.
This is a disfavoured behaviour. 
We stress that the embeddings with $\hat u\simeq\hat u_0$ enter the region 
in which the expansion of the background itself breaks down. One must 
therefore not use this observation to completely rule out the possibility
of a relative minus sign between the DBI and CS action. For a
confirmed answer which of the sign choices is the correct one to preserve 
some supersymmetry, one has to check the kappa symmetry up to order 
$\mathcal{O}(m^2)$. 

Inserting the explicit values $\alpha=\gamma=1$, $\beta=\delta=\frac{2}{3}$
$\tau=-\frac{1}{3}$ for the Polchinski-Strassler background, we find
that the only non-vanishing coefficients are given by
\begin{equation}\label{BuCuPS}
\begin{aligned}
B_u&=-\frac{\zeta^2R^4}{216}\big(
3(m_2+m_3)^2+5(m_2-m_3)^2
-4m_1^2+4m_2m_3(1-\cos2\hat\psi)\big)\col\\
C_u
&=-\frac{\zeta^2R^4}{324}\big(
7(m_2+m_3)^2-11(m_2-m_3)^2
-4m_1^2-36m_2m_3(1-\cos2\hat\psi)\big)\col\\
%C_u^{++}&=0\col\qquad
%C_u^{+-}=0\col\qquad
%C_u^{-+}=0\col\qquad
%C_u^{--}=0\col\qquad
\end{aligned}
\end{equation}
or respectively
\begin{equation}\label{BpsiCpsiPS}
\begin{aligned}
B_\psi
&=-\frac{\zeta^2R^4}{54}m_2m_3\sin2\hat\psi\col\\
C_\psi&=\frac{2\zeta^2R^4}{27}m_2m_3\sin2\hat\psi\col
\end{aligned}
\qquad
\begin{aligned}
C_\psi^{+-}&=\frac{\zeta^2R^4}{54}m_1(m_2+m_3)\col\\
C_\psi^{--}&=-\frac{\zeta^2R^4}{54}m_1(m_2-m_3)
\pnt
\end{aligned}
\end{equation}
%One finds the combinations 
%\begin{equation}
%\begin{aligned}
%B_u+C_u&=-\frac{\zeta^2R^4}{162}\big(
%\frac{23}{4}(m_2+m_3)^2-\frac{7}{4}(m_2-m_3)^2
%-5m_1^2-15m_2m_3(1-\cos2\hat\psi)\big)\col\\
%2B_u+C_u&=-\frac{\zeta^2R^4}{81}\big(
%4(m_2+m_3)^2+(m_2-m_3)^2
%-4m_1^2-6m_2m_3(1-\cos2\hat\psi)\big)\col\\
%2B_\psi+C_\psi
%&=\frac{\zeta^2R^4}{27}m_2m_3\sin2\hat\psi\pnt
%\end{aligned}
%\end{equation}

It is interesting to notice that the identification $\gamma=1$ 
implies that $C_u^I=0$ and $C_\psi^{\pm+}=0$.
This ensures that the radial embedding coordinate $u$ remains
independent of the angles in the $\text{D}7$-brane worldvolume coordinate 
system, regardless of the values of the masses and the other parameters.
Furthermore, the dependence of the angular embedding coordinate $\psi$ on 
the worldvolume angles is also reduced to only two spherical 
harmonics $y_{\pm-}$ in the generic mass case, and their corresponding 
coefficients $C_\psi^{\pm-}$ become independent of the unperturbed angle 
$\hat\psi$.
For $m_2=m_3$ the embedding depends only on $y_{+-}$, and 
for $m_1=0$ in any case $\psi$ does not depend on any of the $SO(4)$
spherical harmonics.

\subsection{The $\mathcal{N}=2$ case with  $m_1=0$ and $m_2=m_3=m$ revisited}

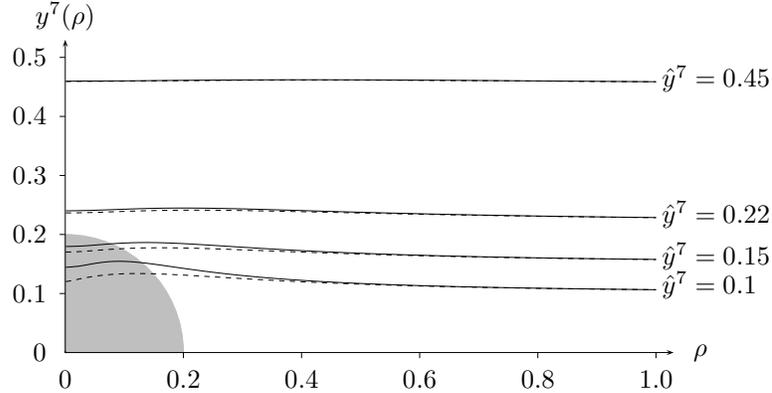
\begin{figure}[t]
\begin{center}
\begin{pspicture}(-0.1,-0.1)(1.1,0.625)
%\psframe(-0.1,-0.1)(1.2,0.625)
\footnotesize
\pscustom[linecolor=lightgray,fillcolor=lightgray,fillstyle=solid]{%
\psline(0,0)(0.2,0)
\psarc(0,0){0.2}{0}{90}
\psline(0,0.2)(0,0)
}
\psaxes[ticksize=2pt,tickstyle=bottom,Dx=0.2,Dy=0.1]{->}(0,0)(1.03,0.53)
\rput(0,0.57){$y^7(\rho)$}
\rput(1.075,0){$\rho$}
\rput[lB](1.01,0.1){$\hat y^7=0.1$}
\rput[lB](1.01,0.15){$\hat y^7=0.15$}
%\rput[lB](1.01,0.2){$\hat y^7=0.2$}
\rput[lB](1.01,0.22){$\hat y^7=0.22$}
\rput[lB](1.01,0.45){$\hat y^7=0.45$}
%\setlength{\unit}{0.0125\textwidth}
%\psset{xunit=\unit,yunit=\unit,runit=\unit}
\readdata{\data}{N2y7num0.1grid.dat}
\psset{linestyle=dashed,dash=2pt 2pt,plotstyle=curve}
\dataplot{\data}
%\listplot{\data}
\readdata{\data}{N2y7num0.15grid.dat}
\dataplot{\data}
%\listplot{\data}
%\readdata{\data}{N2y7num0.2grid.dat}
%\dataplot{\data}
%\listplot{\data}
\readdata{\data}{N2y7num0.22grid.dat}
\dataplot{\data}
%\listplot{\data}
\readdata{\data}{N2y7num0.45grid.dat}
\dataplot{\data}
%\listplot{\data}
\psset{linestyle=solid,dash=2pt 2pt}
\psplot{0.001}{1}{
x 0.1 div dup mul % x=(rho/u)^2
dup dup 1 add % x^2 x^2 1+x^2
ln exch 1 exch div % x^2 ln(1+x^2) 1/x^2
-0.04 mul 5 mul 3 div 0.1 div -0.25 mul % x^2 ln(1+x^2) -B/(4u x^2)
mul exch 1 add % -B/(4u x^2)*ln(1+x^2) 1+x^2
1 exch div % -B/(4u x^2)*ln(1+x^2) 1/(1+x^2)
0.04 mul 22 mul 9 div 0.1 div -0.125 mul % -B/(4u x^2)*ln(1+x^2) -C/(8u(1+x^2))
add 0.1 add}
\psplot{0.001}{1}{
x 0.15 div dup mul % x=(rho/u)^2
dup dup 1 add % x^2 x^2 1+x^2
ln exch 1 exch div % x^2 ln(1+x^2) 1/x^2
-0.04 mul 5 mul 3 div 0.15 div -0.25 mul % x^2 ln(1+x^2) -B/(4u x^2)
mul exch 1 add % -B/(4u x^2)*ln(1+x^2) 1+x^2
1 exch div % -B/(4u x^2)*ln(1+x^2) 1/(1+x^2)
0.04 mul 22 mul 9 div 0.15 div -0.125 mul %-B/(4u x^2)*ln(1+x^2) -C/(8u(1+x^2))
add 0.15 add}
%\psplot{0.001}{1}{
%x 0.2 div dup mul % x=(rho/u)^2
%dup dup 1 add % x^2 x^2 1+x^2
%ln exch 1 exch div % x^2 ln(1+x^2) 1/x^2
%-0.04 mul 5 mul 3 div 0.2 div -0.25 mul % x^2 ln(1+x^2) -B/(4u x^2)
%mul exch 1 add % -B/(4u x^2)*ln(1+x^2) 1+x^2
%1 exch div % -B/(4u x^2)*ln(1+x^2) 1/(1+x^2)
%0.04 mul 22 mul 9 div 0.2 div -0.125 mul %-B/(4u x^2)*ln(1+x^2) -C/(8u(1+x^2))
%add 0.2 add}
\psplot{0.001}{1}{
x 0.22 div dup mul % x=(rho/u)^2
dup dup 1 add % x^2 x^2 1+x^2
ln exch 1 exch div % x^2 ln(1+x^2) 1/x^2
-0.04 mul 5 mul 3 div 0.22 div -0.25 mul % x^2 ln(1+x^2) -B/(4u x^2)
mul exch 1 add % -B/(4u x^2)*ln(1+x^2) 1+x^2
1 exch div % -B/(4u x^2)*ln(1+x^2) 1/(1+x^2)
0.04 mul 22 mul 9 div 0.22 div -0.125 mul %-B/(4u x^2)*ln(1+x^2) -C/(8u(1+x^2))
add 0.22 add}
\psplot{0.001}{1}{
x 0.45 div dup mul % x=(rho/u)^2
dup dup 1 add % x^2 x^2 1+x^2
ln exch 1 exch div % x^2 ln(1+x^2) 1/x^2
-0.04 mul 5 mul 3 div 0.45 div -0.25 mul % x^2 ln(1+x^2) -B/(4u x^2)
mul exch 1 add % -B/(4u x^2)*ln(1+x^2) 1+x^2
1 exch div % -B/(4u x^2)*ln(1+x^2) 1/(1+x^2)
0.04 mul 22 mul 9 div 0.45 div -0.125 mul %-B/(4u x^2)*ln(1+x^2) -C/(8u(1+x^2))
add 0.45 add}
\end{pspicture}
\caption{Analytic (solid line) and exact numerical (dashed line) 
embeddings along $y^7$ for $F=0$ in the 
$\mathcal{N}=2$ background with masses $m_1=0$, $m_2=m_3=m=0.2$
and distinct boundary values $\hat y^7$. 
The grey quarter circle corresponds to $r\le mR^2$, into which the 
background generating $\text{D}3$-branes are expected to expand. 
Lengths and masses are dimensionless and measured in units of $R$ and 
$R^{-1}$ respectively. The dimensionless boundary value $\hat y^7$ determines
the dimensionful quark mass $m_\text{q}$ according to 
$m_\text{q}=\frac{R}{2\pi\alpha'}\hat y^7$.}
\label{fig:N2eqy7embed}
\end{center}
\end{figure}
If the mass parameters are given by $m_1=0$ and $m_2=m_3=m$, the dual 
gauge theory preserves $\mathcal{N}=2$ supersymmetries. 
Adding $\text{D}7$-brane probes as indicated in table \ref{tab:D3D7orient} 
should not break these supersymmetries.
We denote them as $\mathcal{N}=2_\parallel$ embeddings. 
Already at the end of section \ref{sec:gaugefield} we have stressed 
that in \cite{Apreda:2006bu} the $\mathcal{N}=2_\parallel$ embeddings have 
been studied by inserting the solution for the gauge field into the action 
before extracting the equations of motion for the embedding coordinates.  
The respective action is given by \eqref{Sembed} with $F=0$ and 
$\gamma$ and $\delta$ assuming the values given in \eqref{gammadeltaold}.
However, we should consider the gauge field as an independent field and thus 
insert the solution for its field strength $F$ into the equations of motion
for the embedding coordinates. 
In this case we identify $\gamma=\alpha$ and $\delta=\beta$ with 
the explicit values given in \eqref{const1}. 
According to \eqref{BuCudef} and \eqref{BpsiCpsidef} this alters the values 
of $C_u$ and $C_\psi$ w.r.t.\ the ones in \cite{Apreda:2006bu}, 
while $B_u$ and $B_\psi$ are independent of $\gamma$ and $\delta$ and thus 
remain unchanged.
All the other coefficients vanish for $m_1=0$ anyway.
The expressions which substitute the ones in \cite{Apreda:2006bu} 
then read\footnote{Our definitions for $B_\psi$ and $C_\psi$ differs 
from the ones in \cite{Apreda:2006bu}. To match the 
conventions there, we have to multiply our results by a factor 
$\frac{1}{\hat u}$.}
\begin{equation}
\begin{aligned}
C_u
=-\frac{\zeta^2m^2R^4}{81}(-2+9\cos2\hat\psi)\col\qquad
C_\psi
=\frac{2\zeta^2m^2R^4}{27}\sin2\hat\psi
\pnt
\end{aligned}
\end{equation}
Our analytic solutions with $\hat\psi=0$ here and in \cite{Apreda:2006bu} 
are based on $F=0$. As is seen directly 
from \eqref{BuCudef}, with $\hat\psi=0$  
also $C_u$ does not depend on $\gamma$ and $\delta$ and $C_\psi=0$. 
The corresponding embeddings thus coincide for both treatments of gauge field. 

For $\hat\psi=\frac{\pi}{2}$ a difference arises in the 
radial embedding coordinate $u$. While in \cite{Apreda:2006bu} $C_u$ is 
negative for any choice of $\hat\psi$, here it becomes positive for 
$\cos2\hat\psi<\frac{2}{9}$ which in particular is the case for 
$\hat\psi=\frac{\pi}{2}$.
This changes 
the behaviour of the solution. In \cite{Apreda:2006bu} 
for any angle $\hat\psi$ the function $u(\rho)$ is monotonically 
decreasing and hence corresponds to the first case in figure 
\ref{fig:embcases}. 
Here, the function $u(\rho)$ assumes a relative maximum at an  
intermediate value $\rho$ if $\hat\psi$ fulfills 
\begin{equation}
\cos2\hat\psi<-\frac{8}{15}\pnt
\end{equation}
In particular this is the case for $\hat\psi=\frac{\pi}{2}$.

In figure \ref{fig:N2eqy7embed} we compare the analytic solution with the 
numerical one for $F=0$.
The latter is based on the action 
\begin{equation}\label{N2y4action}
\begin{aligned}
S&=-\frac{T_7}{\e^{\hat\phi}}
\int\de^4\xi\de\Omega_3\de\rho\Big[
\rho^3 \sqrt{1+y'^2}\\
&\phantom{{}={}}
%&\phantom{{}={}-\frac{T_7}{\e^{\hat\phi}}\int\de^4\xi\de\Omega_3\de\rho\Big[}
+\frac{\rho^3 m^2 R^4}{36r^4\sqrt{1+y'^2}}\Big(
10\rho^2+14y^2+23\rho^2y'^2+y^2y'^2-26\rho y y'
%\\
%&\phantom{{}={}+\frac{\rho^3m^2R^4}{36r^4\sqrt{1+y'^2}}\Big(} 
-24\sqrt{1+y'^2}(y^2-\rho y y')\Big)
\Big]\col
\end{aligned}
\end{equation}
where $y=y^7$. This result replaces the corresponding one in 
\cite{Apreda:2006bu}. We should stress that even if   
the numerical embeddings are obtained from the action 
\eqref{N2y4action} without making use of the expansion \eqref{yembedexpand}, 
they are not independent of it. The expansion
has already been used in section \ref{sec:gaugefield} 
to obtain $F=0$ which then enters the action \eqref{N2y4action}.\footnote{
An insertion of $F=0$ into the equations of motion is equivalent to 
an insertion directly into the action.}    
For a complete independence of the expansion, one should 
solve the equations of motion for $F$ and for the embedding 
coordinates directly as a coupled system.
We refrain from this more complicated analysis, since the 
similarity of the numerical results and the exact solutions in 
figure \ref{fig:N2eqy7embed} suggests that this should not change the 
numerical result significantly.

\subsection{The $\mathcal{N}=2$ case with $m_1=m_2=m$ and $m_3=0$}

\begin{figure}[t]
\begin{center}
\begin{pspicture}(-0.1,-0.1)(1.2,0.625)
%\psframe(-0.1,-0.1)(1.2,0.625)
\footnotesize
\pscustom[linecolor=lightgray,fillcolor=lightgray,fillstyle=solid]{%
\psline(0,0)(0.2,0)
\psarc(0,0){0.2}{0}{90}
\psline(0,0.2)(0,0)
}
\psaxes[ticksize=2pt,tickstyle=bottom,Dx=0.2,Dy=0.1]{->}(0,0)(1.03,0.53)
\rput(0,0.57){$u(\rho)$}
\rput(1.075,0){$\rho$}
\rput[lB](1.01,0.1){$\hat u=0.1$}
\rput[lB](1.01,0.15){$\hat u=0.15$}
%\rput[lB](1.01,0.2){$\hat u=0.2$}
\rput[lB](1.01,0.22){$\hat u=0.22$}
\rput[lB](1.01,0.45){$\hat u=0.45$}
%\setlength{\unit}{0.0125\textwidth}
%\psset{xunit=\unit,yunit=\unit,runit=\unit}
\psplot{0.001}{1}{
x 0.1 div dup mul % x=(rho/u)^2
dup dup 1 add % x^2 x^2 1+x^2
ln exch 1 exch div % x^2 ln(1+x^2) 1/x^2
-0.04 mul 1 mul 3 div 0.1 div -0.25 mul % x^2 ln(1+x^2) -B/(4u x^2)
mul exch 1 add % -B/(4u x^2)*ln(1+x^2) 1+x^2
1 exch div % -B/(4u x^2)*ln(1+x^2) 1/(1+x^2)
0.04 mul 8 mul 9 div 0.1 div -0.125 mul % -B/(4u x^2)*ln(1+x^2) -C/(8u(1+x^2))
add 0.1 add}
\psplot{0.001}{1}{
x 0.15 div dup mul % x=(rho/u)^2
dup dup 1 add % x^2 x^2 1+x^2
ln exch 1 exch div % x^2 ln(1+x^2) 1/x^2
-0.04 mul 1 mul 3 div 0.15 div -0.25 mul % x^2 ln(1+x^2) -B/(4u x^2)
mul exch 1 add % -B/(4u x^2)*ln(1+x^2) 1+x^2
1 exch div % -B/(4u x^2)*ln(1+x^2) 1/(1+x^2)
0.04 mul 8 mul 9 div 0.15 div -0.125 mul % -B/(4u x^2)*ln(1+x^2) -C/(8u(1+x^2))
add 0.15 add}
\psplot{0.001}{1}{
x 0.22 div dup mul % x=(rho/u)^2
dup dup 1 add % x^2 x^2 1+x^2
ln exch 1 exch div % x^2 ln(1+x^2) 1/x^2
-0.04 mul 1 mul 3 div 0.22 div -0.25 mul % x^2 ln(1+x^2) -B/(4u x^2)
mul exch 1 add % -B/(4u x^2)*ln(1+x^2) 1+x^2
1 exch div % -B/(4u x^2)*ln(1+x^2) 1/(1+x^2)
0.04 mul 8 mul 9 div 0.22 div -0.125 mul % -B/(4u x^2)*ln(1+x^2) -C/(8u(1+x^2))
add 0.22 add}
\psplot{0.001}{1}{
x 0.45 div dup mul % x=(rho/u)^2
dup dup 1 add % x^2 x^2 1+x^2
ln exch 1 exch div % x^2 ln(1+x^2) 1/x^2
-0.04 mul 1 mul 3 div 0.45 div -0.25 mul % x^2 ln(1+x^2) -B/(4u x^2)
mul exch 1 add % -B/(4u x^2)*ln(1+x^2) 1+x^2
1 exch div % -B/(4u x^2)*ln(1+x^2) 1/(1+x^2)
0.04 mul 8 mul 9 div 0.45 div -0.125 mul % -B/(4u x^2)*ln(1+x^2) -C/(8u(1+x^2))
add 0.45 add}
\end{pspicture}
\caption{Radial embedding in the $\mathcal{N}=2$ background with 
masses $m_1=m_2=m=0.2$, $m_3=0$ and distinct boundary values $\hat u$. 
The grey quarter circle corresponds to $r\le mR^2$, into which the 
background generating $\text{D}3$-branes are expected to expand. 
Lengths and masses are dimensionless and measured in units of $R$ and 
$R^{-1}$ respectively. The dimensionless boundary value $\hat u$ determines
the dimensionful quark mass $m_\text{q}$ according to 
$m_\text{q}=\frac{R}{2\pi\alpha'}\hat u$.
}
\label{fig:N2perpembed}
\end{center}
\end{figure}
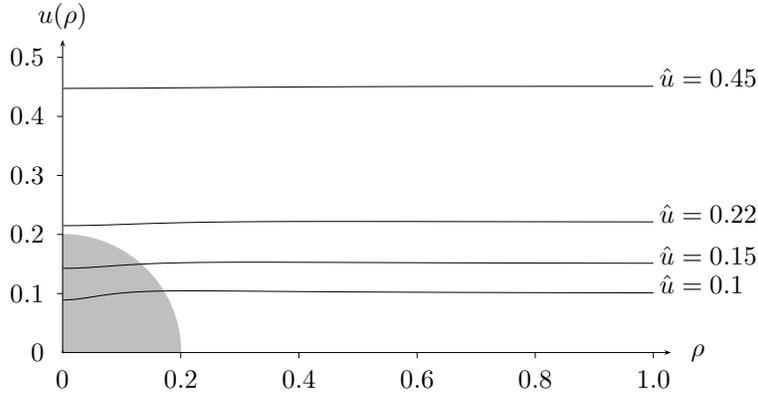
Since we have at hand the expression with generic masses, we can easily
study the case $m_1=m_2=m$ and $m_3=0$ in which 
the Polchinski-Strassler background still preserves $\mathcal{N}=2$ 
supersymmetries, but the embeddings oriented as shown in table 
\ref{tab:D3D7orient} should break (part of) the 
supersymmetries. We denote them as the $\mathcal{N}=2_\perp$ embeddings. 
Due to \eqref{BuCuPS} and \eqref{BpsiCpsiPS} the 
non-vanishing coefficients are given by
\begin{equation}
\begin{aligned}
B_u=-\frac{\zeta^2m^2R^4}{54}\col\qquad
C_u=\frac{4\zeta^2m^2R^4}{81}\col\qquad
C_\psi^{\pm-}=\pm\frac{\zeta^2m^2R^4}{54}
\pnt
\end{aligned}
\end{equation}
They do not depend on the unperturbed angular embedding coordinate $\hat\psi$.
According to \eqref{fregasymp} and figure \ref{fig:embcases}, the relations 
between $B_u$ and $C_u$ tell us that $u(\rho)$ is not monotonic, assuming a 
maximum at an intermediate value $\rho$. The radial embeddings are shown in 
figure \ref{fig:N2perpembed}. 
The angular embedding depends on all three angles in the four 
worldvolume coordinates $y^a$ via the two $SO(4)$ spherical harmonics 
$y_{+-}$ and $y_{--}$ defined in \eqref{SO4shdef}.

\subsection{The $\mathcal{N}=1$ case with equal masses}

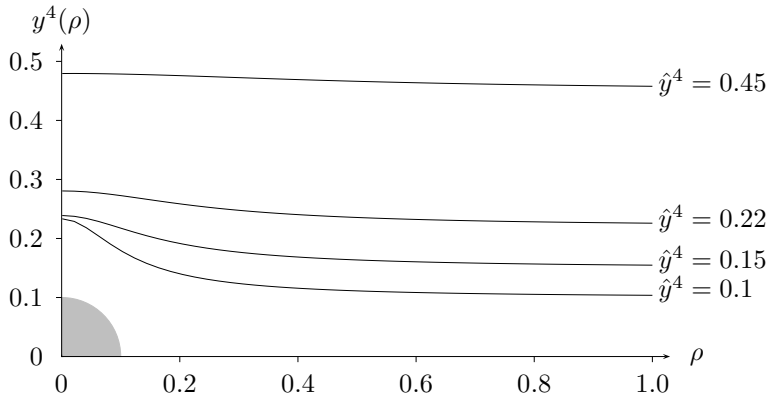
\begin{figure}[t]
\begin{center}
\begin{pspicture}(-0.1,-0.1)(1.2,0.625)
%\psframe(-0.1,-0.1)(1.2,0.625)
\footnotesize
\pscustom[linecolor=lightgray,fillcolor=lightgray,fillstyle=solid]{%
\psline(0,0)(0.1,0)
\psarc(0,0){0.1}{0}{90}
\psline(0,0.1)(0,0)
}
\psaxes[ticksize=2pt,tickstyle=bottom,Dx=0.2,Dy=0.1]{->}(0,0)(1.03,0.53)
\rput(0,0.57){$y^4(\rho)$}
\rput(1.075,0){$\rho$}
\rput[lB](1.01,0.1){$\hat y^4=0.1$}
\rput[lB](1.01,0.15){$\hat y^4=0.15$}
%\rput[lB](1.01,0.2){$\hat y^4=0.2$}
\rput[lB](1.01,0.22){$\hat y^4=0.22$}
\rput[lB](1.01,0.45){$\hat y^4=0.45$}
%\setlength{\unit}{0.0125\textwidth}
%\psset{xunit=\unit,yunit=\unit,runit=\unit}
\psplot{0.001}{1}{
x 0.1 div dup mul % x=(rho/u)^2
dup dup 1 add % x^2 x^2 1+x^2
ln exch 1 exch div % x^2 ln(1+x^2) 1/x^2
-0.04 mul 2 mul 3 div 0.1 div -0.25 mul % x^2 ln(1+x^2) -B/(4u x^2)
mul exch 1 add % -B/(4u x^2)*ln(1+x^2) 1+x^2
1 exch div % -B/(4u x^2)*ln(1+x^2) 1/(1+x^2)
-0.04 mul 4 mul 3 div 0.1 div -0.125 mul % -B/(4u x^2)*ln(1+x^2) -C/(8u(1+x^2))
add 0.1 add}
\psplot{0.001}{1}{
x 0.15 div dup mul % x=(rho/u)^2
dup dup 1 add % x^2 x^2 1+x^2
ln exch 1 exch div % x^2 ln(1+x^2) 1/x^2
-0.04 mul 2 mul 3 div 0.15 div -0.25 mul % x^2 ln(1+x^2) -B/(4u x^2)
mul exch 1 add % -B/(4u x^2)*ln(1+x^2) 1+x^2
1 exch div % -B/(4u x^2)*ln(1+x^2) 1/(1+x^2)
-0.04 mul 4 mul 3 div 0.15 div -0.125 mul %-B/(4u x^2)*ln(1+x^2) -C/(8u(1+x^2))
add 0.15 add}
%\psplot{0.001}{1}{
%x 0.2 div dup mul % x=(rho/u)^2
%dup dup 1 add % x^2 x^2 1+x^2
%ln exch 1 exch div % x^2 ln(1+x^2) 1/x^2
%-0.04 mul 2 mul 3 div 0.2 div -0.25 mul % x^2 ln(1+x^2) -B/(4u x^2)
%mul exch 1 add % -B/(4u x^2)*ln(1+x^2) 1+x^2
%1 exch div % -B/(4u x^2)*ln(1+x^2) 1/(1+x^2)
%-0.04 mul 4 mul 3 div 0.2 div -0.125 mul %-B/(4u x^2)*ln(1+x^2) -C/(8u(1+x^2))
%add 0.2 add}
\psplot{0.001}{1}{
x 0.22 div dup mul % x=(rho/u)^2
dup dup 1 add % x^2 x^2 1+x^2
ln exch 1 exch div % x^2 ln(1+x^2) 1/x^2
-0.04 mul 2 mul 3 div 0.22 div -0.25 mul % x^2 ln(1+x^2) -B/(4u x^2)
mul exch 1 add % -B/(4u x^2)*ln(1+x^2) 1+x^2
1 exch div % -B/(4u x^2)*ln(1+x^2) 1/(1+x^2)
-0.04 mul 4 mul 3 div 0.22 div -0.125 mul %-B/(4u x^2)*ln(1+x^2) -C/(8u(1+x^2))
add 0.22 add}
\psplot{0.001}{1}{
x 0.45 div dup mul % x=(rho/u)^2
dup dup 1 add % x^2 x^2 1+x^2
ln exch 1 exch div % x^2 ln(1+x^2) 1/x^2
-0.04 mul 2 mul 3 div 0.45 div -0.25 mul % x^2 ln(1+x^2) -B/(4u x^2)
mul exch 1 add % -B/(4u x^2)*ln(1+x^2) 1+x^2
1 exch div % -B/(4u x^2)*ln(1+x^2) 1/(1+x^2)
-0.04 mul 4 mul 3 div 0.45 div -0.125 mul %-B/(4u x^2)*ln(1+x^2) -C/(8u(1+x^2))
add 0.45 add}
\end{pspicture}
\caption{Embedding along $y^4$ in the $\mathcal{N}=1$ background with 
masses $m_1=m_2=m_3=m=0.2$ and distinct boundary values $\hat y^4$. 
The grey quarter circle corresponds to $r\le\frac{1}{2}mR^2$, into which the 
background generating $\text{D}3$-branes are expected to expand.
Lengths and masses are dimensionless and measured in units of $R$ and 
$R^{-1}$ respectively. The dimensionless boundary value $\hat y^4$ determines
the dimensionful quark mass $m_\text{q}$ according to 
$m_\text{q}=\frac{R}{2\pi\alpha'}\hat y^4$.
}
\label{fig:N1eqy4embed}
\end{center}
\end{figure}
\begin{figure}[t]
\begin{center}
\begin{pspicture}(-0.1,-0.1)(1.2,0.625)
%\psframe(-0.1,-0.1)(1.2,0.625)
\pscustom[linecolor=lightgray,fillcolor=lightgray,fillstyle=solid]{%
\psline(0,0)(0.3,0)
\psarc(0,0){0.3}{0}{90}
\psline(0,0.3)(0,0)
}
\footnotesize
\psaxes[ticksize=2pt,tickstyle=bottom,Dx=0.2,Dy=0.1]{->}(0,0)(1.03,0.53)
\rput(0,0.57){$y^7(\rho)$}
\rput(1.075,0){$\rho$}
\rput[lB](1.01,0.1){$\hat y^7=0.1$}
\rput[lB](1.01,0.15){$\hat y^7=0.15$}
%\rput[lB](1.01,0.2){$\hat y^7=0.2$}
\rput[lB](1.01,0.22){$\hat y^7=0.22$}
\rput[lB](1.01,0.45){$\hat y^7=0.45$}
%\setlength{\unit}{0.0125\textwidth}
%\psset{xunit=\unit,yunit=\unit,runit=\unit}
\psplot{0.001}{1}{
x 0.1 div dup mul % x=(rho/u)^2
dup dup 1 add % x^2 x^2 1+x^2
ln exch 1 exch div % x^2 ln(1+x^2) 1/x^2
-0.04 mul 4 mul 3 div 0.1 div -0.25 mul % x^2 ln(1+x^2) -B/(4u x^2)
mul exch 1 add % -B/(4u x^2)*ln(1+x^2) 1+x^2
1 exch div % -B/(4u x^2)*ln(1+x^2) 1/(1+x^2)
0.04 mul 8 mul 3 div 0.1 div -0.125 mul % -B/(4u x^2)*ln(1+x^2) -C/(8u(1+x^2))
add 0.1 add}
\psplot{0.001}{1}{
x 0.15 div dup mul % x=(rho/u)^2
dup dup 1 add % x^2 x^2 1+x^2
ln exch 1 exch div % x^2 ln(1+x^2) 1/x^2
-0.04 mul 4 mul 3 div 0.15 div -0.25 mul % x^2 ln(1+x^2) -B/(4u x^2)
mul exch 1 add % -B/(4u x^2)*ln(1+x^2) 1+x^2
1 exch div % -B/(4u x^2)*ln(1+x^2) 1/(1+x^2)
0.04 mul 8 mul 3 div 0.15 div -0.125 mul %-B/(4u x^2)*ln(1+x^2) -C/(8u(1+x^2))
add 0.15 add}
%\psplot{0.001}{1}{
%x 0.2 div dup mul % x=(rho/u)^2
%dup dup 1 add % x^2 x^2 1+x^2
%ln exch 1 exch div % x^2 ln(1+x^2) 1/x^2
%-0.04 mul 4 mul 3 div 0.2 div -0.25 mul % x^2 ln(1+x^2) -B/(4u x^2)
%mul exch 1 add % -B/(4u x^2)*ln(1+x^2) 1+x^2
%1 exch div % -B/(4u x^2)*ln(1+x^2) 1/(1+x^2)
%0.04 mul 8 mul 3 div 0.2 div -0.125 mul %-B/(4u x^2)*ln(1+x^2) -C/(8u(1+x^2))
%add 0.2 add}
\psplot{0.001}{1}{
x 0.22 div dup mul % x=(rho/u)^2
dup dup 1 add % x^2 x^2 1+x^2
ln exch 1 exch div % x^2 ln(1+x^2) 1/x^2
-0.04 mul 4 mul 3 div 0.22 div -0.25 mul % x^2 ln(1+x^2) -B/(4u x^2)
mul exch 1 add % -B/(4u x^2)*ln(1+x^2) 1+x^2
1 exch div % -B/(4u x^2)*ln(1+x^2) 1/(1+x^2)
0.04 mul 8 mul 3 div 0.22 div -0.125 mul %-B/(4u x^2)*ln(1+x^2) -C/(8u(1+x^2))
add 0.22 add}
\psplot{0.001}{1}{
x 0.45 div dup mul % x=(rho/u)^2
dup dup 1 add % x^2 x^2 1+x^2
ln exch 1 exch div % x^2 ln(1+x^2) 1/x^2
-0.04 mul 4 mul 3 div 0.45 div -0.25 mul % x^2 ln(1+x^2) -B/(4u x^2)
mul exch 1 add % -B/(4u x^2)*ln(1+x^2) 1+x^2
1 exch div % -B/(4u x^2)*ln(1+x^2) 1/(1+x^2)
0.04 mul 8 mul 3 div 0.45 div -0.125 mul %-B/(4u x^2)*ln(1+x^2) -C/(8u(1+x^2))
add 0.45 add}
\end{pspicture}
\caption{Embedding along $y^7$ in the $\mathcal{N}=1$ background with 
masses $m_1=m_2=m_3=0.2$ and distinct boundary values $\hat y^7$. 
The grey quarter circle corresponds to $r\le\frac{3}{2}mR^2$, into which the 
background generating $\text{D}3$-branes are expected to expand.
Lengths and masses are dimensionless and measured in units of $R$ and 
$R^{-1}$ respectively. The dimensionless boundary value $\hat y^7$ determines
the dimensionful quark mass $m_\text{q}$ according to 
$m_\text{q}=\frac{R}{2\pi\alpha'}\hat y^7$.
}
\label{fig:N1eqy7embed}
\end{center}
\end{figure}
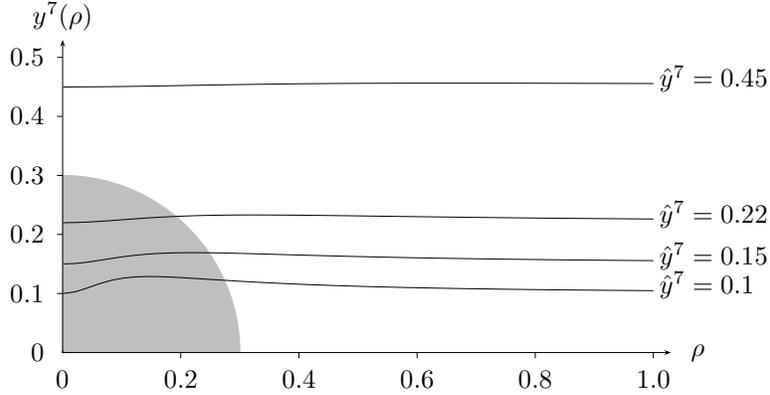

A case of particular interest is the one of $\text{D}7$-brane embeddings
into the $\mathcal{N}=1$ Polchinski-Strassler background with
equal masses. The non-vanishing coefficients of the solution in
\eqref{freg} are again found from  
\eqref{BuCuPS} and \eqref{BpsiCpsiPS}. They read
\begin{equation}
\begin{aligned}
B_u&=-\frac{\zeta^2m^2R^4}{54}(3-\cos2\hat\psi)\col\qquad
C_u&=\frac{\zeta^2m^2R^4}{27}(1-3\cos2\hat\psi)\col
\end{aligned}
\end{equation}
and
\begin{equation}
\begin{aligned}
B_\psi&=-\frac{\zeta^2m^2R^4}{54}\sin2\hat\psi\col\qquad
C_\psi=\frac{2\zeta^2m^2R^4}{27}\sin2\hat\psi\col\qquad
C_\psi^{+-}=\frac{\zeta^2m^2R^4}{27}
\pnt
\end{aligned}
\end{equation}
The function $u(\rho)$ ceases to be monotonic for
\begin{equation}
\cos2\hat\psi<-\frac{1}{5}\pnt
\end{equation}
As in the $\mathcal{N}=2$ case, this in particular happens for
$\hat\psi=\frac{\pi}{2}$. Embeddings with constant angular direction $\psi$ 
do not exist at all in the $\mathcal{N}=1$ case. However, the embeddings with 
$\hat\psi=0$ or $\hat\psi=\frac{\pi}{2}$ are still peculiar, since
they are directed along or respectively perpendicular to the  
principal axis with length determined by $m_1$ 
of the polarization ellipsoid of the $\text{D}3$-branes.

The explicit expression for the 
radial embedding coordinate in the case $\hat\psi=0$ where $u=y^4$ and 
$2B_u=C_u$ reads
\begin{equation}
\begin{aligned}
u&=\hat u\Big(1+\frac{\zeta^2m^2R^4}{108}\Big(
\frac{1}{\rho^2}\ln\frac{\hat r^2}{\hat u^2}+\frac{1}{\hat r^2}\Big)\Big)
\col
%u&=\hat u\Big(1-\frac{\zeta^2m^2R^4}{54}\Big(\frac{1}{\hat r^2}
%-\frac{1}{\rho^2}\ln\frac{\hat r^2}{\hat u^2}\Big)\Big)\col\\
\end{aligned}
\end{equation}
while for $\hat\psi=\frac{\pi}{2}$ where $u=y^7$ we find
\begin{equation}
\begin{aligned}
u&=\hat u\Big(1+\frac{\zeta^2m^2R^4}{54}\Big(
\frac{1}{\rho^2}\ln\frac{\hat r^2}{\hat u^2}-\frac{1}{\hat r^2}\Big)\Big)
\col\quad
%u&=\hat u\Big(1-\frac{\zeta^2m^2R^4}{54}\Big(\frac{1}{\hat r^2}
%-\frac{1}{\rho^2}\ln\frac{\hat r^2}{\hat u^2}\Big)\Big)\col\\
\end{aligned}
\end{equation}
which has the property that $u(0)=\hat u$ as follows from 
\eqref{fregasymp} with the relation $2B_u=-C_u$ in this case. 
We have printed the corresponding $y^4$ embedding in figure 
\ref{fig:N1eqy4embed} and the $y^7$ embedding in figure \ref{fig:N1eqy7embed}. 
The background generating $\text{D}3$-branes extend in these directions
with two different radii \cite{Apreda:2006bu}.

\begin{figure}[t]
\begin{center}
\leavevmode
\put(150,215){$\psi(\rho,\phi_1)$}
\put(300,63){$y^5=y^8$}
\put(243,172){$y^6=-y^9$}
%\fbox{
\epsfig{file=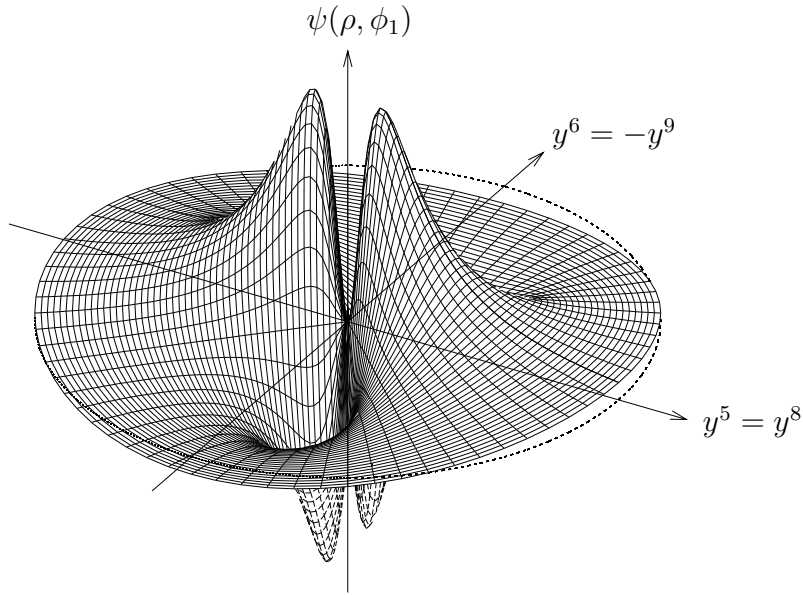,
bbllx=200,bblly=170,bburx=530,bbury=400}
%}
\caption{Form of the angular embedding in the $\mathcal{N}=1$ background with 
masses $m_1=m_2=m_3=m$ in the $2$-dimensional subplane defined by 
$y^5=y^8$, $y^6=-y^9$.}
\label{fig:N1eqy4psiembed}
\end{center}
\end{figure}

For the angular embeddings with $\hat\psi=0$ or $\hat\psi=\frac{\pi}{2}$ 
we find
$B_\psi=C_\psi=0$ like in the corresponding $\mathcal{N}=2$ cases. 
Only $C_\psi^{+-}$ is non-zero. Defining polar coordinates for 
the four worldvolume coordinates $y^a$ as in \eqref{realcoordangles}, the 
corresponding spherical harmonic $y_{+-}$ only depends on two combinations 
of the three worldvolume angles. 
Up to the respective constant boundary values $\hat\psi=0$ or 
$\hat\psi=\frac{\pi}{2}$, the angular embedding is identical for both cases. 
We find
\begin{equation}
\begin{aligned}
\psi=\hat\psi
+\frac{\zeta^2m^2R^4}{216}\Big(
\frac{2}{\rho^2}\Big(1-\frac{\hat u^2}{\rho^2}\ln\frac{\hat r^2}{\hat u^2}\Big)
-\frac{1}{\hat r^2}\Big)y_{+-}
\pnt
\end{aligned}
\end{equation}
In the $2$-dimensional subplane given by 
$y^5=y^8=\frac{\rho}{\sqrt{2}}\cos\phi_1$,
$y^6=-y^9=\frac{\rho}{\sqrt{2}}\sin\phi_1$ in the coordinates 
\eqref{realcoordangles}, in which according to
\eqref{SO4shangles} $y_{+-}=\cos2\phi_1$, the angular embedding is shown in 
figure \ref{fig:N1eqy4psiembed}.

%\subsection{The $\mathcal{N}=1$ case with $m_1=\mu m$, $m_2=m$ and $m_3=0$}

%Since we have at hand the expression with generic masses, it appears also 
%to be interesting to study the case $m_1=\mu m$, $m_2=m$ and $m_3=0$. 
%Due to \eqref{BuCuPS} and \eqref{BpsiCpsiPS} the non-vanishing coefficents 
%are given by
%\begin{equation}
%\begin{aligned}
%B_u=-\frac{\zeta^2m^2R^4}{54}(2-\mu^2)\col\qquad
%C_u=\frac{2\zeta^2R^4}{81}(1+\mu^2)\col\qquad
%C_\psi^{\pm-}=\pm\frac{\zeta^2m^2R^4}{54}\mu
%\pnt
%\end{aligned}
%\end{equation}
%The function $u(\rho)$ is non-monotonic for $\mu^2>\frac{4}{5}$, i.e.\ also 
%in the case with $\mu=1$ in which $m_1=m_2=m$ and $m_3=0$.
%In this case the Polchinski-Strassler background itself preserves 
%$\mathcal{N}=2$ supersymmetries. We also see from \eqref{BuBumrel} that 
%for $\mu=1$ a change of the relative sign between the DBI and CS action 
%does not yield a monotonic function $u(\rho)$.  
%%\begin{equation}
%%\begin{aligned}
%%B_u^{(-)}=\frac{\zeta^2m^2R^4}{54}(1-2\mu^2)\col\qquad
%%C_u^{(-)}
%%=\frac{\zeta^2R^4}{81}(1+\mu^2)\col\qquad
%%C_\psi^{(-)\pm-}=\mp\frac{\zeta^2m^2R^4}{54}\mu
%%\pnt
%%\end{aligned}
%%\end{equation}
%With the altered sign, the embedding is non-monotonic for $\mu^2<\frac{5}{4}$.

\subsection{Error estimates}

\setlength{\extrarowheight}{-1pt}
\renewcommand{\arraystretch}{1.5}
\begin{table}[t]
\begin{center}
\begin{tabular}{|c||c|c|c||c|c||c|c|c|}
\hline
& \multicolumn{3}{c||}{$\mathcal{N}=2_\parallel$} 
& \multicolumn{2}{c||}{$\mathcal{N}=2_\perp$}
& \multicolumn{3}{c|}{$\mathcal{N}=1$} \\
\hline
$\hat\psi$ 
& $0$ & \multicolumn{2}{c||}{$\frac{\pi}{2}$} 
& \multicolumn{2}{c||}{} 
& $0$ & \multicolumn{2}{c|}{$\frac{\pi}{2}$} \\
\hline
\hline
$\frac{B_u}{m^2R^4}$ 
& $-1$ & \multicolumn{2}{c||}{$-\frac{5}{3}$} 
& \multicolumn{2}{c||}{$-\frac{1}{3}$}
& $-\frac{2}{3}$ & \multicolumn{2}{c|}{$-\frac{4}{3}$} \\
\hline
$\frac{C_u}{m^2R^4}$ 
& $-\frac{14}{9}$ & \multicolumn{2}{c||}{$\frac{22}{9}$}
& \multicolumn{2}{c||}{$\frac{8}{9}$}
& $-\frac{4}{3}$ & \multicolumn{2}{c|}{$\frac{8}{3}$} \\
\hline
\hline
$\frac{\rho_\text{e}}{\hat u}$ 
& $0$ & $0$ & $0.912$ 
& $0$ & $2.140$
& $0$ & $0$ & $1.471$ \\
\hline
$\frac{\hat u\tilde u(\rho_\text{e})}{m^2R^4}$ 
& $\frac{4}{9}$ & $\frac{1}{9}$ & $0.136$ 
& $-\frac{1}{36}$ & $0.011$
& $\frac{1}{3}$ & $0$ & $0.072$ \\
\hline
\hline
$\frac{\rho_\text{t}}{\hat u}$ 
& $0.640$ & $0.383$ & $1.814$ 
& $0.501$ & $3.273$
& $0.630$ & $0.462$ & $2.432$ \\
\hline
$\frac{\hat u^2u'(\rho_\text{t})}{m^2R^4}$ 
& $-0.226$ & $0.045$ & $-0.036$ 
& $0.040$ & $-0.002$
& $-0.174$ & $-0.090$ & $-0.016$ \\
\hline
\end{tabular}
\caption{The three different types of radial embeddings with their values of 
the constants $B_u$ and $C_u$ in units of $m^2R^4$ and the corresponding 
extrema $\rho_\text{e}$ and turning 
points $\rho_\text{t}$ in units of $\hat u$. The correction 
$\tilde u(\rho_\text{e})$ is measured in units $\frac{m^2R^4}{\hat u}$ and 
the first derivative 
$u'(\rho_\text{t})$ in units of $\frac{m^2R^4}{\hat u^2}$.}
\label{fig:turninpoints}
\end{center}
\end{table}
With the expansion \eqref{yembedexpand}
we have found the regular analytic solutions \eqref{freg}
for the embedding coordinates. In the following we will analyse in which 
regimes these solutions with underlying action \eqref{Sembed}
are good approximations to the exact solutions which we can only find 
numerically from the corresponding action \eqref{Sexpand}.
We recall that by exact solutions we 
mean the exact solutions in the order $\mathcal{O}(m^2)$ 
Polchinski-Strassler background. One should keep in mind that 
even these solutions are limited to the regime in which a 
perturbative expansion of the background around $\AdS_5\times\text{S}^5$
is justified. The embeddings should avoid the deep interior of the space in 
which the extension of the background generating $\text{D}3$-brane sources
becomes important. This requires $r\gtrsim mR^2$. 
The embeddings that are attracted by the origin of the space 
should have a boundary condition $\hat u\gtrsim mR^2$, while the ones 
that are repulsed can stay away from the
interior of the space also for $\hat u\ll mR^2$ .

To derive from the action \eqref{Sexpand}
the result \eqref{Sembed} we have neglected terms that are beyond linear order 
in $\tilde y^m$ and in the derivatives $\partial_a\tilde y^m$. 
This allows us to estimate an upper bound for the difference between 
the exact numerical embedding and the corresponding analytic solution.
It should be given by 
$\max\big(\big|\frac{\tilde y}{\hat y}\big|,|\partial\tilde y|\big)$. 
In table \ref{fig:turninpoints} we show the extrema $\rho_\text{e}$ and 
turning points  $\rho_\text{t}$ of the radial embedding coordinates in 
the previously discussed special cases.
The presented normalized expressions are independent of the explicit values 
of $m$, $R$ and $\hat u$.
To find the relative deviation from the exact result for the radial 
embedding on first has to select the maximum value of 
$\big|\frac{\tilde u(\rho_\text{e})}{\hat u}\big|$ and 
$|u'(\rho_\text{t})|$. 
For $\hat u=mR^2$ this directly gives the respective
upper bound on the relative deviation. 
For $\hat u\neq mR^2$ we also have to 
restore the normalization by multiplying with $\frac{m^2R^4}{\hat u^2}$.
For the $y^4$ embedding in the $\mathcal{N}=2_\parallel$ case the deviation is 
quite substantial with $\frac{4}{9}\simeq 44.4\%$. Doubling $\hat u$ brings
it already down to $11.1\%$. 
For $\hat u=mR^2$ the $y^7$ embedding only deviates $13.6\%$ from the exact 
solution. 
In all cases, the non-monotonic embeddings are much more accurately described 
by the analytic solution than the monotonic ones. 
Furthermore, the situation improves in the $\mathcal{N}=2_\perp$ case and 
in the $\mathcal{N}=1$ background with equal masses.

We should stress here that $\hat u=mR^2$ in general yields embeddings that 
run into a regime where the error from the perturbative expansion of the 
background itself should already be substantial. The corresponding analytic
as well as the exact solutions should not be trusted carelessly.
Only in the case of monotonically decreasing embeddings
the exact solutions are substantially superior to the analytic ones, since 
they avoid the region of small $r$ even for $\hat u\ll mR^2$ and hence 
can be trusted. The analytic solution does not hold for $\hat u\ll mR^2$.
The situation is different for the non-monotonic embeddings. 
The ones that avoid the region of small $r$ obey $\hat u\gtrsim mR^2$ and thus 
are accurately described also by the analytic solution.
A comparison with figure \ref{fig:N2eqy7embed} furthermore
suggests that the above described procedure provides quite appropriate
estimates of the deviation.

\section{Holographic renormalization}
\label{sec:holoren}

The non-constant boundary behaviour \eqref{fregasymp} of the found embeddings
might imply the presence of a VEV for the fermion bilinear (quark condensate)
in the dual gauge theory which would break supersymmetry. In the holographic 
gravity description the VEV is determined by varying the on-shell action 
w.r.t.\ the boundary value $\hat u$. This procedure requires holographic 
renormalization \cite{Skenderis:2002wp,deHaro:2000xn,Karch:2005ms} to cancel 
the occurring divergences by appropriate counterterms. 
By also including appropriate finite 
counterterms the renormalized on-shell action can be made vanishing. 
In particular, a quark condensate is hence absent and 
our found embeddings of the form \eqref{freg} are at least consistent 
with supersymmetry. Surprisingly, the procedure works for the action
\eqref{Sembed} independent of the concrete values for the introduced 
constants. Therefore, also the case with an alternative relative sign choice 
between the DBI and CS action is covered, implying that the procedure does not 
provide further information for finally fixing this sign. 

In appendix \ref{app:expaction} we derive the explicit form of the 
expanded action \eqref{Sembed}, which is then given by \eqref{Lexppi}.
The terms of relevance for the holographic renormalization procedure 
read
\begin{equation}
\begin{aligned}
S
&=-\frac{T_7}{\e^{\hat\phi}}\int\de\xi^4\de\Omega_3\de\rho\rho^3\Big[
1%+\partial_A\tilde u\partial_{\bar A}\tilde u
%+\hat u^2\partial_A\tilde\psi\partial_{\bar A}\tilde\psi
+\frac{B_u}{2}\frac{\hat u^2}{\hat r^4}\\
%&\phantom{{}={}-\frac{T_7}{\e^{\hat\phi}}
%\int\de\xi^4\de\Omega_3\de\rho\rho^3\Big[}
&\qquad\qquad\quad
+\frac{\zeta^2}{216}\hat Z\Big(
\frac{5}{3}M^2(\hat r^2+\hat u^2)%\\
%&\phantom{{}={}-\frac{T_7}{\e^{\hat\phi}}
%\int\de\xi^4\de\Omega_3\de\rho\rho^3\Big[
%+\frac{\zeta^2}{216}Z\Big(}
+3m_1((m_2+m_3)y_{++}-(m_2-m_3)y_{-+})\rho^2\Big)\Big]
\col
\end{aligned}
\end{equation}
where we remind that $M^2$ is the sum of all three mass squares as defined in
\eqref{T3abssquare}.
We transform to a new coordinate $\chi$ that parameterizes the 
radial direction. The relations read
\begin{equation}
\hat r=\frac{1}{\sqrt{\chi}}\col\qquad
%\frac{\rho\de\rho}{\hat r}=-\frac{\de\chi}{2\sqrt{\chi}^3}\col\qquad
\rho^2=\frac{1}{\chi}-\hat u^2\col\qquad
\rho\de\rho=-\frac{\de\chi}{2\chi^2}\col\qquad
\partial_\rho=-2\rho\chi^2\partial_\chi
\pnt
\end{equation}
Performing the radial integration over the interval 
$\varepsilon\le\chi\le\frac{1}{\hat u^2}$, the 
regularized on-shell action is given by
\begin{equation}\label{Sreg}
\begin{aligned}
S_\text{reg}
&=-\frac{T_7}{2\e^{\hat\phi}}\int\de\xi^4\de\Omega_3
\Big[\frac{1}{2\varepsilon^2}+\frac{\hat u^4}{2}-\frac{\hat u^2}{\varepsilon}
-B_u\frac{\hat u^2}{2}(\ln\varepsilon\hat u^2+1-\varepsilon\hat u^2)\\
&\phantom{{}={}-\frac{T_7}{2\e^{\hat\phi}}\int\de\xi^4\de\Omega_3\Big[}
+\frac{\zeta^2R^4}{216}
\frac{5}{3}M^2
\Big(\frac{1}{\varepsilon}-2\hat u^2+\varepsilon\hat u^4\Big)\Big]
\pnt
\end{aligned}
\end{equation}
From \eqref{freg} we derive the solution for the radial embedding coordinate
in the variable $\chi$. It is given by
\begin{equation}
\begin{aligned}
u&=\hat u-\frac{\hat u}{8}\chi\Big(
-B_f\frac{2}{1-\chi\hat u^2}\ln\chi\hat u^2
+C_f
+C_f^I\Big(1
-\frac{2}{1-\chi\hat u^2}
-\frac{2\chi\hat u^2}{(1-\chi\hat u^2)^2}\ln\chi\hat u^2\Big)y_I\Big)
\pnt
\end{aligned}
\end{equation}
We evaluate the above relation at $\chi=\varepsilon$, expand it up to 
order $\mathcal{O}(\varepsilon^2)$, and invert it to express the 
boundary value $\hat u$ in terms of the value $u_\varepsilon=u(\varepsilon)$.
The inverted relation then reads
\begin{equation}
\begin{aligned}
\hat u
&=u_\varepsilon\Big(1
+\frac{\varepsilon}{8}\big(
-2B_u(1+\varepsilon u_\varepsilon^2)\ln\varepsilon u_\varepsilon^2
+C_u
-C_u^I\big(1
+2\varepsilon u_\varepsilon^2(1+\ln\varepsilon u_\varepsilon^2)\big)y_I\big)
\Big)\pnt
\end{aligned}
\end{equation}
This result is inserted into the regularized on-shell action \eqref{Sreg}.
The terms that contain a single spherical harmonic $y_I$ drop out when the 
angle integration is performed. Thus we obtain
\begin{equation}
\begin{aligned}
S_\text{reg}
&=-\frac{T_7}{\e^{\hat\phi}}\frac{\Omega_3}{2}\int\de\xi^4
\Big[\frac{1}{2\varepsilon^2}
+\frac{u_\varepsilon^4}{2}-\frac{u_\varepsilon^2}{\varepsilon}
-(2B_u+C_u)\frac{u_\varepsilon^2}{4}
(1-\varepsilon u_\varepsilon^2)\\
&\phantom{{}={}-\frac{T_7}{\e^{\hat\phi}}\frac{\Omega_3}{2}\int\de\xi^4\Big[}
+\frac{\zeta^2R^4}{216}\frac{5}{3}M^2
\Big(\frac{1}{\varepsilon}-2u_\varepsilon^2+\varepsilon u_\varepsilon^4\Big)
\Big]
\col
\end{aligned}
\end{equation}
where $\Omega_3$ is the volume of the unit $\text{S}^3$.
In contrast to \eqref{Sreg}, which is a functional of the boundary value 
$\hat u$, the above result depends on the data 
$u_\varepsilon$ at the regulator hypersurface at $\chi=\varepsilon$.
By this change of variables the logarithmic term that is present in 
\eqref{Sreg} cancels out. 
With the local counterterm action given by
\begin{equation}
\begin{aligned}
S_\text{ct}
&=\frac{T_7}{\e^{\hat\phi}}\frac{\Omega_3}{2}\int\de\xi^4
\Big[\frac{1}{2\varepsilon^2}
+\frac{u_\varepsilon^4}{2}-\frac{u_\varepsilon^2}{\varepsilon}
-(2B_u+C_u)\frac{u_\varepsilon^2}{4}%\\
%&\phantom{{}={}-\frac{T_7}{\e^{\hat\phi}}\frac{\Omega_3}{2}\int\de\xi^4\Big[}
+\frac{\zeta^2R^4}{216}\frac{5}{3}M^2
\Big(\frac{1}{\varepsilon}-2u_\varepsilon^2\Big)
\Big]
\end{aligned}
\end{equation}
we can then make the subtracted action
$S_\text{sub}=S_\text{reg}+S_\text{ct}$ vanish. To this purpose we have 
also included finite counterterms in the above expression.
The explicit form of the combination $2B_u+C_u$ found from \eqref{BuCudef} 
is given by
\begin{equation}\label{BuCucomb}
\begin{aligned}
2B_u+C_u
%&=\e^{\hat\phi}\frac{\alpha^2}{18R^4}\Big(
%(m_2+m_3)^2\Big(\tau-\frac{5}{9}\Big)
%+(m_2-m_3)^2\Big(\gamma-\delta-\frac{5}{9}\Big)\\
%&\phantom{{}={}\e^{\hat\phi}\frac{\alpha^2}{18R^4}\big(}
%+m_1^2\Big(\gamma+\delta-\tau-\frac{10}{9}\Big)
%+2m_2m_3(\gamma-\delta-\tau)(1-\cos2\hat\psi)\Big)\\
&=\frac{\zeta^2R^4}{18}\Big(
(m_2+m_3)^2\tau
+(m_2-m_3)^2(\gamma-\delta)\\
&\phantom{{}={}\frac{\zeta^2R^4}{18}\Big(}
+m_1^2(\gamma+\delta-\tau)
+2m_2m_3(\gamma-\delta-\tau)(1-\cos2\hat\psi)%\\
%&\phantom{{}={}\frac{\zeta^2R^4}{18}\Big(}
-\frac{10}{9}M^2\Big)%\\
%
%B_u+C_u&=\frac{\zeta^2R^4}{36}\Big(
%(m_2+m_3)^2\Big(\tau-\frac{17}{18}\Big)
%-(m_2-m_3)^2\Big(\alpha-\beta-2(\gamma-\delta)-\frac{1}{18}\Big)\\
%&\phantom{{}={}\frac{\zeta^2R^4}{36}\Big(}
%-m_1^2\Big(\alpha+\beta-2(\gamma+\delta)+\tau+\frac{8}{9}\Big)\\
%&\phantom{{}={}\frac{\zeta^2R^4}{36}\Big(}
%-2m_2m_3(\alpha-\beta-2(\gamma-\delta)+\tau-1)
%(1-\cos2\hat\psi)\Big)
\pnt
\end{aligned}
\end{equation}
The counter term action is then explicitly given by
\begin{equation}\label{Sctexpl}
\begin{aligned}
S_\text{ct}
&=\frac{T_7}{\e^{\hat\phi}}\frac{\Omega_3}{2}\int\de\xi^4
\Big[\frac{1}{2\varepsilon^2}
+\frac{u_\varepsilon^4}{2}-\frac{u_\varepsilon^2}{\varepsilon}
+\frac{\zeta^2R^4}{216}\frac{5}{3\varepsilon}M^2
\\
&\phantom{{}={}\frac{T_7}{\e^{\hat\phi}}\frac{\Omega_3}{2}\int\de\xi^4\Big[}
-\frac{\zeta^2R^4}{72}\Big(
(m_2+m_3)^2\tau
+(m_2-m_3)^2(\gamma-\delta)\\
&\phantom{{}={}\frac{T_7}{\e^{\hat\phi}}\frac{\Omega_3}{2}\int\de\xi^4\Big[
-\frac{\zeta^2R^4}{72}\Big(}
+m_1^2(\gamma+\delta-\tau)
+2m_2m_3(\gamma-\delta-\tau)(1-\cos2\hat\psi)\Big)u_\varepsilon^2
\Big]
\pnt
\end{aligned}
\end{equation} 
For any values for the constants and mass parameters
that enter the action \eqref{Sembed} we can therefore obtain 
$S_\text{sub}=0$ at least up to order $\mathcal{O}(m^2)$ and hence 
show that no quark condensate can be present up to this order.  

We should remark that in the $\mathcal{N}=2$ Polchinski-Strassler 
background with the parameters given by \eqref{const1} and 
\eqref{gammadeltaold}
our result \eqref{Sctexpl} should reduce to the one found in 
\cite{Apreda:2006bu}.
However, our equation (F.1) in \cite{Apreda:2006bu} contains an 
error. It only affects the embedding with $\hat\psi=\frac{\pi}{2}$, 
since it is caused by a wrong sign in front of a term which is proportional to
$1-\cos2\hat\psi$. To correct this mistake, 
one has to replace $-(\frac{1}{3}+\cos2\hat\psi)$ in the second line of 
(F.1) by $-(\frac{7}{3}-\cos2\hat\psi)$. In equation (F.3) one then has to set 
$c_0=-\frac{10}{3}$. This mistake has no further effect, since the 
statement $c_0+c_1-\frac{5}{3}=0$, which is essential for the procedure to 
succeed,
is in fact only fulfilled by the corrected numerical value. The finite 
counterterm then depends on $\hat\psi$, as is also seen from the above 
result \eqref{Sctexpl}.
%In the case $m_2=m_3=m$, $m_1=0$ this becomes
%\begin{equation}
%\begin{aligned}
%S_\text{ct}
%&=\frac{T_7}{2}\e^{\hat\phi}\int\de\xi^4\de\Omega_3
%%\Big[\frac{1}{2\varepsilon^2}
%+\frac{u_\varepsilon^4}{2}-\frac{u_\varepsilon^2}{\varepsilon}
%+\frac{\zeta^2m^2R^4}{108}\frac{5}{3\varepsilon}\\
%&\phantom{{}={}\frac{T_7}{2}\e^{\hat\phi}\int\de\xi^4\de\Omega_3\Big[}
%-\frac{\zeta^2m^2R^4}{36}\Big(
%2\tau+(\gamma-\delta-\tau)(1-\cos2\hat\psi)\Big)u_\varepsilon^2
%\Big]
%\pnt
%\end{aligned}
%\end{equation}
%To make contact with the analysis in \cite{Apreda:2006bu} we
%set $\gamma=\frac{1}{3}$, $\delta=\frac{4}{3}$, $\tau=-\frac{1}{3}$ and 
%find
%\begin{equation}
%\begin{aligned}
%S_\text{ct}
%&=\frac{T_7}{2}\e^{\hat\phi}\int\de\xi^4\de\Omega_3
%\Big[\frac{1}{2\varepsilon^2}
%+\frac{u_\varepsilon^4}{2}-\frac{u_\varepsilon^2}{\varepsilon}
%+\frac{\zeta^2m^2R^4}{108}\frac{5}{3\varepsilon}
%+\frac{\zeta^2m^2R^4}{54}(2-\cos2\hat\psi)u_\varepsilon^2
%\Big]
%\pnt
%\end{aligned}
%\end{equation}

\section{Conclusions}

In this paper we have analyzed the embedding of $\text{D}7$-brane probes into 
the Polchinski-Strassler background at order $\mathcal{O}(m^2)$, 
keeping the three mass perturbation parameters general. 
To this order we have seen that all embeddings are consistent 
with a vanishing gauge field strength $F=0$ on their worldvolumes. 
Thereby, the expansion of the 
embedding coordinates $y^m$ into a constant unperturbed embedding $\hat y^m$ 
in $\AdS_5\times\text{S}^5$ and a non-constant correction $\tilde y^m$ of 
order $\mathcal{O}(m^2)$ decoupled the differential equations for 
$F$ and $y^m$. This expansion resembled the perturbative expansion
in which the known part of the Polchinski-Strassler background itself is given.
It also allowed us to find analytic solutions for the expanded embedding.

If the additional constant $\gamma$ 
introduced into the action assumes its value $\gamma=1$, 
the radial embedding coordinate
$u$ is a function of only the worldvolume radial direction $\rho$ for all 
values of the mass parameters and for all choices of the embedding angle 
$\hat\psi$. The angular embedding coordinate $\psi$ itself
depends on the three worldvolume angles if the mass parameter $m_1$ 
associated to the embedding directions $z^1$, $\bar z^1$ is non-zero. 
For $\gamma=1$ the angular dependence is 
encoded in only two $l=2$ $SO(4)$ spherical harmonics, and their 
coefficients do not depend 
on the boundary value $\hat\psi$ of the embedding angle. 
Moreover, the angular dependence reduces to only one spherical harmonic 
if the two masses $m_a$, $a=2,3$ that correspond to 
the worldvolume directions $z^a$, $\bar z^a$ are equal.
A complete independence from the worldvolume angles as found in the 
$\mathcal{N}=2$ case \cite{Apreda:2006bu} cannot be reached in the 
Polchinski-Strassler case if all 
masses are different from zero, even if they are equal.  
This would require that the parameters in the action \eqref{Sembed} 
fulfilled $\gamma+\delta+\tau=1$, such that with $\gamma=1$ embeddings 
with arbitrary $\hat\psi$ would not depend on the worldvolume angles. 
If  $\gamma+\delta+\tau=1$ but $\gamma\neq1$ at least the embeddings with 
$\hat\psi=0,\frac{\pi}{2}$ became angle independent. 
The angle dependence of the embeddings is understandable if one 
remembers that for $m_1=m_2=m_3=m$ the background generating $\text{D}3$-branes
are polarized into an ellipsoid with distinct lengths of its 
principal axes. This breaks the $SO(4)$ rotational symmetry 
in the worldvolume directions of the embedded
$\text{D}7$-brane \cite{Apreda:2006bu}. 
Surprisingly, this does not affect the radial embedding coordinate $u$ which 
with $\gamma=1$ in all cases only depends on $\rho$.

It would be interesting to find an interpretation for  the condition 
$\gamma+\delta+\tau=0$ for which angle independent embeddings can be found.
The above given relation might be fulfilled 
in a more symmetric background in which the $\text{D}3$-branes are 
polarized not into an ellipsoid but into a sphere. In this context 
one could analyze the polarization in 
presence of a non-vanishing gaugino mass \cite{Taylor-Robinson:2001pp}.
Furthermore, it appears to be interesting to study the embedding of a stack of 
coincident $\text{D}7$-branes with 
non-vanishing worldvolume instantons along the lines
of \cite{Guralnik:2004ve,Erdmenger:2005bj,Arean:2007nh}. The backreaction 
of the instanton gauge field strength on the embeddings could also influence 
their worldvolume angle dependence. There might exist particular cases
in which the embeddings do not depend on these angles.

We have then discussed 
the monotony properties of the analytic radial solutions $u(\rho)$ and 
found four types of 
embeddings. For all of these the radial coordinate in six dimensions $r(\rho)$ 
is a monotonically increasing function at least for sufficiently large 
boundary value $\hat u$, such that the embeddings are physical 
\cite{Babington:2003vm}.

With $F=0$ we have revisited the $y^7$ embedding in the $\mathcal{N}=2$
case with our modified treatment of the gauge field. This changed the 
monotonically decreasing $y^7(\rho)$ found in \cite{Apreda:2006bu} 
into a non-monotonic one that still led to a monotonically increasing 
$r(\rho)$. We compared the analytic solution with the numerical exact solution 
and found agreement. 

Furthermore, we have discussed in brief the case of a
$\text{D}7$-brane probe with a different orientation in the 
 $\mathcal{N}=2$ background and the $y^4$ and $y^7$ embeddings in
the $\mathcal{N}=1$ case with equal masses. 
We then proposed error estimates for the corresponding analytic solutions.

In a last step we have applied the method of holographic renormalization 
to the action \eqref{Sembed} of a $\text{D}7$-brane 
probe. In the general case this demonstrated that the embeddings did 
not induce a non-vanishing quark condensate in 
the dual boundary theory, since the subtracted action could be made vanish
by adding appropriate finite counterterms. 

Our final remark concerns the meson spectra. We believe that qualitatively 
for all the discussed embeddings the corresponding spectra will show 
mass gaps, and the squared meson mass $M^2(m_\text{q}^2)$ 
should be a (nearly) linear function of the squared quark mass $m_\text{q}^2$.
In case of the $\mathcal{N}=2_\parallel$ embeddings we have already given 
indications for this behaviour in \cite{Apreda:2006bu}.

\section*{Acknowledgements}
\addcontentsline{toc}{section}{Acknowledgements}
I am very grateful to Dmitri Sorokin for intensive enlightening 
discussions and to Ralf Blumenhagen and Andreas Karch for useful
email correspondences. I also thank Riccardo Apreda for providing 
some of his {\tt mathematica} routines for cross-checks. 
Furthermore I could benefit a lot from discussions with
Dietmar Klemm, Luca Martucci, David Mateos,
Patrick Meessen, Alfonso Ramallo, 
Alberto Santambrogio, Brian Wecht and Alberto Zaffaroni.
This work is supported by INFN. 

\appendix

\section{Generalization to arbitrary masses}
\label{app:bases}

Here we present the necessary expressions to compute the equations of motion 
and the action explicitly. They are the generalizations to three arbitrary 
masses of the corresponding expressions in the $\mathcal{N}=2$ 
Polchinski-Strassler background with $m_1=0$, $m_2=m_3=m$ 
computed in \cite{Apreda:2006bu}, where all necessary conventions can be found.
We again work in the complex basis \eqref{cplxbasis}. 
%Using these transformations, one finds that in the real basis 
%the corresponding components of the tensor $T_3$ defined in \eqref{Tcplx}
%are given by
%\begin{equation}\label{Tinrealcoord}
%\begin{aligned}
%%T_{456}&=\phantom{-}\frac{1}{2 \sqrt{2}}(m_1+m_2+m_3)  \phantom{\epsilon_{pqr}} = \phantom{-}i   T_{789} \col \\
%T_{p+3\,q+3\,r+3}
%&=\phantom{-} \frac{1}{2\sqrt{2}}\epsilon_{pqr}
%(m_1+m_2+m_3)=\phantom{-} iT_{p+6\,q+6\,r+6}
%\col\\
%T_{p+3\,q+3\,r+6}
%&=-\frac{i}{2\sqrt{2}}\epsilon_{pqr}
%(m_1+m_2-m_3)=-iT_{p+6\,q+6\,r+3}
%\col\\
%T_{p+3\,q+6\,r+3}
%&=-\frac{i}{2\sqrt{2}}\epsilon_{pqr}
%(m_1-m_2+m_3)=-iT_{p+6\,q+3\,r+6}
%\col\\
%T_{p+6\,q+3\,r+3}
%&=\phantom{-}\frac{i}{2\sqrt{2}}\epsilon_{pqr}
%(m_1-m_2-m_3)=-iT_{p+3\,q+6\,r+6}
%\pnt
%\end{aligned}
%\end{equation}
%In the complex basis,
Inserting the components \eqref{Tcplx} of $T_3$ in this basis
into \eqref{tildeC2B2inS2}, we find for 
the components of $\tilde C_2$ and $B$ the explicit expressions
\begin{equation}\label{tildeC2B2cplx}
\begin{aligned}
\tilde C_{pq}&=\e^{-\hat\phi}\frac{\zeta}{6}Z\epsilon_{rpq}
m_r\bar z^r\col\\
\tilde C_{p\bar q}&=\e^{-\hat\phi}\frac{\zeta}{6}Z\epsilon_{rpq}
(m_p\bar z^r+m_qz^r)\col\\ 
\tilde C_{\bar pq}&=\e^{-\hat\phi}\frac{\zeta}{6}Z\epsilon_{rpq}
(m_q\bar z^r+m_pz^r)\col\\ 
\tilde C_{\bar p\bar q}&=\e^{-\hat\phi}\frac{\zeta}{6}Z\epsilon_{rpq}
m_rz^r\col
\end{aligned}
\qquad
\begin{aligned}
B_{pq}&=-i\frac{\zeta}{6}Z\epsilon_{rpq}
m_r\bar z^r\col\\
B_{p\bar q}&=i\frac{\zeta}{6}Z\epsilon_{rpq}
(m_p\bar z^r-m_qz^r)\col\\ 
B_{\bar pq}&=i\frac{\zeta}{6}Z\epsilon_{rpq}
(m_q\bar z^r-m_pz^r)\col\\ 
B_{\bar p\bar q}&=i\frac{\zeta}{6}Z\epsilon_{rpq}
m_rz^r\pnt
\end{aligned}
\end{equation}

If the $\text{D}7$-brane is embedded as shown in table \ref{tab:D3D7orient}, 
the directions $z^a$, $\bar z^a$ $a=2,3$ run along its
worldvolume, while $z^m$, $\bar z^m$, $m=1$ are perpendicular to it. 
The following expressions are defined in the four 
flat parallel directions, and hence all inner products 
(denoted by $\cdot$) and Hodge star operators (denoted by $\star_4$) are 
understood to be the ones in the $4$-dimensional flat space spanned by
$z^a$, $\bar z^a$, $a=2,3$.

With the definition of the sum of the squared masses in \eqref{T3abssquare}
the inner products in four dimensions thus become 
\begin{equation}
\begin{aligned}
\tilde C_2\cdot\tilde C_2
&=\tilde C_{ab}\tilde C_{\bar a\bar b}
+\tilde C_{a\bar b}\tilde C_{\bar ab}
%\\
%&=\frac{\alpha^2}{36R^8}Z^2\epsilon_{mab}\epsilon_{nab}
%(m_mm_n\bar z^mz^n+(m_a\bar z^m+m_b z^m)(m_b\bar z^n+m_a z^n))\\
=\e^{-2\hat\phi}\frac{\zeta^2}{18}Z^2
(M^2z^m\bar z^m+m_am_b(z^mz^m+\bar z^m\bar z^m))\col\\
%\tilde C_2\cdot B
%&=\tilde C_{ab}B_{\bar a\bar b}
%+\tilde C_{\bar a\bar b}B_{ab}
%+\tilde C_{a\bar b}B_{\bar ab}
%+\tilde C_{b\bar a}B_{\bar ba}
%%\\
%%&=i\e^{\hat\phi}\frac{\alpha^2}{36R^8}Z^2\epsilon_{mab}\epsilon_{nab}
%%(m_mm_n(\bar z^mz^n-z^m\bar z^n)+(m_a\bar z^m+m_bz^m)(m_b\bar z^n-m_az^n)\\
%%&\phantom{={}i\e^{\hat\phi}\frac{\alpha^2}{36R^8}Z^2
%%\epsilon_{mab}\epsilon_{nab}(}
%%+(m_b\bar z^m+m_az^m)(m_a\bar z^n-m_bz^n)\\
%=-i\e^{-\hat\phi}\frac{\zeta^2}{18}Z^2
%m_am_b(z^mz^m-\bar z^m\bar z^m)\col\\
B\cdot B
&=B_{ab}B_{\bar a\bar b}
+B_{a\bar b}B_{\bar ab}
%\\
%&=\e^{2\hat\phi}\frac{\alpha^2}{36R^8}Z^2\epsilon_{mab}\epsilon_{nab}
%(m_mm_n\bar z^mz^n-(m_a\bar z^m-m_b z^m)(m_b\bar z^n-m_a z^n))\\
=\frac{\zeta^2}{18}Z^2
(M^2z^m\bar z^m-m_am_b(z^mz^m+\bar z^m\bar z^m))
%\pnt
\end{aligned}
\end{equation}
and
\begin{equation}
\begin{aligned}
\tilde C_2\cdot\star_4\tilde C_2
&=-\tilde C_{ab}\tilde C_{\bar a\bar b}
+\tilde C_{a\bar a}\tilde C_{b\bar b}
+\tilde C_{a\bar b}\tilde C_{\bar ab}
\\
%&=-\frac{\alpha^2}{36R^8}Z^2\epsilon_{mab}\epsilon_{nab}
%(m_mm_n\bar z^mz^n-(m_a\bar z^m+m_bz^m)(m_b\bar z^n+m_az^n))\\
&=-\e^{-2\hat\phi}\frac{\zeta^2}{18}Z^2
((m_m^2-m_a^2-m_b^2)z^m\bar z^m-m_am_b(z^mz^m+\bar z^m\bar z^m))
\col\\
%\tilde C_2\cdot\star_4B
%&=-\frac{1}{2}(\tilde C_{ab}B_{\bar a\bar b}
%+\tilde C_{\bar a\bar b}B_{ab})+\tilde C_{a\bar a}B_{b\bar b}
%+\tilde C_{a\bar b}B_{\bar ab}\\
%%&=i\e^{\hat\phi}\frac{\alpha^2}{36R^8}Z^2\epsilon_{mab}\epsilon_{nab}
%%(m_mm_n(\bar z^mz^n-z^m\bar z^n)+(m_a\bar z^m+m_bz^m)(m_b\bar z^m-m_az^m)\\
%%&\phantom{=i\e^{\hat\phi}\frac{\alpha^2}{36R^8}Z^2
%%\epsilon_{mab}\epsilon_{nab}(}
%%+(m_b\bar z^m+m_az^m)(m_a\bar z^m-m_bz^m))\\
%&=i\e^{-\hat\phi}\frac{\zeta^2}{18}Z^2
%m_am_b(z^mz^m-\bar z^m\bar z^m)\col\\
B\cdot\star_4B
&=-B_{ab}B_{\bar a\bar b}
+B_{a\bar a}B_{b\bar b}
+B_{a\bar b}B_{\bar ab}\\
%&=-\e^{2\hat\phi}\frac{\alpha^2}{36R^8}Z^2\epsilon_{mab}\epsilon_{nab}
%(m_mm_n\bar z^mz^n
%+(m_a\bar z^m-m_bz^m)(m_b\bar z^m-m_az^m))\\
&=-\frac{\zeta^2}{18}Z^2
((m_m^2-m_a^2-m_b^2)z^m\bar z^m+m_am_b(z^mz^m+\bar z^m\bar z^m))
\col
\end{aligned}
\end{equation}
where on the r.h.s.\ of the above expressions the indices 
$a,b,m\in\{1,2,3\}$ take fixed distinct values. 

A combination that appears in the action \eqref{Sembed} and in the 
equations of motion \eqref{tildeyeom} for the embedding coordinates is then 
determined as
\begin{equation}\label{BBCCsum}
\begin{aligned}
{}&\frac{1}{2}Z^{-1}\big((\alpha-\beta\star_4)B\cdot B+
\tau\e^{2\hat\phi}\star_4\tilde C_2\cdot\tilde C_2\big)\\
&\qquad=\frac{\zeta^2}{36}Z\big(
(\alpha+\beta-\tau)m_m^2+(\alpha-\beta+\tau)(m_a^2+m_b^2))z^m\bar z^m\\
&\qquad\phantom{{}={}\frac{\zeta^2}{36}Z\big(}
-(\alpha-\beta-\tau)m_am_b(z^mz^m+\bar z^m\bar z^m)\big)
\col
\end{aligned}
\end{equation}
where $\star_4$ is understood to act on the first form on its right.
%We should remark that the terms depending on $m_m$ drop out in this 
%combination.

Furthermore, one needs similar expressions where not all components are 
summed. They read
\begin{equation}\label{odformprodcplx}
\begin{aligned}
{}&Z^{-1}\big(
(\gamma-\delta\star_4)(B_{ab} B_{m\bar b}+ B_{a\bar b} B_{mb})
+\tau\e^{2\hat\phi}\star_4(\tilde C_{ab}\tilde C_{m\bar b}
+\tilde C_{a\bar b}\tilde C_{mb})\big)\\
&\qquad=-\frac{\zeta^2}{36}Z\big(
((\gamma-\delta+\tau)m_a^2+(\gamma+\delta-\tau)m_m^2)\bar z^a\bar z^m\\
&\qquad\phantom{{}={}-\frac{\zeta^2}{36}Z\big(}
-m_b((\gamma+\delta+\tau)m_mz^a\bar z^m
+(\gamma-\delta-\tau)m_a\bar z^az^m)
\big)
\col\\
{}&Z^{-1}\big(
(\gamma-\delta\star_4)(B_{\bar ab} B_{m\bar b}+ B_{\bar a\bar b} B_{mb})
+\tau\e^{2\hat\phi}\star_4(\tilde C_{\bar ab}\tilde C_{m\bar b}
+\tilde C_{\bar a\bar b}\tilde C_{mb})\big)\\
&\qquad=-\frac{\zeta^2}{36}Z\big(
2\gamma m_am_m\bar z^az^m
-(\gamma-\delta-\tau)m_b(m_az^az^m+m_m\bar z^a\bar z^m)\\
&\qquad\phantom{{}={}-\frac{\zeta^2}{36}Z\big(}
+(\gamma-\delta+\tau)m_b^2z^a\bar z^m
\big)
\col
\end{aligned}
\end{equation}
where on the l.h.s.\ one first has to act with $\star_4$ on the first 
tensor on its right, and then extract the required components.
While on the l.h.s.\ a sum over $b$ is understood and $a$, $m$ take fixed 
values, on the r.h.s.\ all indices $a$, $b$, $m$ take the 
corresponding fixed distinct values. 

The tensors \eqref{IWdef} for the corrected metric \eqref{metriccorr} read in 
complex coordinates
\begin{equation}\label{Icplxbasis}
I_{pq}=-\frac{\bar z^p\bar z^q}{10z\bar z}\col\qquad
I_{p\bar q}=\frac{1}{5}\Big(\delta_{p\bar q}-\frac{\bar z^pz^q}{2z\bar z}\Big)
\col\qquad
%I_{\bar pq}=\frac{1}{5}\Big(\delta_{\bar pq}-\frac{z^p\bar z^q}{2z\bar z}\Big)
%\col\qquad
%I_{\bar p\bar q}=-\frac{z^pz^q}{10z\bar z}\col\\
\end{equation}
and
\begin{equation}
\begin{aligned}\label{Wcplxbasis}
W_{pq}
&=\frac{1}{4M^2z\bar z}(2\delta_{p\bar q}m_pm_r\bar z^r\bar z^r
-(m_p^2+m_q^2)\bar z^p\bar z^q)+\frac{\bar z^p\bar z^q}{10z\bar z}\col\\
W_{p\bar q}
%&=\frac{1}{4M^2z\bar z}(\delta_{p\bar q}M^2z\bar z
%-(M^2-m_p^2-m_q^2)\bar z^pz^q-2m_pm_qz^p\bar z^q)
%-\frac{1}{5}\Big(\delta_{p\bar q}-\frac{\bar z^p z^q}{2z\bar z}\Big)\\
&=\frac{1}{20}\Big(\delta_{p\bar q}-3\frac{\bar z^p z^q}{z\bar z}\Big)
+\frac{1}{4M^2z\bar z}((m_p^2+m_q^2)\bar z^pz^q-2m_pm_qz^p\bar z^q)
%\col\\
%W_{\bar pq}
%&=\frac{1}{4M^2z\bar z}(\delta_{\bar pq}M^2z\bar z
%-(M^2-m_p^2-m_q^2)z^p\bar z^q
%-2m_pm_q\bar z^pz^q)
%-\frac{1}{5}\Big(\delta_{\bar pq}-\frac{z^p\bar z^q}{2z\bar z}\Big)\\
%&=\frac{1}{20}\Big(\delta_{\bar pq}-3\frac{z^p\bar z^q}{z\bar z}\Big)
%+\frac{1}{4M^2z\bar z}((m_p^2+m_q^2)z^p\bar z^q-2m_pm_q\bar z^pz^q)
%\col\\
%W_{\bar p\bar q}
%&=\frac{1}{4M^2z\bar z}(2\delta_{\bar pq}m_pm_rz^rz^r-(m_p^2+m_q^2)z^pz^q)
%+\frac{z^pz^q}{10z\bar z}
\col
\end{aligned}
\end{equation}
where we have introduced $z\bar z=z^p\bar z^p$ to abbreviate the expression 
summed over $p$.
The remaining components are obtained by complex conjugation from the 
above expressions.
Taking the traces of the corrections in \eqref{metriccorr}
w.r.t.\ to the four-dimensional subspace, i.e.\ summing over $a=2,3$, 
thereby using that in the complex basis the radii defined in \eqref{rhordef}
become
\begin{equation}\label{cplxrhordef}
 r=\sqrt{2z\bar z}=\sqrt{\rho^2+u^2}\col\qquad
\rho=\sqrt{2z^a\bar z^a}\col\qquad u=\sqrt{2z^m\bar z^m}\col
\end{equation}
we find
\begin{equation}\label{trtildegtexpl}
\begin{aligned}
\tilde g_{aa}
%&=p\frac{1}{5}\Big(4-\frac{\rho^2}{r^2}\Big)
%+q\frac{\rho^2}{r^2}+w\frac{1}{10}\Big(2-3\frac{\rho^2}{r^2}\Big)\\
%&=\frac{1}{5}(4p+w)
%+\frac{\rho^2}{r^2}\frac{1}{10}(-2p+10q-3w)\\
&=\frac{1}{10}(6p+10q-w)
+\frac{u^2}{r^2}\frac{1}{10}(2p-10q+3w)
\pnt
\end{aligned}
\end{equation}
Using that according to \eqref{h0pqrel} the trace over the first four 
directions is given by $Z\tilde g_{\mu\mu}=4h_0Z=q-p$, the required 
combination of traces becomes
\begin{equation}\label{subtrtildegsum}
\begin{aligned}
Z^{-\frac{1}{2}}(Z\tilde g_{\mu\mu}+\tilde g_{aa})
%&=Z^{-\frac{1}{2}}\Big(q-p+\frac{1}{10}(6p+10q-\omega)
%+\frac{u^2}{r^2}\frac{1}{10}(2p-10q+3\omega)\Big)\\
%&=Z^{-\frac{1}{2}}\Big(\frac{1}{10}(-4p+20q-\omega)
%+\frac{u^2}{r^2}\frac{1}{10}(2p-10q+3\omega)\Big)\\
%&=\frac{1}{10}\Big(Z^{-\frac{1}{2}}(-4p+20q-\omega)
%+\frac{u^2}{R^2}(2p-10q+3\omega)\Big)\\
%&=\frac{1}{10R^2}\big(r^2(-4p+20q-\omega)+u^2(2p-10q+3\omega)\big)\\
&=\frac{1}{10R^2}\big(\rho^2(-4p+20q-\omega)+u^2(-2p+10q+2\omega)\big)
\pnt
\end{aligned}
\end{equation}
Furthermore, the required off-diagonal elements read
\begin{equation}\label{odtildegcplx}
\begin{aligned}
\tilde g_{am}
%&=\frac{1}{10z\bar z}(5q-p)\bar z^a\bar z^m+w W_{am}\\
&=\frac{1}{10z\bar z}(5q-p+w)\bar z^a\bar z^m
-\frac{w}{4M^2z\bar z}(m_a^2+m_m^2)\bar z^a\bar z^m
\col\\
%\tilde g_{a\bar m}
%&=\frac{1}{10z\bar z}(5q-p)\bar z^az^m+w W_{a\bar m}\\
%&=\frac{1}{20z\bar z}(10q-2p-3w)\bar z^az^m
%+\frac{w}{4M^2z\bar z}((m_a^2+m_m^2)\bar z^az^m-2m_am_mz^a\bar z^m)
%\col\\
\tilde g_{\bar am}
%&=\frac{1}{10z\bar z}(5q-p)z^a\bar z^m+w W_{\bar am}\\
&=\frac{1}{20z\bar z}(10q-2p-3w)z^a\bar z^m
+\frac{w}{4M^2z\bar z}((m_a^2+m_m^2)z^a\bar z^m-2m_am_m\bar z^az^m)
\col\\
%\tilde g_{\bar a\bar m}
%&=\frac{1}{10z\bar z}(5q-p)z^az^m+w W_{\bar a\bar m}\\
%&=\frac{1}{10z\bar z}(5q-p+w)z^az^m
%-\frac{w}{4M^2z\bar z}(m_a^2+m_m^2)z^az^m
%\pnt
\end{aligned}
\end{equation}
where the missing combinations are obtained by complex conjugation.

\section{Perturbative expansion of the equations of motion}
\label{app:eomex}

In this section we derive the explicit form of the equations of motion 
\eqref{tildeyeom} for the expanded embedding coordinates 
\eqref{yembedexpand}.
In the complex basis \eqref{cplxbasis} 
the equations of motion \eqref{tildeyeom} with $F=0$ are given by
\begin{equation}\label{eomcplx}
\begin{aligned}
{}&2\partial_a\partial_{\bar a}\tilde{\bar z}^m
+\partial_a\big(Z^{-\frac{1}{2}}\tilde g_{m\bar a}\big)
+\partial_{\bar a}\big(Z^{-\frac{1}{2}}\tilde g_{ma}\big)\\
&
+\partial_a\big(Z^{-1}\big(
(\gamma-\delta\star_4)(B_{\bar ab}B_{m\bar b}+B_{\bar a\bar b}B_{mb}) 
+\tau\e^{2\hat\phi}((\star_4\tilde C)_{\bar ab}\tilde C_{m\bar b}
+(\star_4\tilde C)_{\bar a\bar b}\tilde C_{mb})\big)\big)\\
&
+\partial_{\bar a}\big(Z^{-1}\big(
(\gamma-\delta\star_4)(B_{ab}B_{m\bar b}+B_{a\bar b}B_{mb})
+\tau\e^{2\hat\phi}((\star_4\tilde C)_{ab}\tilde C_{m\bar b}
+(\star_4\tilde C)_{a\bar b}\tilde C_{mb})\big)\big)\\
&=
\preparderiv{\tilde z^m}\Big(
\tilde\phi+\frac{1}{2}Z^{\frac{1}{2}}\tilde g_{\mu\mu}
+\frac{1}{2}Z^{-\frac{1}{2}}\tilde g_{aa}%\\
%&\phantom{{}={}\preparderiv{\tilde z^m}\Big(}
+\frac{1}{2}Z^{-1}\big(
(\alpha-\beta\star_4)B\cdot B
+\tau\e^{2\hat\phi}\star_4\tilde C_2\cdot
\tilde C_2\big)\Big)\Big|_{\tilde z^m=\tilde{\bar z}^m=0}
\end{aligned}
\end{equation}
and by its complex conjugate.
The individual expressions that enter the above equation are given by the  
derivatives of the results computed in Appendix \ref{app:bases}.
From \eqref{BBCCsum} we find with the definition of $\rho$, $u$ and $r$ 
in \eqref{rhordef} and in \eqref{cplxrhordef}
\begin{equation}\label{mderformprodcplx}
\begin{aligned}
{}&\partial_m\Big(\frac{1}{2}Z^{-1}\big(
(\alpha-\beta\star_4)B\cdot B
+\tau\e^{2\hat\phi}\star_4\tilde C_2\cdot\tilde C_2\big)\Big)\\
&\qquad=\frac{\zeta^2}{36}Z\Big(
(\alpha+\beta-\tau)m_m^2+(\alpha-\beta+\tau)(m_a^2+m_b^2))
\Big(1-2\frac{u^2}{r^2}\Big)\bar z^m\\
&\qquad\phantom{{}={}\frac{\zeta^2}{36}Z\Big(}
-2(\alpha-\beta-\tau)m_am_b\Big(\Big(1-\frac{u^2}{r^2}\Big)z^m
-2\frac{(\bar z^m)^3}{r^2}\Big)\Big)
\col
\end{aligned}
\end{equation}
where on the r.h.s.\ the indices $a$, $b$, and $m$ take fixed distinct values. 
The divergence of \eqref{odformprodcplx} reads
\begin{equation}\label{mdivformprodcplx}
\begin{aligned}
{}&\partial_a\big(Z^{-1}\big(
(\gamma-\delta\star_4)(B_{\bar ab} B_{m\bar b}+ B_{\bar a\bar b} B_{mb})
+\tau\e^{2\hat\phi}\star_4(\tilde C_{\bar ab}\tilde C_{m\bar b}
+\tilde C_{\bar a\bar b}\tilde C_{mb})\big)\big)
+(a\leftrightarrow\bar a)\\
%&
%+\partial_{\bar a}\Big(Z^{-1}\Big(
%(\gamma-\delta\star_4)(B_{ab} B_{m\bar b}+ B_{a\bar b} B_{mb})
%+\tau\e^{2\hat\phi}\star_4(\tilde C_{ab}\tilde C_{m\bar b}
%+\tilde C_{a\bar b}\tilde C_{mb})\Big)\Big)\\
&=\frac{\zeta^2}{18}Z\Big(
-\big(((\gamma-\delta+\tau)(m_a^2+m_b^2)+(\gamma+\delta-\tau)m_m^2)\bar z^m%\\
%&\phantom{{}={}\frac{\zeta^2}{18}Z\Big({}-{}\big(}
-2(\gamma-\delta-\tau)m_bm_az^m\big)\frac{u^2}{r^2}\\
&\phantom{{}={}\frac{\zeta^2}{18}Z\Big(}
+\frac{2}{r^2}m_m\big(2\gamma(m_a\bar z^a\bar z^a+m_b\bar z^b\bar z^b)z^m\\
&\phantom{{}={}\frac{\zeta^2}{18}Z\Big(
+\frac{2}{r^2}m_m\big(}
-(m_b((\gamma+\delta+\tau)z^az^a+(\gamma-\delta-\tau)\bar z^a\bar z^a)\\
&\phantom{{}={}\frac{\zeta^2}{18}Z\Big(
+\frac{2}{r^2}m_m\big({}-{}(}
+m_a((\gamma+\delta+\tau)z^bz^b+(\gamma-\delta-\tau)\bar z^b\bar z^b))
\bar z^m\big)\Big)
\col
\end{aligned}
\end{equation}
where on the l.h.s.\ we have abbreviated the second term which is found from 
the first one by exchanging the summation indices $a$ and $\bar a$. 
While $a$ and $b$ are summed over on the l.h.s.,
on the r.h.s.\ $a$, $b$, $m$ take fixed distinct values.
The gradient of the dilaton as given in \eqref{tildephi}
becomes in complex coordinates
\begin{equation}\label{mdertildephicplx}
\begin{aligned}
\partial_m\tilde\phi
&=\frac{\zeta^2}{18}Z\Big(
-\frac{2}{r^2}
(m_am_b\bar z^m\bar z^m+m_am_m(z^bz^b+\bar z^b\bar z^b)
+m_bm_m(z^az^a+\bar z^a\bar z^a))\bar z^m\\
&\phantom{{}={}\frac{\zeta^2}{18}Z\Big(}
+m_am_b\Big(1-\frac{u^2}{r^2}\Big)z^m\Big)
%\col\\
%\partial_{\bar m}\tilde\phi
%&=\e^{\hat\phi}\frac{\alpha^2}{18R^8}Z\Big(
%-\frac{1}{z\bar z}
%(m_am_bz^mz^m+m_am_n(z^bz^b+\bar z^b\bar z^b)
%+m_bm_n(z^az^a+\bar z^a\bar z^a))z^m\\
%&\phantom{{}={}\e^{\hat\phi}\frac{\alpha^2}{18R^8}Z\Big(}
%+m_am_b\Big(1-\frac{u^2}{r^2}\Big)\bar z^m\Big)
\col
\end{aligned}
\end{equation}
where $a$, $b$, $m$ take fixed distinct values.
The derivative of the subtraces of the corrections to the metric
in \eqref{subtrtildegsum} become
\begin{equation}\label{mdertrtildegcplx}
\begin{aligned}
\frac{1}{2}\partial_m
\big(Z^{-\frac{1}{2}}(Z\tilde g_{\mu\mu}+\tilde g_{aa})\big)
%&=\frac{1}{10R^2}\Big(
%2(4p-20q+w)\frac{\rho^2}{r^2}
%-(2p-10q-2w)\Big(1-2\frac{u^2}{r^2}\Big)\Big)\bar z^m\\
&=\frac{1}{5R^2}\Big(3p-15q+2w-(2p-10q+3w)\frac{u^2}{r^2}\Big)
\bar z^m%\col\\
%\frac{1}{2}\partial_{\bar m}
%\big(Z^{-\frac{1}{2}}(Z\tilde g_{\mu\mu}+\tilde g_{aa})\big)
%&=\frac{1}{10R^2}\Big(
%2(4p-20q+w)\frac{\rho^2}{r^2}
%-(2p-10q-2w)\Big(1-2\frac{u^2}{r^2}\Big)\Big)z^m\\
%&=\frac{1}{5R^2}\Big(3p-15q+2w-(2p-10q+3w)\frac{u^2}{r^2}\Big)
%z^m
\pnt
\end{aligned}
\end{equation}
Finally, the derivatives of the off-diagonal elements 
of the corrections to the metric in \eqref{odtildegcplx}
are found to be given by
\begin{equation}\label{mdivtildegcplx}
\begin{aligned}
{}&\partial_a(Z^{-\frac{1}{2}}\tilde g_{\bar am})
+\partial_{\bar a}(Z^{-\frac{1}{2}}\tilde g_{am})\\
%&\qquad=\frac{1}{5R^2}(20q-4p-\omega)\Big(1-\frac{\rho^2}{r^2}\Big)\bar z^m
%+\frac{4\omega}{M^2R^2r^2}m_m(m_a\bar z^a\bar z^a+m_b\bar z^b\bar z^b)z^m\\
&\qquad=\frac{1}{5R^2}(20q-4p-\omega)\frac{u^2}{r^2}\bar z^m
+\frac{4\omega}{M^2R^2r^2}m_m(m_a\bar z^a\bar z^a+m_b\bar z^b\bar z^b)z^m
\pnt
\end{aligned}
\end{equation}
While $a$ is a summation index on the l.h.s., $a$, $b$ and $m$ take fixed 
distinct values on the r.h.s.
Inserting the above equations into \eqref{eomcplx}, we obtain
\begin{equation}
\begin{aligned}
2\partial_a\partial_{\bar a}\tilde{\bar z}^m
&=-\frac{2}{r^2}\Big(\frac{\zeta^2}{9}Z\gamma+\frac{2\omega}{M^2R^2}\Big)
m_m(m_a\bar z^a\bar z^a+m_b\bar z^b\bar z^b)z^m\\
&\phantom{{}={}}+\frac{\zeta^2}{36}Z\Big(
\frac{4}{r^2}m_m\big(m_b((\gamma+\delta+\tau-1)z^az^a
+(\gamma-\delta-\tau-1)\bar z^a\bar z^a)\\
&\phantom{{}={}+\frac{\zeta^2}{36}Z\Big(
\frac{4}{r^2}m_m\big(}
+m_a((\gamma+\delta+\tau-1)z^bz^b
+(\gamma-\delta-\tau-1)\bar z^b\bar z^b)\big)\bar z^m\\
&\phantom{{}={}+\frac{\zeta^2}{36}Z\Big(}
+m_m^2\Big(\alpha+\beta-\tau-2(\alpha+\beta-\gamma-\delta)
\frac{u^2}{r^2}\Big)\bar z^m
\\
&\phantom{{}={}+\frac{\zeta^2}{36}Z\Big(}
+(m_a^2+m_b^2)
\Big(\alpha-\beta+\tau-2(\alpha-\beta-\gamma+\delta)
\frac{u^2}{r^2}\Big)\bar z^m\\
&\phantom{{}={}+\frac{\zeta^2}{36}Z\Big(}
-2m_am_b\Big(\Big(\alpha-\beta-\tau-1
-(\alpha-\beta-2(\gamma-\delta)+\tau-1)\frac{u^2}{r^2}\Big)z^m\\
&\phantom{{}={}+\frac{\zeta^2}{36}Z\Big(-2m_am_b\Big(}
-2(\alpha-\beta-\tau-1)\frac{(\bar z^m)^3}{r^2}\Big)\Big)\\
&\phantom{{}={}}
+\frac{1}{5R^2}\Big(3p-15q+2\omega+(2p-10q-2\omega)\frac{u^2}{r^2}\Big)
\bar z^m
\col
\end{aligned}
\end{equation}
where on the l.h.s.\ $a$ is a summation index. On the r.h.s.\ $a$, $b$ and
$m$ take fixed distinct values, and the expressions have to be evaluated 
using the unperturbed 
embedding coordinates $\hat z^m$ and $\hat{\bar z}^m$ as required by
\eqref{eomcplx}.

We use the explicit values for $p$, $q$ and $\omega$ 
given in \eqref{wpqh0} to compute the combinations
\begin{equation}
\begin{aligned}
3p-15q+2\omega
%&=\e^{\hat\phi}\frac{\alpha^2}{R^6}\frac{M^2}{3}Z
%\Big(-\frac{3}{16}-\frac{15}{432}-\frac{2}{6}\Big)
%=\e^{\hat\phi}\frac{\alpha^2}{R^6}\frac{M^2}{3}Z
%\Big(-\frac{3}{16}-\frac{5}{144}-\frac{1}{3}\Big)\\
&=-\frac{5\zeta^2}{27}M^2R^2Z
\col\qquad%\\
2p-10q-2\omega
%&=\e^{\hat\phi}\frac{\alpha^2}{R^6}\frac{M^2}{3}Z
%\Big(-\frac{2}{16}-\frac{10}{432}+\frac{2}{6}\Big)
%=\e^{\hat\phi}\frac{\alpha^2}{R^6}\frac{M^2}{3}Z
%\Big(-\frac{1}{8}-\frac{5}{216}+\frac{1}{3}\Big)\\
&=\frac{5\zeta^2}{81}M^2R^2Z
\pnt
\end{aligned}
\end{equation}
They are used to obtain the result
\begin{equation}\label{tildebarzeom}
\begin{aligned}
2\partial_a\partial_{\bar a}\tilde{\bar z}^m
&=\frac{\zeta^2}{36}\hat Z\Big(
\frac{4}{\hat r^2}m_m\big(2(1-\gamma)(m_a\bar z^a\bar z^a+m_b\bar z^b\bar z^b)
\hat z^m\\
&\phantom{{}={}\frac{\zeta^2}{36}\hat Z\Big(
\frac{4}{\hat r^2}m_m\big(}
+(m_b((\gamma+\delta+\tau-1)z^az^a
+(\gamma-\delta-\tau-1)\bar z^a\bar z^a)\\
&\phantom{{}={}\frac{\zeta^2}{36}\hat Z\Big(
\frac{4}{\hat r^2}m_m\big({}+{}(}
+m_a((\gamma+\delta+\tau-1)z^bz^b
+(\gamma-\delta-\tau-1)\bar z^b\bar z^b))
\hat{\bar z}^m\big)\\
&\phantom{{}={}\frac{\zeta^2}{36}\hat Z\Big(}
+m_m^2\Big(\alpha+\beta-\tau-\frac{4}{3}
-2\Big(\alpha+\beta-\gamma-\delta-\frac{2}{9}\Big)
\frac{\hat u^2}{\hat r^2}\Big)\hat{\bar z}^m
\\
&\phantom{{}={}\frac{\zeta^2}{36}\hat Z\Big(}
+(m_a^2+m_b^2)
\Big(\alpha-\beta+\tau-\frac{4}{3}
-2\Big(\alpha-\beta-\gamma+\delta-\frac{2}{9}\Big)
\frac{\hat u^2}{\hat r^2}\Big)\hat{\bar z}^m\\
&\phantom{{}={}\frac{\zeta^2}{36}Z\Big(}
-2m_am_b\Big(\Big(\alpha-\beta-\tau-1
-(\alpha-\beta-2(\gamma-\delta)+\tau-1)\frac{\hat u^2}{\hat r^2}\Big)
\hat z^m\\
&\phantom{{}={}\frac{\zeta^2}{36}\hat Z\Big(-2m_am_b\Big(}
-2(\alpha-\beta-\tau-1)\frac{(\hat{\bar z}^m)^3}{\hat r^2}\Big)\Big)
\col
\end{aligned}
\end{equation}
where the quantities that carry a `hat' are related to or respectively 
evaluated with the unperturbed part of the embedding.
Multiplying both sides with $\hat z^m$, adding and subtracting the complex
conjugate of the result, and making use of the decomposition
\begin{equation}\label{cplxdecomp}
f\phi+g\bar\phi=\frac{f+g}{2}(\phi+\bar\phi)+\frac{f-g}{2}(\phi-\bar\phi)\col
\end{equation}
valid for arbitrary $f$, $g$ and $\phi$, 
we obtain the two equations
\begin{equation}\label{cplxpluseom}
\begin{aligned}
{}&2(\hat z^m\partial_a\partial_{\bar a}\tilde{\bar z}^m
+\hat{\bar z}^m\partial_a\partial_{\bar a}\tilde z^m)\\
&=\frac{\zeta^2}{36}\hat Z\hat u^2\Big(
\frac{2}{\hat r^2}(1-\gamma)m_m\Big(
(m_a(z^az^a+\bar z^a\bar z^a)+m_b(z^bz^b+\bar z^b\bar z^b))
\Big(\frac{\hat z^m}{\hat{\bar z}^m}+\frac{\hat{\bar z}^m}{\hat z^m}\Big)\\
&\phantom{{}={}\frac{\zeta^2}{36}\hat Z\hat u^2\Big(
\frac{2}{\hat r^2}(1-\gamma)m_m\Big(}
-(m_a(z^az^a-\bar z^a\bar z^a)+m_b(z^bz^b-\bar z^b\bar z^b))
\Big(\frac{\hat z^m}{\hat{\bar z}^m}-\frac{\hat{\bar z}^m}{\hat z^m}\Big)\\
&\phantom{{}={}\frac{\zeta^2}{36}\hat Z\hat u^2\Big(
\frac{2}{\hat r^2}(1-\gamma)m_m\Big(}
-2(m_b(z^az^a+\bar z^a\bar z^a)
+m_a(z^bz^b+\bar z^b\bar z^b))\Big)\\
&\phantom{{}={}\frac{\zeta^2}{36}\hat Z\hat u^2\Big(}
+m_m^2\Big(\alpha+\beta-\tau-\frac{4}{3}
-2\Big(\alpha+\beta-\gamma-\delta-\frac{2}{9}\Big)
\frac{\hat u^2}{\hat r^2}\Big)
\\
&\phantom{{}={}\frac{\zeta^2}{36}\hat Z\hat u^2\Big(}
+(m_a^2+m_b^2)
\Big(\alpha-\beta+\tau-\frac{4}{3}
-2\Big(\alpha-\beta-\gamma+\delta-\frac{2}{9}\Big)
\frac{\hat u^2}{\hat r^2}\Big)\\
&\phantom{{}={}\frac{\zeta^2}{36}\hat Z\hat u^2\Big(}
-m_am_b\Big(\alpha-\beta-\tau-1
-2(\alpha-\beta-\gamma+\delta-1)\frac{\hat u^2}{\hat r^2}\Big)
\Big(\frac{\hat z^m}{\hat{\bar z}^m}+\frac{\hat{\bar z}^m}{\hat z^m}\Big)
\Big)\\
\end{aligned}
\end{equation}
and
\begin{equation}\label{cplxminuseom}
\begin{aligned}
{}&2(\hat z^m\partial_a\partial_{\bar a}\tilde{\bar z}^m
-\hat{\bar z}^m\partial_a\partial_{\bar a}\tilde z^m)\\
&=\frac{\zeta^2}{36}\hat Z\hat u^2\Big(
\frac{2}{\hat r^2}m_m\Big(
(1-\gamma)\Big((m_a(z^az^a+\bar z^a\bar z^a)+m_b(z^bz^b+\bar z^b\bar z^b))
\Big(\frac{\hat z^m}{\hat{\bar z}^m}-\frac{\hat{\bar z}^m}{\hat z^m}\Big)\\
&\phantom{{}={}\frac{\zeta^2}{36}\hat Z\hat u^2\Big(
\frac{2}{\hat r^2}m_m\Big((1-\gamma)\Big(}
-(m_a(z^az^a-\bar z^a\bar z^a)+m_b(z^bz^b-\bar z^b\bar z^b))
\Big(\frac{\hat z^m}{\hat{\bar z}^m}+\frac{\hat{\bar z}^m}{\hat z^m}\Big)\Big)
\\
&\phantom{{}={}\frac{\zeta^2}{36}\hat Z\hat u^2\Big(
\frac{2}{\hat r^2}m_m\Big(}
+2(\delta+\tau)(m_b(z^az^a-\bar z^a\bar z^a)
+m_a(z^bz^b-\bar z^b\bar z^b))\Big)\\
&\phantom{{}={}\frac{\zeta^2}{36}\hat Z\hat u^2\Big(}
-m_am_b\Big(\alpha-\beta-\tau-1
+2(\gamma-\delta-\tau)\frac{\hat u^2}{\hat r^2}\Big)
\Big(\frac{\hat z^m}{\hat{\bar z}^m}-\frac{\hat{\bar z}^m}{\hat z^m}\Big)
\Big)
\pnt
\end{aligned}
\end{equation}

The linear combinations that appear in the above expressions on the l.h.s.\
are directly related to the radial and angular coordinate in a polar 
coordinate system. Expanding the embedding coordinates in the complex basis 
up to linear order in the corrections $\tilde u$ and $\tilde\psi$ as
\begin{equation}\label{cylembedcoord}
\begin{aligned}
\sqrt{2}z^m&=u\e^{i\psi}=(\hat u+\tilde u)\e^{i(\hat\psi+\tilde\psi)}
=(\hat u+\tilde u+i\hat u\tilde\psi)\e^{i\hat\psi}
\col\\
\sqrt{2}\bar z^m&=u\e^{-i\psi}=(\hat u+\tilde u)\e^{-i(\hat\psi+\tilde\psi)}
=(\hat u+\tilde u-i\hat u\tilde\psi)\e^{-i\hat\psi}
\col
\end{aligned}
\end{equation}
we find that the required linear combinations are given by
\begin{equation}\label{barztildezlincomb}
\begin{aligned}
\hat z^m\partial_a\partial_{\bar a}\tilde{\bar z}^m
+\hat{\bar z}^m\partial_a\partial_{\bar a}\tilde z^m
&=\hat u\partial_a\partial_{\bar a}\tilde u
\col\\
\hat z^m\partial_a\partial_{\bar a}\tilde{\bar z}^m
-\hat{\bar z}^m\partial_a\partial_{\bar a}\tilde z^m
&=-i\hat u^2\partial_a\partial_{\bar a}\tilde\psi
\pnt
\end{aligned}
\end{equation}
Furthermore, with
$\hat u^2=2\hat z^m\hat{\bar z}^m$,
$\frac{\hat z^m}{\hat{\bar z}^m}=\e^{2i\hat\psi}$
the combinations that appear on the r.h.s.\ of \eqref{cplxpluseom}
and \eqref{cplxminuseom} can be expressed in 
terms of the angle $\hat\psi$ as
\begin{equation}
\frac{\hat z^m}{\hat{\bar z}^m}+\frac{\hat{\bar z}^m}{\hat z^m}=2\cos2\hat\psi
\col\qquad
\frac{\hat z^m}{\hat{\bar z}^m}-\frac{\hat{\bar z}^m}{\hat z^m}=2i\sin2\hat\psi
\pnt
\end{equation}

The combinations of the coordinates $z^a$ and $\bar z^a$ 
are abbreviated in terms of 
four of the in total nine $l=2$ $SO(4)$ spherical harmonics which are
defined as
\begin{equation}\label{SO4shdef}
\begin{aligned}
y_{++}&=\frac{z^az^a+z^bz^b+\bar z^a\bar z^a+\bar z^b\bar z^b}{\rho^2}
=\frac{y^5y^5+y^6y^6-y^8y^8-y^9y^9}{\rho^2}
\col\\
y_{+-}&=-i\frac{z^az^a+z^bz^b-\bar z^a\bar z^a-\bar z^b\bar z^b}{\rho^2}
=2\frac{y^5y^8+y^6y^9}{\rho^2}
\col\\
y_{-+}&=\frac{z^az^a-z^bz^b+\bar z^a\bar z^a-\bar z^b\bar z^b}{\rho^2}
=\frac{y^5y^5-y^6y^6-y^8y^8+y^9y^9}{\rho^2}
\col\\
y_{--}&=-i\frac{z^az^a-z^bz^b-\bar z^a\bar z^a+\bar z^b\bar z^b}{\rho^2}
=2\frac{y^5y^8-y^6y^9}{\rho^2}
\col
\end{aligned}
\end{equation}
where in the second equalities we have fixed the indices to
$a=2$, $b=3$ and $m=1$. This corresponds to an embedding of
the $\text{D}7$-brane as shown in table
\ref{tab:D3D7orient}.
Using the following parameterization of the real coordinates
\begin{equation}\label{realcoordangles}
\begin{aligned}
y^5=\rho\cos\theta\cos\phi_1
\col\quad
y^6=\rho\cos\theta\sin\phi_1
\col\quad
y^8=\rho\sin\theta\cos\phi_2
\col\quad
y^9=\rho\sin\theta\sin\phi_2
\col
\end{aligned}
\end{equation}
where $0\le\theta\le\frac{\pi}{2}$ and $0\le\phi_{1,2}\le2\pi$, 
the spherical harmonics become
\begin{equation}\label{SO4shangles}
\begin{aligned}
y_{++}&=\cos2\theta
\col\\
y_{+-}&=\sin2\theta\cos(\phi_1-\phi_2)
\col\\
y_{-+}&=\cos^2\theta\cos2\phi_1-\sin^2\theta\cos2\phi_2
\col\\
y_{--}&=\sin2\theta\cos(\phi_1+\phi_2)
\pnt
\end{aligned}
\end{equation}
The equations of motion then read
\begin{equation}\label{eomexpl}
\begin{aligned}
%{}&\frac{2}{\hat u}(\hat z^m\partial_a\partial_{\bar a}\tilde{\bar z}^m
%+\hat{\bar z}^m\partial_a\partial_{\bar a}\tilde z^m)\\
2\partial_a\partial_{\bar a}\tilde u
&=\frac{\zeta^2}{36}\hat Z\hat u\Big(
2\frac{\rho^2}{\hat r^2}(1-\gamma)m_1\big(
-(m_2+m_3) y_{++}(1-\cos2\hat\psi)\\
&\phantom{{}={}\frac{\zeta^2}{36}\hat Z\hat u\Big(
2\frac{\rho^2}{\hat r^2}(1-\gamma)m_1\big(}
+(m_2-m_3)y_{-+}(1+\cos2\hat\psi)\\
&\phantom{{}={}\frac{\zeta^2}{36}\hat Z\hat u\Big(
2\frac{\rho^2}{\hat r^2}(1-\gamma)m_1\big(}
+((m_2+m_3)y_{+-}+(m_2-m_3)y_{--})\sin2\hat\psi\big)\\
&\phantom{{}={}\frac{\zeta^2}{36}\hat Z\hat u\Big(}
+m_1^2\Big(\alpha+\beta-\tau-\frac{4}{3}
-2\Big(\alpha+\beta-\gamma-\delta-\frac{2}{9}\Big)
\frac{\hat u^2}{\hat r^2}\Big)
\\
&\phantom{{}={}\frac{\zeta^2}{36}\hat Z\hat u\Big(}
+(m_2^2+m_3^2)
\Big(\alpha-\beta+\tau-\frac{4}{3}
-2\Big(\alpha-\beta-\gamma+\delta-\frac{2}{9}\Big)
\frac{\hat u^2}{\hat r^2}\Big)\\
&\phantom{{}={}\frac{\zeta^2}{36}\hat Z\hat u\Big(}
-2m_2m_3\Big(\alpha-\beta-\tau-1
-2(\alpha-\beta-\gamma+\delta-1)\frac{\hat u^2}{\hat r^2}\Big)\cos2\hat\psi
\Big)\\
%
%{}&\frac{2i}{\hat u^2}(\hat z^m\partial_a\partial_{\bar a}\tilde{\bar z}^m
%-\hat{\bar z}^m\partial_a\partial_{\bar a}\tilde z^m)\\
2\partial_a\partial_{\bar a}\tilde\psi
&=\frac{\zeta^2}{36}\hat Z\Big(
2\frac{\rho^2}{\hat r^2}m_1\Big(
(\gamma-1)((m_2+m_3)y_{++}+(m_2-m_3)y_{-+})\sin2\hat\psi\\
&\phantom{{}={}\frac{\zeta^2}{36}\hat Z\Big(
2\frac{\rho^2}{\hat r^2}m_1\Big(}
-(m_2+m_3)y_{+-}(\delta+\tau+(\gamma-1)\cos2\hat\psi)\\
&\phantom{{}={}\frac{\zeta^2}{36}\hat Z\Big(
2\frac{\rho^2}{\hat r^2}m_1\Big(}
+(m_2-m_3)y_{--}(\delta+\tau-(\gamma-1)\cos2\hat\psi)\Big)\\
&\phantom{{}={}\frac{\zeta^2}{36}\hat Z\Big(}
+2m_2m_3\Big(\alpha-\beta-\tau-1
+2(\gamma-\delta-\tau)\frac{\hat u^2}{\hat r^2}\Big)\sin2\hat\psi
\Big)
\pnt
\end{aligned}
\end{equation}
For both, $u$ and $\psi$ they have the same structure which is compactly 
summarized as
\begin{equation}\label{app:eomimpl}
\begin{aligned}
2\partial_a\partial_{\bar a}f
&=\frac{n_f}{\hat r^4}\Big(B_f+C_f\frac{\hat u^2}{\hat r^2}
-(C_f^{++}y_{++}+C_f^{+-}y_{+-}+C_f^{-+}y_{-+}+C_f^{--}y_{--})
\frac{\rho^2}{\hat r^2}\Big)%\\
%&=\frac{n_f}{\hat r^4}\Big(B_f+C_f
%-(C_f+C_f^{++}y_{++}+C_f^{+-}y_{+-}+C_f^{-+}y_{-+}+C_f^{--}y_{--})
%\frac{\rho^2}{\hat r^2}\Big)
\col
\end{aligned}
\end{equation}
where $f=u$ or $f=\psi$, $n_u=\hat u$ or respectively $n_\psi=1$,
and the constants are given by 
\begin{equation}\label{BuCudef}
\begin{aligned}
B_u&=\frac{\zeta^2R^4}{216}\big(
(m_2+m_3)^2(6\tau-1)+(m_2-m_3)^2(6(\alpha-\beta)-7)\\
&\phantom{{}={}\frac{\zeta^2R^4}{216}\big(}
+2m_1^2(3(\alpha+\beta-\tau)-4)
+12m_2m_3(\alpha-\beta-\tau-1)(1-\cos2\hat\psi)\big)\col\\
C_u
&=-\frac{\zeta^2R^4}{108}\Big(
(m_2+m_3)^2\frac{7}{3}
+(m_2-m_3)^2\Big(6(\alpha-\beta-\gamma+\delta)-\frac{11}{3}\Big)
\\
&\phantom{{}={}+\frac{\zeta^2R^4}{108}\big(}
+2m_1^2\Big(3(\alpha+\beta-\gamma-\delta)-\frac{2}{3}\Big)
\\
&\phantom{{}={}+\frac{\zeta^2R^4}{108}\big(}
+12m_2m_3(\alpha-\beta-\gamma+\delta-1)
(1-\cos2\hat\psi)\Big)\col\\
C_u^{++}
&=-\frac{\zeta^2R^4}{18}
m_1(m_2+m_3)(\gamma-1)(1-\cos2\hat\psi)\col\\
C_u^{+-}
&=\frac{\zeta^2R^4}{18}
m_1(m_2+m_3)(\gamma-1)\sin2\hat\psi\col\\
C_u^{-+}
&=\frac{\zeta^2R^4}{18}
m_1(m_2-m_3)(\gamma-1)(1+\cos2\hat\psi)\col\\
C_u^{--}
&=\frac{\zeta^2R^4}{18}
m_1(m_2-m_3)(\gamma-1)\sin2\hat\psi\col\\
\end{aligned}
\end{equation}
or respectively by 
\begin{equation}\label{BpsiCpsidef}
\begin{aligned}
B_\psi
&=\frac{\zeta^2R^4}{18}
m_2m_3(\alpha-\beta-\tau-1)\sin2\hat\psi\col\\
C_\psi
&=\frac{\zeta^2R^4}{9}
m_2m_3(\gamma-\delta-\tau)\sin2\hat\psi\col\\
C_\psi^{++}
&=-\frac{\zeta^2R^4}{18}
m_1(m_2+m_3)(\gamma-1)\sin2\hat\psi\col\\
C_\psi^{+-}
&=\frac{\zeta^2R^4}{18}
m_1(m_2+m_3)(\gamma+\delta+\tau-1-(\gamma-1)(1-\cos2\hat\psi))\col\\
C_\psi^{-+}
&=-\frac{\zeta^2R^4}{18}
m_1(m_2-m_3)(\gamma-1)\sin2\hat\psi\\
C_\psi^{--}
&=\frac{\zeta^2R^4}{18}
m_1(m_2-m_3)(\gamma-\delta-\tau-1-(\gamma-1)(1-\cos2\hat\psi))
\pnt
\end{aligned}
\end{equation}

\section{Explicit expansion of the action}
\label{app:expaction}

In this section we derive the explicit expression for the action 
\eqref{Sembed}. We need the corrections to the dilaton 
\eqref{tildephi} in complex coordinates, as well as the combination 
of the subtraces \eqref{subtrtildegsum} of the corrections to the metric.
Furthermore, we need the form combination \eqref{BBCCsum}.
The components of the correction to the metric \eqref{odtildegcplx}
and of the form combination \eqref{odformprodcplx} have to be contracted
with the derivatives of the embedding coordinates.
Finally, \eqref{wpqh0} serves to replace the parameters in the metric 
by their explicit values. 

The combination of non-derivative terms is then found to be given by
\begin{equation}\label{ytermscplx}
\begin{aligned}
{}&\tilde\phi
+\frac{1}{2}Z^{-\frac{1}{2}}(Z\tilde g_{\mu\mu}+\tilde g_{aa})
+\frac{1}{2}\e^{-\hat\phi}Z^{-1}(\alpha-\beta\star_4)B\cdot B
+\frac{\tau}{2}\e^{\hat\phi}Z^{-1}\star_4\tilde C_2\cdot\tilde C_2\\
%{}&=\e^{\hat\phi}\frac{\alpha^2}{108R^8}Z\Big(
%3m_am_b(z^mz^m+\bar z^m\bar z^m)+3m_am_m(z^bz^b+\bar z^b\bar z^b)
%+3m_bm_m(z^az^a+\bar z^a\bar z^a)\\
%&\phantom{{}={}\e^{\hat\phi}\frac{\alpha^2}{108R^8}Z\Big(}
%+\frac{M^2}{3}\Big(\frac{5}{2}\rho^2-u^2\Big)\\
%&\phantom{{}={}\e^{\hat\phi}\frac{\alpha^2}{108R^8}Z\Big(}
%+((3(\alpha+\beta)-3\tau)m_m^2+(3(\alpha-\beta)+3\tau)(m_a^2+m_b^2))
%z^m\bar z^m\\
%&\phantom{{}={}\e^{\hat\phi}\frac{\alpha^2}{108R^8}Z\Big(}
%-(3(\alpha-\beta)-3\tau)m_am_b(z^mz^m+\bar z^m\bar z^m)\Big)
%\col\\
{}&=\frac{\zeta^2}{36}Z\Big(
-(\alpha-\beta-\tau-1)m_am_b(z^mz^m+\bar z^m\bar z^m)\\
&\phantom{{}={}\frac{\zeta^2}{36}Z\Big(}
+m_am_m(z^bz^b+\bar z^b\bar z^b)
+m_bm_m(z^az^a+\bar z^a\bar z^a)
+\frac{5}{18}M^2\rho^2\\
&\phantom{{}={}\frac{\zeta^2}{36}Z\Big(}
+\frac{1}{18}((9(\alpha+\beta-\tau)-2)m_m^2+(9(\alpha-\beta+\tau)-2)
(m_a^2+m_b^2))u^2\Big)
\pnt
\end{aligned}
\end{equation}
While on the l.h.s.\ the indices $\mu$, $a$ are summed over, on the
r.h.s.\ $a$, $b$, $m$ take fixed distinct values. 
After some manipulations thereby using also \eqref{cplxdecomp} to 
obtain simple linear combinations of the derivative terms,
the remaining contributions to the action combine as 
\begin{equation}\label{partialytermscplx}
\begin{aligned}
{}&Z^{-\frac{1}{2}}\partial_a\tilde y^m\tilde g_{am}
+2\e^{-\hat\phi}Z^{-1}\big(
(\gamma-\delta\star_4)B\cdot\partial\tilde yB
+\tau\e^{2\hat\phi}\star_4\tilde C_2\cdot\partial\tilde y\tilde C_2
\big)\\
&\qquad=-\frac{\zeta^2}{72}Z\Big(
\Big(\Big(\gamma-\delta+\tau-\frac{5}{9}\Big)(m_a^2+m_b^2)
+\Big(\gamma+\delta-\tau-\frac{5}{9}\Big)m_m^2\Big)\\
&\qquad\phantom{{}={}-\frac{\zeta^2}{72}Z\Big(}
(z^a\partial_a+\bar z^a\partial_{\bar a})
(\hat z^m\tilde{\bar z}^m+\hat{\bar z}^m\tilde z^m)\\
&\qquad\phantom{{}={}-\frac{\zeta^2}{72}Z\Big(}
+((\gamma-\delta+\tau-1)(m_a^2-m_b^2)+(\gamma+\delta-\tau-1)m_m^2)\\
&\qquad\phantom{{}={}-\frac{\zeta^2}{72}Z\Big({}+{}}
(z^a\partial_a-\bar z^a\partial_{\bar a})
(\hat z^m\tilde{\bar z}^m-\hat{\bar z}^m\tilde z^m)\\
&\qquad\phantom{{}={}-\frac{\zeta^2}{72}Z\Big(}
-2(\gamma-\delta-\tau)m_am_b
(z^a\partial_a+\bar z^a\partial_{\bar a})
(\hat z^m\tilde z^m+\hat{\bar z}^m\tilde{\bar z}^m)\\
&\qquad\phantom{{}={}-\frac{\zeta^2}{72}Z\Big(}
-2m_mm_b(\gamma(z^a\partial_{\bar a}+\bar z^a\partial_a)
(\hat z^m\tilde{\bar z}^m+\hat{\bar z}^m\tilde z^m)\\
&\qquad\phantom{{}={}-\frac{\zeta^2}{72}Z\Big(-2m_mm_b(}
-(\delta+\tau)(z^a\partial_{\bar a}-\bar z^a\partial_a)
(\hat z^m\tilde{\bar z}^m-\hat{\bar z}^m\tilde z^m))\\
&\qquad\phantom{{}={}-\frac{\zeta^2}{72}Z\Big(}
+4(\gamma-1)m_am_m(z^a\partial_{\bar a}(\hat{\bar z}^m\tilde{\bar z}^m)
+\bar z^a\partial_a(\hat z^m\tilde z^m))
\Big)
\col
\end{aligned}
\end{equation}
where on the l.h.s.\ for compactness we have used real coordinates and 
all indices are summed over independently, 
while on the r.h.s. which is expressed in complex coordinates 
the summation runs over $a$ and $b$, such that $a$, $b$ and $m$ take distinct
values.

To obtain the action as an expansion in terms of the corrections which 
appear in the decomposition \eqref{yembedexpand}, we have to 
expand the warp factor up to linear order in $\tilde z^m$ as
\begin{equation}
Z
%=\frac{R^4}{(\rho^2+2(\hat z^m\hat{\bar z}^m+\hat z^m\tilde{\bar z}^m
%+\hat{\bar z}^m\tilde z^m))^2}
=\hat Z\Big(1-\frac{4}{\hat r^2}(\hat z^m\tilde{\bar z}^m
+\hat{\bar z}^m\tilde z^m)\Big)\col\qquad
\hat Z=\frac{R^4}{\hat r^4}
\col
\end{equation}
where a `hat' indicates that the corresponding expression has to be evaluated
with the unperturbed value $\hat z^m$.
Introducing polar coordinates as in \eqref{cylembedcoord} and 
embedding the $\text{D}7$-brane as in table \ref{tab:D3D7orient}, 
the Lagrangian is then given by
\begin{equation}
{\scriptsize
\begin{aligned}
-\frac{\e^{\hat\phi}}{T_7}\mathcal{L}
&=1+\partial_a\tilde u\partial_{\bar a}\tilde u
+\hat u^2\partial_a\tilde\psi\partial_{\bar a}\tilde\psi\\
&\phantom{{}={}}+\frac{\zeta^2}{72}\hat Z\Big(
\big((m_2-m_3)^2(\alpha-\beta-1)+(m_2+m_3)^2\tau\big)\hat u\Big(
\hat u+2\Big(1-2\frac{\hat u^2}{\hat r^2}\Big)\tilde u\Big)\\
&\phantom{{}={}+\frac{\zeta^2}{72}Z\Big(}
-((m_2-m_3)^2(\gamma-\delta)+(m_2+m_3)^2\tau)
(z^a\partial_a+\bar z^a\partial_{\bar a})\hat u\tilde u\\
&\phantom{{}={}+\frac{\zeta^2}{72}Z\Big(}
+\frac{1}{3}(m_2^2+m_3^2)\Big(\frac{2}{3}\hat u^2+\frac{5}{3}\hat r^2
-2\Big(1+\frac{4}{3}\frac{\hat u^2}{\hat r^2}\Big)\hat u\tilde u
+\frac{5}{3}(z^a\partial_a+\bar z^a\partial_{\bar a})\hat u\tilde u
\Big)\\
&\phantom{{}={}+\frac{\zeta^2}{72}Z\Big(}
+m_1^2\Big(
\Big(\alpha+\beta-\tau-\frac{7}{9}\Big)\hat u^2
+\frac{5}{9}\hat r^2%\\
%&\phantom{{}={}+\frac{\zeta^2}{72}Z\Big(+m_1^2\Big(}
+2\Big(\alpha+\beta-\tau-\frac{4}{3}-2\Big(\alpha+\beta-\tau-\frac{7}{9}\Big)
\frac{\hat u^2}{\hat r^2}\Big)\hat u\tilde u\\
&\phantom{{}={}+\frac{\zeta^2}{72}Z\Big(+m_1^2\Big(}
-\Big(\gamma+\delta-\tau-\frac{5}{9}\Big)
(z^a\partial_a+\bar z^a\partial_{\bar a})\hat u\tilde u\Big)\\
&\phantom{{}={}+\frac{\zeta^2}{72}Z\Big(}
+i((m_2^2-m_3^2)(\gamma-\delta+\tau-1)
(z^2\partial_2-\bar z^2\partial_{\bar 2}
-z^3\partial_3+\bar z^3\partial_{\bar 3})\\
&\phantom{{}={}+\frac{\zeta^2}{72}Z\Big(+i(}
+m_1^2(\gamma+\delta-\tau-1)
(z^a\partial_a-\bar z^a\partial_{\bar a}))\hat u^2\tilde\psi\\
&\phantom{{}={}+\frac{\zeta^2}{72}Z\Big(}
+2m_2m_3\Big((\alpha-\beta-\tau-1)\hat u
\Big(\hat u(1-\cos2\hat\psi)%\\
%&\phantom{{}={}+\frac{\zeta^2}{72}Z\Big(
%+2m_2m_3\Big((\alpha-\beta-\tau-1)\hat u\Big(}
+2\Big(1-2\frac{\hat u^2}{\hat r^2}\Big)(1-\cos2\hat\psi)\tilde u
+2\hat u\sin2\hat\psi\,\tilde\psi\Big)\\
&\phantom{{}={}+\frac{\zeta^2}{72}Z\Big(+2m_2m_3(}
-(\gamma-\delta-\tau)(z^a\partial_a+\bar z^a\partial_{\bar a})
\hat u((1-\cos2\hat\psi)\tilde u+\hat u\sin2\hat\psi\,\tilde\psi)\Big)\\
&\phantom{{}={}+\frac{\zeta^2}{72}Z\Big(}
+m_1(m_2+m_3)\Big(
(\hat r^2-\hat u^2)y_{++}\Big(1-\frac{4}{\hat r^2}\hat u\tilde u\Big)\\
&\phantom{{}={}+\frac{\zeta^2}{72}Z\Big(+m_1(m_2+m_3)\Big(}
+\hat u(z^a\partial_{\bar a}+\bar z^a\partial_a)
(\tilde u
+(\gamma-1)((1-\cos2\hat\psi)\tilde u+\hat u\sin2\hat\psi\,\tilde\psi))\\
&\phantom{{}={}+\frac{\zeta^2}{72}Z\Big(+m_1(m_2+m_3)\Big(}
+i\hat u(z^a\partial_{\bar a}-\bar z^a\partial_a)
((\gamma+\delta+\tau-1)\hat u\tilde\psi%\\
%&\phantom{{}={}+\frac{\zeta^2}{72}Z\Big(+m_1(m_2+m_3)\Big(
%+i\hat u(z^a\partial_{\bar a}-\bar z^a\partial_a)(}
+(\gamma-1)(\sin2\hat\psi\,\tilde u-(1-\cos2\hat\psi)\hat u\tilde\psi))\Big)\\
&\phantom{{}={}+\frac{\zeta^2}{72}Z\Big(}
-m_1(m_2-m_3)\Big(
(\hat r^2-\hat u^2)y_{-+}
\Big(1-\frac{4}{\hat r^2}\hat u\tilde u\Big)\\
&\phantom{{}={}+\frac{\zeta^2}{72}Z\Big(-m_1(m_2-m_3)\Big(}
+\hat u(z^2\partial_{\bar 2}+\bar z^2\partial_2
-z^3\partial_{\bar 3}-\bar z^3\partial_3)%\\
%&\phantom{{}={}+\frac{\zeta^2}{72}Z\Big(-m_1(m_2-m_3)\Big(\hat u}
((2\gamma-1)\tilde u
-(\gamma-1)((1-\cos2\hat\psi)\tilde u+\hat u\sin2\hat\psi\,\tilde\psi))\\
&\phantom{{}={}+\frac{\zeta^2}{72}Z\Big(-m_1(m_2-m_3)\Big(}
-i\hat u(z^2\partial_{\bar 2}-\bar z^2\partial_2
-z^3\partial_{\bar 3}+\bar z^3\partial_3)%\\
%&\phantom{{}={}+\frac{\zeta^2}{72}Z\Big(-m_1(m_2-m_3)\Big(-i\hat u}
((\gamma-\delta-\tau-1)\hat u\tilde\psi\\
%&\phantom{{}={}+\frac{\zeta^2}{72}Z\Big(-m_1(m_2-m_3)\Big(-i\hat u((}
&\phantom{{}={}+\frac{\zeta^2}{72}Z\Big(-m_1(m_2-m_3)\Big(
-i\hat u(z^2\partial_{\bar 2}-\bar z^2\partial_2
-z^3\partial_{\bar 3}+\bar z^3\partial_3)(}
+(\gamma-1)(\sin2\hat\psi\,\tilde u-(1-\cos2\hat\psi)\hat u\tilde\psi)
\Big)\Big)
\col
\end{aligned}}
\end{equation}
where a summation over $a=2,3$ on both sides is understood.
Finally, we partially integrate all terms which contain derivatives of 
$\tilde u$ and $\tilde\psi$. We require the expressions 
\begin{equation}
\begin{aligned}
(\partial_az^a+\partial_{\bar a}\bar z^a)\hat Z
&=4\hat Z\frac{\hat u^2}{\hat r^2}\col\\
-\frac{i}{2}(\partial_az^a-\partial_{\bar a}\bar z^a)\hat Z
&=0\col
\end{aligned}
\qquad
\begin{aligned}
(\partial_{\bar a}z^a+\partial_a\bar z^a)\hat Z
&=-4\hat Z\frac{\rho^2}{\hat r^2}y_{++}\col\\
-i(\partial_{\bar a}z^a-\partial_a\bar z^a)\hat Z
&=-4\hat Z\frac{\rho^2}{\hat r^2}y_{+-}\col\\
\end{aligned}
\end{equation}
and
\begin{equation}
\begin{aligned}
(\partial_{\bar 2}z^2-\partial_{\bar 3}z^3+\partial_2\bar z^2
-\partial_3\bar z^3)\hat Z
&=-4\hat Z\frac{\rho^2}{\hat r^2}y_{-+}\col\\
-i(\partial_{\bar 2}z^2-\partial_{\bar 3}z^3
-\partial_2\bar z^2+\partial_3\bar z^3)\hat Z
&=-4\hat Z\frac{\rho^2}{\hat r^2}y_{--}\col
\end{aligned}
\end{equation}
where the derivatives act on all functions on the right, and the 
$SO(4)$ spherical harmonics that appear on the r.h.s.\ are defined in 
\eqref{SO4shdef}.
The Lagrangian can then be cast into the compact form
\begin{equation}\label{Lexppi}
\begin{aligned}
-\frac{\e^{\hat\phi}}{T_7}\mathcal{L}
&=1+\partial_a\tilde u\partial_{\bar a}\tilde u
+\hat u^2\partial_a\tilde\psi\partial_{\bar a}\tilde\psi
+\frac{B_u}{2}\frac{\hat u^2}{\hat r^4}\\
&\phantom{{}={}}+\frac{\zeta^2}{216}\hat Z\Big(
\frac{5}{3}M^2(\hat r^2+\hat u^2)%\\
%&\phantom{{}={}+\e^{\hat\phi}\frac{\alpha^2}{216R^8}Z\Big(}
+3m_1((m_2+m_3)y_{++}-(m_2-m_3)y_{-+})\rho^2\Big)\\
&\phantom{{}={}}
+\Big(B_u+C_u\frac{\hat u^2}{\hat r^2}
-(C_u^{++}y_{++}+C_u^{+-}y_{+-}+C_u^{-+}y_{-+}+C_u^{--}y_{--})
\frac{\rho^2}{\hat r^2}\Big)
\frac{\hat u}{\hat r^4}\tilde u\\
&\phantom{{}={}}
+\Big(B_\psi+C_\psi\frac{\hat u^2}{\hat r^2}
-(C_\psi^{++}y_{++}+C_\psi^{+-}y_{+-}+C_\psi^{-+}y_{-+}+C_\psi^{--}y_{--})
\frac{\rho^2}{\hat r^2}\Big)
\frac{\hat u^2}{\hat r^4}\tilde \psi\\
&\phantom{{}={}}+\text{total derivatives}
\col
\end{aligned}
\end{equation}
where $M^2$ can be found in \eqref{T3abssquare}, and the constants are 
defined in \eqref{BuCudef} and \eqref{BpsiCpsidef}. 
It is easy to verify that 
the equations of motion derived from this Lagrangian coincide with the 
ones given in \eqref{eomimpl} and \eqref{eomexpl}.

\section{Analytic solution of the equations of motion}
\label{app:anasol}

The equations of motion \eqref{eomimpl} and \eqref{eomexpl} have the 
structure
\begin{equation}\label{eomstruc}
\begin{aligned}
2\partial_a\partial_{\bar a}f
&=\frac{n_f}{\hat r^4}\Big(B_f+C_f\frac{\hat u^2}{\hat r^2}
-C_f^Iy_I\frac{\rho^2}{\hat r^2}\Big)
=\frac{n_f}{\hat r^4}\Big(B_f+C_f
-(C_f+C_f^Iy_I)\frac{\rho^2}{\hat r^2}\Big)
\col
\end{aligned}
\end{equation}
where $n_u=\hat u$, $n_\psi=1$, and we sum over $I$ which distinguishes 
the level $l=2$ $SO(4)$ spherical harmonics defined in \eqref{SO4shdef}.

To find a solution of the above equation, we recall the action of 
the Laplace operator in $d$-dimensional flat space when it acts on a function 
$f(\rho,\theta_i)=f^I(\rho)Y_I(\theta_i)$, $i=1,\dots,d-1$, where 
$\rho$ is the radial coordinate and $\theta_i$ denote the angle coordinates.
The spherical harmonics $Y_I$ (representations of $SO(d)$) carry 
labels $I=(l,m_1\dots m_{d-2})$, including also the case $l=0$. 
The Laplace operator acts as
\begin{equation}
\partial_a\partial_af
=\sum_I\Big(\frac{1}{\rho^{d-1}}\partial_\rho(\rho^{d-1}\partial_\rho)
-\frac{l(l+d-2)}{\rho^2}\Big)f^IY_I\col
\end{equation}
where on the l.h.s.\ we sum over $a=1,\dots,d$. We separate from the 
radial dependent coefficients of the 
level $l$ spherical harmonics a factor $\rho^l$ by rewriting 
$f^I=\rho^lh_f^I$. This yields the relation
\begin{equation}
\begin{aligned}
\Big(\frac{1}{\rho^{d-1}}\partial_\rho(\rho^{d-1}\partial_\rho)
-\frac{l(l+d-2)}{\rho^2}\Big)f^I
%&=\frac{1}{\rho^{d-1}}\partial_\rho(\rho^{l+d-2}(lh^I+\rho h'^I))
%-l(l+d-2)\rho^{l-2}h^I\\
%&=\rho^{l-1}\big((2l+d-1)h'^I+\rho h''^I\big)\\
&=\frac{1}{\rho^{l+d-1}}\partial_\rho
\big(\rho^{2l+d-1}\partial_\rho h_f^I\big)
\col
\end{aligned}
\end{equation}
i.e.\ rewritten in terms of the functions $h_f^I$, 
the Laplace operator only generates
first and second derivative of $h_f^I$ in a nested manner, but does not leave
the $h_f^I$ without derivatives.

Using this transformation in the special case of $d=4$, the equations of 
motion \eqref{eomstruc} are rewritten as
\begin{equation}
\frac{1}{\rho^{l+3}}\partial_\rho
\big(\rho^{2l+3}\partial_\rho h_f^I\big)
=\frac{n_f}{\hat r^4}\Big(B_f^I+C_f^I-C_f^I\frac{\rho^2}{\hat r^2}\Big)\pnt
\end{equation} 
We have split it into individual equations for each value of $I$. 
For the coefficients $h_f(\rho)$ of the constant spherical harmonic ($l=0$) 
and for $h_f^{l=2}(\rho)$ for the level $l=2$ spherical
harmonics we respectively find
\begin{equation}
\begin{aligned}
\frac{1}{\rho^3}\partial_\rho(\rho^3\partial_\rho h_f)
&=\frac{n_f}{\hat r^4}\Big(B_f+C_f-C_f\frac{\rho^2}{\hat r^2}\Big)\col\\
\frac{1}{\rho^5}\partial_\rho(\rho^7\partial_\rho h_f^{l=2})
&=-n_fC_f^{l=2}\frac{\rho^2}{\hat r^6}
\pnt
\end{aligned}
\end{equation} 
Performing the first step of integration, the results read
\begin{equation}
\begin{aligned}
\rho^3\partial_\rho h_f
&=\frac{n_f}{2}\Big((B_f-C_f)\frac{\hat u^2}{\hat r^2}
+B_f\ln\hat r^2+C_f\frac{\hat u^4}{2\hat r^4}-2A_f\Big)\col\\
\rho^7\partial_\rho h_f^{l=2}
&=-\frac{n_f}{2}\Big(C_f^{l=2}\Big(\rho^2
-3\frac{\hat u^4}{\hat r^2}
+\frac{\hat u^6}{2\hat r^4}
-3\hat u^2\ln\hat r^2\Big)+6A_f^{l=2}\hat u^2\Big)
\col
\end{aligned}
\end{equation}
where $A_f$ and $A_f^{l=2}$ denote integration constants.
After the second integration we obtain
\begin{equation}
\begin{aligned}
h_f
&=\hat h_f-\frac{n_f}{2}\Big(
(2B_f-C_f)\frac{1}{4\rho^2}
+B_f\frac{1}{2\rho^2}\ln\hat r^2
+C_f\frac{1}{4\hat r^2}
-A_f\frac{1}{\rho^2}\Big)\col\\
h_f^{l=2}
&=\hat h_f^{l=2}+\frac{n_f}{2\rho^2}\Big(
C_f^{l=2}\Big(\frac{1}{2\rho^2}
\Big(1-\frac{\hat u^2}{\rho^2}\ln\hat r^2\Big)
-\frac{5}{12}\frac{\hat u^2}{\rho^4}
-\frac{1}{4\hat r^2}\Big)
+A_f^{l=2}\frac{\hat u^2}{\rho^4}\Big)
\col
\end{aligned}
\end{equation}
where $\hat h_f$ and $\hat h_f^{l=2}$ are the corresponding integration 
constants. In general the above given functions diverge in the limit 
$\rho\to0$. However, for appropriately chosen constants 
\begin{equation}
\begin{aligned}
A_f=\frac{1}{2}B_f(1+\ln\hat u^2)-\frac{1}{4}C_f\col\qquad
A_f^{l=2}=\frac{1}{12}C_f^{l=2}(5+6\ln\hat u^2)\col
\end{aligned}
\end{equation}
the functions become regular at $\rho=0$, i.e.\ the $\text{D}7$-brane 
embeddings are regular at $\rho=0$ in this case. 
The coefficient functions assume the form
\begin{equation}
\begin{aligned}
h_f
&=\hat h_f-\frac{n_f}{8}\Big(
B_f\frac{2}{\rho^2}\ln\frac{\hat r^2}{\hat u^2}
+C_f\frac{1}{\hat r^2}
\Big)\col\\
h_f^{l=2}
&=\hat h_f^{l=2}+\frac{n_f}{8\rho^2}C_f^{l=2}\Big(
\frac{2}{\rho^2}
\Big(1-\frac{\hat u^2}{\rho^2}\ln\frac{\hat r^2}{\hat u^2}\Big)
-\frac{1}{\hat r^2}\Big)
\pnt
\end{aligned}
\end{equation}
In the full solution $h_f^{l=2}$ is multiplied by $\rho^2$. 
For $\rho\to\infty$ it is therefore only regular if $\hat h_f^{l=2}=0$.
The everywhere regular solution of \eqref{eomstruc}
hence depends on a single integration constant $\hat f=\hat h_f$.  
Its final form and its first and second derivative 
are given by
\begin{equation}\label{app:freg}
\begin{aligned}
f&=\hat f-\frac{n_f}{8}\Big(
B_f\frac{2}{\rho^2}\ln\frac{\hat r^2}{\hat u^2}
+C_f\frac{1}{\hat r^2}
-C_f^I\Big(
\frac{2}{\rho^2}
\Big(1-\frac{\hat u^2}{\rho^2}\ln\frac{\hat r^2}{\hat u^2}\Big)
-\frac{1}{\hat r^2}\Big)y_I\Big)\\
\partial_\rho f
&=\frac{n_f}{4}\Big(
B_f\frac{2}{\rho}\Big(\frac{1}{\rho^2}\ln\frac{\hat r^2}{\hat u^2}
-\frac{1}{\hat r^2}\Big)
+C_f\frac{\rho}{\hat r^4}
-C_f^I\Big(
\frac{2}{\rho^3}\Big(
1-2\frac{\hat u^2}{\rho^2}\ln\frac{\hat r^2}{\hat u^2}
+\frac{\hat u^2}{\hat r^2}\Big)
-\frac{\rho}{\hat r^4}\Big)y_I\Big)\\
\partial_\rho^2f
&=\frac{n_f}{4}\Big(
B_f\frac{2}{\rho^2}\Big(-\frac{3}{\rho^2}\ln\frac{\hat r^2}{\hat u^2}
+\frac{3}{\hat r^2}+2\frac{\rho^2}{\hat r^4}\Big)
+C_f\frac{1}{\hat r^4}\Big(1-4\frac{\rho^2}{r^2}\Big)\\
&\phantom{{}={}\frac{n_f}{4}\Big(}
-C_f^I\Big(
-\frac{20}{\rho^4}\Big(
1-\frac{\hat u^2}{\rho^2}\ln\frac{\hat r^2}{\hat u^2}\Big)
+\frac{1}{\rho^2\hat r^2}\Big(10+3\frac{\rho^2}{\hat r^2}\Big)
+4\frac{\rho^2}{\hat r^6}\Big)\Big)y_I\Big)
\pnt
\end{aligned}
\end{equation}
For a closer analysis it is necessary to understand the asymptotic behaviour 
of the above expressions. 
In the limits $\rho\ll\hat u$ and $\rho\gg\hat u$ we find
\begin{equation}\label{app:fregasymp}
\begin{aligned}
f&=
\begin{cases}
\hat f-\frac{n_f}{8\hat u^2}(2B_f+C_f)& \rho\to0 \\
\hat f-\frac{n_f}{8\rho^2}\big(
2B_f\ln\frac{\rho^2}{\hat u^2}+C_f-C_f^Iy_I\big) & 
\rho\to\infty
\end{cases}\col\\
\partial_\rho f&=
\begin{cases}
%\frac{n_f}{12\hat u^2}(3(B_f+C_f)+C_f^Iy_I)\rho 
0 & \rho\to0 \\
\frac{n_f}{4\rho^3}\big(
2B_f(\ln\frac{\rho^2}{\hat u^2}-1)+C_f-C_f^Iy_I\big) & 
\rho\to\infty
\end{cases}\col\\
\partial_\rho^2f&=
\begin{cases}
\frac{n_f}{12\hat u^2}(3(B_f+C_f)+C_f^Iy_I) & \rho\to0 \\
\frac{n_f}{4\rho^4}(
2B_f(-3\ln\frac{\rho^2}{\hat u^2}+5)-3C_f+3C_f^Iy_I) 
& \rho\to\infty
\end{cases}\pnt
\end{aligned}
\end{equation}
In particular, $\hat f$ is the constant value of $f$ at the boundary 
at $\rho\to\infty$.

\footnotesize
\bibliographystyle{utphys}
\addcontentsline{toc}{section}{Bibliography}
\bibliography{references}

\providecommand{\href}[2]{#2}\begingroup\raggedright\begin{thebibliography}{10}

\bibitem{Maldacena:1997re}
J.~M. Maldacena, ``{The large $N$ limit of superconformal field theories and
  supergravity},'' {\em Adv. Theor. Math. Phys.} {\bf 2} (1998) 231--252,
\href{http://www.arXiv.org/abs/hep-th/9711200}{{\tt hep-th/9711200}}.
%%CITATION = HEP-TH 9711200;%%.

\bibitem{Witten:1998zw}
E.~Witten, ``{Anti-de Sitter space, thermal phase transition, and confinement
  in gauge theories},'' {\em Adv. Theor. Math. Phys.} {\bf 2} (1998) 505--532,
\href{http://www.arXiv.org/abs/hep-th/9803131}{{\tt hep-th/9803131}}.
%%CITATION = HEP-TH 9803131;%%.

\bibitem{Klebanov:2000hb}
I.~R. Klebanov and M.~J. Strassler, ``{Supergravity and a confining gauge
  theory: Duality cascades and chiSB-resolution of naked singularities},'' {\em
  JHEP} {\bf 08} (2000) 052,
\href{http://www.arXiv.org/abs/hep-th/0007191}{{\tt hep-th/0007191}}.
%%CITATION = HEP-TH 0007191;%%.

\bibitem{Constable:1999ch}
N.~R. Constable and R.~C. Myers, ``{Exotic scalar states in the AdS/CFT
  correspondence},'' {\em JHEP} {\bf 11} (1999) 020,
\href{http://www.arXiv.org/abs/hep-th/9905081}{{\tt hep-th/9905081}}.
%%CITATION = HEP-TH 9905081;%%.

\bibitem{Maldacena:2000yy}
J.~M. Maldacena and C.~Nunez, ``{Towards the large $N$ limit of pure
  $\mathcal{N}=1$ super Yang Mills},'' {\em Phys. Rev. Lett.} {\bf 86} (2001)
  588--591,
\href{http://www.arXiv.org/abs/hep-th/0008001}{{\tt hep-th/0008001}}.
%%CITATION = HEP-TH 0008001;%%.

\bibitem{Polchinski:2000uf}
J.~Polchinski and M.~J. Strassler, ``{The string dual of a confining
  four-dimensional gauge theory},''
\href{http://www.arXiv.org/abs/hep-th/0003136}{{\tt hep-th/0003136}}.
%%CITATION = HEP-TH 0003136;%%.

\bibitem{Myers:1999ps}
R.~C. Myers, ``{Dielectric-branes},'' {\em JHEP} {\bf 12} (1999) 022,
\href{http://www.arXiv.org/abs/hep-th/9910053}{{\tt hep-th/9910053}}.
%%CITATION = HEP-TH 9910053;%%.

\bibitem{Apreda:2006bu}
R.~Apreda, J.~Erdmenger, D.~L{\"u}st, and C.~Sieg, ``{Adding flavour to the
  Polchinski-Strassler background},'' {\em JHEP} {\bf 01} (2007) 079,
\href{http://www.arXiv.org/abs/hep-th/0610276}{{\tt hep-th/0610276}}.
%%CITATION = HEP-TH 0610276;%%.

\bibitem{Freedman:2000xb}
D.~Z. Freedman and J.~A. Minahan, ``{Finite temperature effects in the
  supergravity dual of the $\mathcal{N}=1^*$ gauge theory},'' {\em JHEP} {\bf
  01} (2001) 036,
\href{http://www.arXiv.org/abs/hep-th/0007250}{{\tt hep-th/0007250}}.
%%CITATION = HEP-TH 0007250;%%.

\bibitem{LopesCardoso:2004ni}
G.~Lopes~Cardoso, G.~Curio, G.~Dall'Agata, and D.~L{\"u}st, ``{Gaugino
  condensation and generation of supersymmetric $3$-form flux},'' {\em JHEP}
  {\bf 09} (2004) 059,
\href{http://www.arXiv.org/abs/hep-th/0406118}{{\tt hep-th/0406118}}.
%%CITATION = HEP-TH 0406118;%%.

\bibitem{Taylor-Robinson:2001pp}
M.~Taylor-Robinson, ``{Anomalies, counterterms and the $\mathcal{N}=0$
  Polchinski-Strassler solutions},''
\href{http://www.arXiv.org/abs/hep-th/0103162}{{\tt hep-th/0103162}}.
%%CITATION = HEP-TH 0103162;%%.

\bibitem{Karch:2002sh}
A.~Karch and E.~Katz, ``{Adding flavor to AdS/CFT},'' {\em JHEP} {\bf 06}
  (2002) 043,
\href{http://www.arXiv.org/abs/hep-th/0205236}{{\tt hep-th/0205236}}.
%%CITATION = HEP-TH 0205236;%%.

\bibitem{Kruczenski:2003be}
M.~Kruczenski, D.~Mateos, R.~C. Myers, and D.~J. Winters, ``{Meson spectroscopy
  in AdS/CFT with flavour},'' {\em JHEP} {\bf 07} (2003) 049,
\href{http://www.arXiv.org/abs/hep-th/0304032}{{\tt hep-th/0304032}}.
%%CITATION = HEP-TH 0304032;%%.

\bibitem{Kruczenski:2003uq}
M.~Kruczenski, D.~Mateos, R.~C. Myers, and D.~J. Winters, ``{Towards a
  holographic dual of large-N(c) QCD},'' {\em JHEP} {\bf 05} (2004) 041,
\href{http://www.arXiv.org/abs/hep-th/0311270}{{\tt hep-th/0311270}}.
%%CITATION = HEP-TH 0311270;%%.

\bibitem{Sakai:2004cn}
T.~Sakai and S.~Sugimoto, ``{Low energy hadron physics in holographic QCD},''
  {\em Prog. Theor. Phys.} {\bf 113} (2005) 843--882,
\href{http://www.arXiv.org/abs/hep-th/0412141}{{\tt hep-th/0412141}}.
%%CITATION = HEP-TH 0412141;%%.

\bibitem{Casero:2005se}
R.~Casero, A.~Paredes, and J.~Sonnenschein, ``{Fundamental matter, meson
  spectroscopy and non-critical string / gauge duality},'' {\em JHEP} {\bf 01}
  (2006) 127,
\href{http://www.arXiv.org/abs/hep-th/0510110}{{\tt hep-th/0510110}}.
%%CITATION = HEP-TH 0510110;%%.

\bibitem{Sakai:2003wu}
T.~Sakai and J.~Sonnenschein, ``{Probing flavored mesons of confining gauge
  theories by supergravity},'' {\em JHEP} {\bf 09} (2003) 047,
\href{http://www.arXiv.org/abs/hep-th/0305049}{{\tt hep-th/0305049}}.
%%CITATION = HEP-TH 0305049;%%.

\bibitem{Kuperstein:2004hy}
S.~Kuperstein, ``{Meson spectroscopy from holomorphic probes on the warped
  deformed conifold},'' {\em JHEP} {\bf 03} (2005) 014,
\href{http://www.arXiv.org/abs/hep-th/0411097}{{\tt hep-th/0411097}}.
%%CITATION = HEP-TH 0411097;%%.

\bibitem{Ouyang:2003df}
P.~Ouyang, ``{Holomorphic D7-branes and flavored $\mathcal{N}=1$ gauge
  theories},'' {\em Nucl. Phys.} {\bf B699} (2004) 207--225,
\href{http://www.arXiv.org/abs/hep-th/0311084}{{\tt hep-th/0311084}}.
%%CITATION = HEP-TH 0311084;%%.

\bibitem{Evans:2005ti}
N.~Evans, J.~P. Shock, and T.~Waterson, ``{D7 brane embeddings and chiral
  symmetry breaking},'' {\em JHEP} {\bf 03} (2005) 005,
\href{http://www.arXiv.org/abs/hep-th/0502091}{{\tt hep-th/0502091}}.
%%CITATION = HEP-TH 0502091;%%.

\bibitem{Arean:2007nh}
D.~Arean, A.~V. Ramallo, and D.~Rodriguez-Gomez, ``{Holographic flavor on the
  Higgs branch},''
\href{http://www.arXiv.org/abs/hep-th/0703094}{{\tt hep-th/0703094}}.
%%CITATION = HEP-TH 0703094;%%.

\bibitem{Arean:2004mm}
D.~Arean, D.~E. Crooks, and A.~V. Ramallo, ``{Supersymmetric probes on the
  conifold},'' {\em JHEP} {\bf 11} (2004) 035,
\href{http://www.arXiv.org/abs/hep-th/0408210}{{\tt hep-th/0408210}}.
%%CITATION = HEP-TH 0408210;%%.

\bibitem{Nunez:2003cf}
C.~Nunez, A.~Paredes, and A.~V. Ramallo, ``{Flavoring the gravity dual of
  $\mathcal{N}=1$ Yang-Mills with probes},'' {\em JHEP} {\bf 12} (2003) 024,
\href{http://www.arXiv.org/abs/hep-th/0311201}{{\tt hep-th/0311201}}.
%%CITATION = HEP-TH 0311201;%%.

\bibitem{Babington:2003vm}
J.~Babington, J.~Erdmenger, N.~J. Evans, Z.~Guralnik, and I.~Kirsch, ``{Chiral
  symmetry breaking and pions in non-supersymmetric gauge / gravity duals},''
  {\em Phys. Rev.} {\bf D69} (2004) 066007,
\href{http://www.arXiv.org/abs/hep-th/0306018}{{\tt hep-th/0306018}}.
%%CITATION = HEP-TH 0306018;%%.

\bibitem{Bertolini:2001qa}
M.~Bertolini, P.~Di~Vecchia, M.~Frau, A.~Lerda, and R.~Marotta,
  ``{$\mathcal{N}=2$ gauge theories on systems of fractional D3/D7 branes},''
  {\em Nucl. Phys.} {\bf B621} (2002) 157--178,
\href{http://www.arXiv.org/abs/hep-th/0107057}{{\tt hep-th/0107057}}.
%%CITATION = HEP-TH 0107057;%%.

\bibitem{Wang:2003yc}
X.-J. Wang and S.~Hu, ``{Intersecting branes and adding flavors to the
  Maldacena-Nunez background},'' {\em JHEP} {\bf 09} (2003) 017,
\href{http://www.arXiv.org/abs/hep-th/0307218}{{\tt hep-th/0307218}}.
%%CITATION = HEP-TH 0307218;%%.

\bibitem{Casero:2006pt}
R.~Casero, C.~Nunez, and A.~Paredes, ``{Towards the string dual of
  $\mathcal{N}=1$ SQCD-like theories},'' {\em Phys. Rev.} {\bf D73} (2006)
  086005,
\href{http://www.arXiv.org/abs/hep-th/0602027}{{\tt hep-th/0602027}}.
%%CITATION = HEP-TH 0602027;%%.

\bibitem{Burrington:2004id}
B.~A. Burrington, J.~T. Liu, L.~A. Pando~Zayas, and D.~Vaman, ``{Holographic
  duals of flavored $\mathcal{N}=1$ super Yang-Mills: Beyond the probe
  approximation},'' {\em JHEP} {\bf 02} (2005) 022,
\href{http://www.arXiv.org/abs/hep-th/0406207}{{\tt hep-th/0406207}}.
%%CITATION = HEP-TH 0406207;%%.

\bibitem{Kirsch:2005uy}
I.~Kirsch and D.~Vaman, ``{The D3/D7 background and flavor dependence of Regge
  trajectories},'' {\em Phys. Rev.} {\bf D72} (2005) 026007,
\href{http://www.arXiv.org/abs/hep-th/0505164}{{\tt hep-th/0505164}}.
%%CITATION = HEP-TH 0505164;%%.

\bibitem{Paredes:2006wb}
A.~Paredes, ``{On unquenched $\mathcal{N}=2$ holographic flavor},'' {\em JHEP}
  {\bf 12} (2006) 032,
\href{http://www.arXiv.org/abs/hep-th/0610270}{{\tt hep-th/0610270}}.
%%CITATION = HEP-TH 0610270;%%.

\bibitem{Benini:2006hh}
F.~Benini, F.~Canoura, S.~Cremonesi, C.~Nunez, and A.~V. Ramallo, ``{Unquenched
  flavors in the Klebanov-Witten model},'' {\em JHEP} {\bf 02} (2007) 090,
\href{http://www.arXiv.org/abs/hep-th/0612118}{{\tt hep-th/0612118}}.
%%CITATION = HEP-TH 0612118;%%.

\bibitem{Cherkis:2002ir}
S.~A. Cherkis and A.~Hashimoto, ``{Supergravity solution of intersecting branes
  and AdS/CFT with flavor},'' {\em JHEP} {\bf 11} (2002) 036,
\href{http://www.arXiv.org/abs/hep-th/0210105}{{\tt hep-th/0210105}}.
%%CITATION = HEP-TH 0210105;%%.

\bibitem{Bandos:2006wb}
I.~Bandos and D.~Sorokin, ``{Aspects of D-brane dynamics in supergravity
  backgrounds with fluxes, kappa-symmetry and equations of motion. IIB},'' {\em
  Nucl. Phys.} {\bf B759} (2006) 399--446,
\href{http://www.arXiv.org/abs/hep-th/0607163}{{\tt hep-th/0607163}}.
%%CITATION = HEP-TH 0607163;%%.

\bibitem{Martucci:2006ij}
L.~Martucci, ``{D-branes on general $\mathcal{N}=1$ backgrounds:
  Superpotentials and D-terms},'' {\em JHEP} {\bf 06} (2006) 033,
\href{http://www.arXiv.org/abs/hep-th/0602129}{{\tt hep-th/0602129}}.
%%CITATION = HEP-TH 0602129;%%.

\bibitem{Skenderis:2002wp}
K.~Skenderis, ``{Lecture notes on holographic renormalization},'' {\em Class.
  Quant. Grav.} {\bf 19} (2002) 5849--5876,
\href{http://www.arXiv.org/abs/hep-th/0209067}{{\tt hep-th/0209067}}.
%%CITATION = HEP-TH 0209067;%%.

\bibitem{deHaro:2000xn}
S.~de~Haro, S.~N. Solodukhin, and K.~Skenderis, ``{Holographic reconstruction
  of spacetime and renormalization in the AdS/CFT correspondence},'' {\em
  Commun. Math. Phys.} {\bf 217} (2001) 595--622,
\href{http://www.arXiv.org/abs/hep-th/0002230}{{\tt hep-th/0002230}}.
%%CITATION = HEP-TH 0002230;%%.

\bibitem{Karch:2005ms}
A.~Karch, A.~O'Bannon, and K.~Skenderis, ``{Holographic renormalization of
  probe D-branes in AdS/CFT},'' {\em JHEP} {\bf 04} (2006) 015,
\href{http://www.arXiv.org/abs/hep-th/0512125}{{\tt hep-th/0512125}}.
%%CITATION = HEP-TH 0512125;%%.

\bibitem{Guralnik:2004ve}
Z.~Guralnik, S.~Kovacs, and B.~Kulik, ``{Holography and the Higgs branch of
  $\mathcal{N}=2$ SYM theories},'' {\em JHEP} {\bf 03} (2005) 063,
\href{http://www.arXiv.org/abs/hep-th/0405127}{{\tt hep-th/0405127}}.
%%CITATION = HEP-TH 0405127;%%.

\bibitem{Erdmenger:2005bj}
J.~Erdmenger, J.~Grosse, and Z.~Guralnik, ``{Spectral flow on the Higgs branch
  and AdS/CFT duality},'' {\em JHEP} {\bf 06} (2005) 052,
\href{http://www.arXiv.org/abs/hep-th/0502224}{{\tt hep-th/0502224}}.
%%CITATION = HEP-TH 0502224;%%.

\end{thebibliography}\endgroup

\end{document}